\documentclass[5p,times]{elsarticle}
\usepackage{hyperref}

\biboptions{sort&compress}

\usepackage{subfigure}
\usepackage{amssymb}
\usepackage[figuresright]{rotating}

\usepackage{xspace}
\usepackage{array,tabularx}

\usepackage{lipsum}
\usepackage{multicol}

\newcommand\blfootnote[1]{%
  \begingroup
  \renewcommand\thefootnote{}\footnote{#1}%
  \addtocounter{footnote}{-1}%
  \endgroup
}

\makeatletter
\newcommand{\s}[1]{%
\@ifnextchar[{\s@i{#1}}{\s@i{#1}[{}]}%
}
\def\s@i#1[#2]{%
\@ifnextchar[{\s@ii{#1}[{#2}]}{\s@ii{#1}[{#2}][{}]}%
}
\def\s@ii#1[#2][#3]{%
\ensuremath{\tilde{#1}_{#2}^{#3}}%
}
\makeatother

\usepackage[usenames]{xcolor}

\newcommand{\MET}{\ensuremath{E_{T}^{\mbox{\footnotesize miss}}}\xspace}
\newcommand{\pt}{\ensuremath{p_{T}}\xspace}
\newcommand{\METvec}{\ensuremath{\vec{p}_{T}^{\mbox{\tiny ~miss}}}\xspace}
\newcommand{\ptvec}{\ensuremath{\vec{p}_{T}}\xspace}
\newcommand{\St}{\ensuremath{S_{T}}\xspace}
\newcommand{\Ht}{\ensuremath{H_{T}}\xspace}
\newcommand{\mt}{\ensuremath{m_{T}}\xspace}
\newcommand{\meff}{\ensuremath{m_{\mbox{\footnotesize eff}}}\xspace}
\newcommand{\alphat}{\ensuremath{\alpha_{T}}\xspace}
\newcommand{\chione}{\ensuremath{\tilde{\chi}^0_1}\xspace}
\newcommand{\chitwo}{\ensuremath{\tilde{\chi}^0_2}\xspace}
\newcommand{\chithree}{\ensuremath{\tilde{\chi}^0_3}\xspace}
\newcommand{\chifour}{\ensuremath{\tilde{\chi}^0_4}\xspace}
\newcommand{\chaone}{\ensuremath{\tilde{\chi}^\pm_1}\xspace}
\newcommand{\chaoneT}{\ensuremath{\tilde{\chi}^\mp_1}\xspace}
\newcommand{\Grav}{\ensuremath{\tilde{G}}\xspace}

\newcommand{\SM}{standard model\xspace}
\newcommand{\ATLAS}{ATLAS\xspace}
\newcommand{\CMS}{CMS\xspace}

\newcommand{\second}{\ensuremath{2^{\mbox{\footnotesize nd}}}\xspace}

\newcommand{\Tone}{ {\footnotesize \ensuremath{\tilde{g}\tilde{g}\to q\bar{q}q\bar{q}\tilde{\chi}^0_1\tilde{\chi}^0_1}\xspace}}
\newcommand{\Ttwo}{ {\footnotesize \ensuremath{\tilde{q}\tilde{q}\to q\bar{q}\tilde{\chi}^0_1\tilde{\chi}^0_1}\xspace}}
\newcommand{\Tonett}{ {\footnotesize \ensuremath{\tilde{g}\tilde{g}\to t\bar{t}t\bar{t}\tilde{\chi}^0_1\tilde{\chi}^0_1}\xspace}}
\newcommand{\Ttt}{    {\footnotesize \ensuremath{\tilde{t}\tilde{t}\to X}\xspace}}
\newcommand{\TfiveGG}{ {\footnotesize \ensuremath{\tilde{g}\tilde{g}\to qq\gamma\gamma\Grav\Grav}\xspace}}
\newcommand{\TfiveWG}{ {\footnotesize \ensuremath{\tilde{g}\tilde{g}\to qqW^\pm\gamma\Grav\Grav}\xspace}}
\newcommand{\TChiNg}{ {\footnotesize \ensuremath{\chaone\tilde{\chi}^{0,\pm}_1\to Z\gamma\Grav\Grav+\mbox{soft jets}}\xspace}}
\newcommand{\TChiWg}{ {\footnotesize \ensuremath{\chaone\chione\to W^\pm\gamma\Grav\Grav}\xspace}}
\newcommand{\ttwottwo}{ {\footnotesize \ensuremath{\tilde{t}_2\tilde{t}_2~\mbox{SMS}}}}
\newcommand{\tonetone}{ {\footnotesize \ensuremath{\tilde{t}_1\tilde{t}_1\to ttHZ\Grav\Grav}\xspace}}
\newcommand{\chicha}{ {\footnotesize \ensuremath{\chaone\chaone, \chaone\chitwo}\xspace}}
\newcommand{\sldecay}{ {\footnotesize \ensuremath{\tilde{\chi}^{0,\pm}\to l^\pm\tilde{\nu}, \nu\tilde{l}^\pm}\xspace}}
\newcommand{\Vdecay }{ {\footnotesize \ensuremath{\tilde{\chi}^{0,\pm}\to \chione+W^\pm,Z^0,H}\xspace}}

\newcommand{\Rsfof }{\ensuremath{R_{\mbox{\tiny SF/OF }}}\xspace}

\begin{document}

\begin{frontmatter}





\title{ \vspace{1cm} Experimental Status of Supersymmetry after the LHC Run-I}
\author{Christian Autermann}
\ead{auterman@cern.ch}
\address{I. Phys. Inst. B, RWTH Aachen University, Germany}

%\maketitle
\begin{abstract} 
The \ATLAS and \CMS experiments at the Large Hadron Collider (LHC) at CERN have searched for signals of new physics, in particular for supersymmetry. The data collected until 2012 at center-of-mass energies of $7$~and $8$\,TeV and integrated luminosities of $5$\,fb$^{-1}$ and $20$\,fb$^{-1}$, respectively, agree with the expectation from \SM processes. Constraints on supersymmetry have been calculated and interpreted in different models. Limits on supersymmetry particle masses at the TeV scale have been derived and interpreted generally in the context of simplified model spectra. The constrained minimal supersymmetric \SM is disfavored by the experimental results. Natural supersymmetry scenarios with low supersymmetry particle masses remain possible in multiple regions, for example in those with compressed spectra, that are difficult to access experimentally. The upgraded LHC operating at $\sqrt{s}=13$\,TeV is gaining sensitivity to the remaining unexplored SUSY parameter space.
\end{abstract}
%\begin{keyword}Supersymmetry\sep Limits\sep Searches\sep LHC\sep MSSM\sep BSM
%\end{keyword}

\end{frontmatter}

\blfootnote{\\ \href{http://dx.doi.org/10.1016/j.ppnp.2016.06.001}{http://dx.doi.org/10.1016/j.ppnp.2016.06.001}\\
0146-6410/\copyright\ 2016 Elsevier B.V. All rights reserved.\\[1.5ex]
Published in Progress in Particle and Nuclear Physics 90 (2016) 125-155.}

%%\linenumbers

%\eject
%\pagebreak
%\begin{figure*}
%\begin{minipage}[h]{\textwidth}
 \tableofcontents
 \clearpage
%\end{minipage}
%\end{figure*}
%\begin{@twocolumnfalse}
%\tableofcontents
%\end{@twocolumnfalse}
%\pagebreak

%\linenumbers

\section{Introduction}
\label{sec:introduction}
%% Brief intro
A major motivation for the largest experiment ever built, the Large Hadron Collider (LHC) at CERN, is the search for new physics beyond the \SM. In particular after the \SM (SM) Higgs boson has been discovered~\cite{Aad:2012tfa,Chatrchyan:2012xdj}, the experiments focus on searches for new physics which is expected to explain some of the open questions of the \SM, like the so-called gauge hierarchy problem or the nature of the dark matter in the universe. While the \SM is a remarkably successful theory, it has multiple free parameters with values constrained only by experimental observations. A grand unified theory (GUT) could reduce the number of free parameters by virtue of a larger symmetry.  

In this article the experimental results of searches for supersymmetry (SUSY) at the LHC with the \ATLAS and the \CMS experiments are discussed. Supersymmetry is one of the most popular theories for physics beyond the \SM, and can solve some of the open questions. The implications for the excluded SUSY mass-ranges and the still allowed SUSY phase space regions are reviewed. The experimental methods are summarized, focusing on the published results with the data\-set collected until 2012 at a center-of-mass energy of $8$\,TeV, corresponding to an integrated luminosity of about $20$\,fb$^{-1}$ per experiment. During the writing of this article the first results using data recorded at $13$~TeV become available.

%Brief theoretical intro of SUSY
Supersymmetry~\cite{Ramond,Ramond:1971kx,Golfand,Volkov,Wess:1974tw,Freedman:1976xh,Deser:1976eh,Ferrara:1976fu,Fayet,Kane} is a space-time symmetry developed since the 1970s, relating fermions and bosons. SUSY multiplets contain particles differing in spin by $1/2$, but having otherwise the same properties, as for example the Yukawa coupling $\lambda_S=|\lambda_f|^2$ to the Higgs field. The masses of the superpartners differ, because SUSY is a broken symmetry. The minimal supersymmetric standard model (MSSM)~\cite{PrimerMartin,Feng:2009te} contains chiral supermultiplets, e.g. a spin-$1/2$ fermion and two scalar bosons. The fermion has two spin helicity states, therefore two real scalar bosons have the same number of degrees of freedom. Other vector supermultiplets contain spin-$1$ vector bosons and spin-$1/2$ fermions, both having two helicity states. In extended supersymmetry $N>1$ models the supermultiplets are enlarged. These extended theories are not considered in the following interpretations of the experimental results.

\subsection{The particle content of supersymmetry}
%particle content
The MSSM is minimal with respect to the field content by which the \SM is extended. The particle content in the MSSM more than doubles the number of SM particles. No known particle of the \SM can be the superpartner of another SM particle, and in contrast to the \SM two electroweak Higgs doublets are necessary to keep the theory free of anomalies and to generate the masses of up-type and down-type fermions. In the \SM the masses are generated by Yukawa couplings to the Higgs field $\Phi$ and $\Phi^*$, but complex-conjugate fields are not allowed in the superpotential. Therefore, at least two Higgs doublets $H_u$ and $H_d$ are required in supersymmetric theories, that have together eight degrees of freedom. When the $Z^0$ and $W^\pm$ bosons have acquired mass, the remaining five degrees of freedom generate the spin-$0$ Higgs bosons $h$, $H$, $A$, and $H^\pm$. 

The Higgs bosons and all other particles of the \SM get supersymmetric partner ``sparticles''. The sparticle names refer to the SM partner with prefix ``s'' for bosonic superpartners and suffix ``ino'' for fermionic superpartners. The supersymmetric particle content is summarized in Tab.~\ref{tab:susy}. 
\begin{table*}[htb]
\label{tab:susy}
\begin{center}
\begin{tabular}{rlcrl}
\multicolumn{2}{c}{Gauge eigenstates}&Spin&\multicolumn{2}{c}{Mass eigenstates}\\ \hline ~\\[0.25ex]
bino, wino, higgsinos & $\tilde{B},\tilde{W}^0,\tilde{H}_u^0,\tilde{H}_d^0$ & $\frac{1}{2}$ & $\chione, \chitwo, \chithree, \chifour$ & neutralinos \\[1.5ex]
wino, higgsinos & $\tilde{W}^+,\tilde{H}_u^+$   & $\frac{1}{2}$ & $\tilde{\chi}^+_1,\tilde{\chi}^+_2$ & charginos \\
                & $\tilde{W}^-,\tilde{H}_d^-$   & $\frac{1}{2}$ & $\tilde{\chi}^-_1,\tilde{\chi}^-_2$ &   \\[1.5ex]
gluinos         & $\tilde{g}$                   & $\frac{1}{2}$ & \multicolumn{2}{l}{no mixing} \\[1.5ex]
selectron, smuon, stau  & $\tilde{e}_{L,R},\tilde{\mu}_{L,R},\tilde{\tau}_{L,R}$ & $0$ & $\tilde{e}_{L,R},\tilde{\mu}_{L,R},\tilde{\tau}_{1},\tilde{\tau}_2$ & sleptons \\
sneutrinos      & $\tilde{\nu}_{e},\tilde{\nu}_{\mu},\tilde{\nu}_{\tau}$ & $0$ & $\tilde{\nu}_{e},\tilde{\nu}_{\mu},\tilde{\nu}_{\tau}$    & sleptons or sneutrinos \\[1.5ex]
sup, scharm, stop & $\tilde{u}_{L,R},\tilde{c}_{L,R},\tilde{t}_{L,R}$ & $0$ & $\tilde{u}_{L,R},\tilde{c}_{L,R},\tilde{t}_{1},\tilde{t}_2$  & squarks \\
sdown, sstrange, sbottom & $\tilde{d}_{L,R},\tilde{s}_{L,R},\tilde{b}_{L,R}$ & $0$ & $\tilde{d}_{L,R},\tilde{s}_{L,R},\tilde{b}_{1},\tilde{b}_2$  &  \\[1.5ex]
\end{tabular}
\caption{Supersymmetric partner particles content in the minimal supersymmetric standard model}
\end{center}
\end{table*}
The superpartner gauge eigenstates of each line can mix, i.e. the mass eigenstates are linear combinations of the gauge eigenstates. The neutral gauginos $\tilde{B},\tilde{W}^0$, i.e. the SUSY partners of the \SM U(1) and the neutral SU(2) gauge bosons respectively, and the neutral higgsinos mix to form four neutralinos $\tilde{\chi}^0_{1,2,3,4}$, where the mass increases with respect to the lower index. Similarly, the charginos $\tilde{\chi}^\pm_{1,2}$ are mixings of the charged gauge eigenstates $\tilde{W}^\pm$ and the charged higgsinos. The gluinos $\tilde{g}$ do not mix, and for the sleptons and quarks the mixing of the first and second generation sparticles is usually assumed to be small.  The so-called left- and right-handed third generation squarks, e.g. $\tilde{t}_{L,R}$ where the name refers to the chirality of the \SM spin-$1/2$ partner, mix to form two mass-eigenstates $\tilde{t}_{1}$ and $\tilde{t}_{2}$, and similarly for the sbottom $\tilde{b}_{L,R}$ and the stau $\tilde{\tau}_{L,R}$. 

%gravity
If supersymmetry is imposed as a local symmetry then gravity is naturally included. This constitutes another theoretical motivation for SUSY. In this case, the MSSM can contain the spin-$2$ graviton and its supersymmetric partner the spin-$3/2$ gravitino $\tilde{G}$. The graviton is massless and the gravitational coupling is suppressed by the Planck mass. 

\subsection{The hierarchy problem and the SUSY particle mass scale}
%susy to the rescue
Supersymmetry solves the hierarchy problem associated with the higher order corrections to the Higgs mass-squared parameter $m_H^2$. The spin-$0$ Higgs boson receives quadratically divergent higher-order quantum corrections $\Delta m_H^2$, due to loops of every particle that couples to the Higgs field, e.g. for a scalar with mass $m_S$
\begin{equation}
\label{eq:hierarchy}
\Delta m_H^2 \quad\propto\quad \lambda_S \Lambda_{\mbox{\footnotesize UV}}^2 - 2\lambda_S m_S^2 \ln\left(\frac{\Lambda_{\mbox{\footnotesize UV}}}{m_S}\right).
\end{equation}
The cutoff scale $\Lambda_{\mbox{\footnotesize UV}}$ can be as large as the Planck scale $M_P$, where gravitational effects are no longer negligible. The large difference to the electroweak scale is referred to as the hierarchy problem~\cite{Barbieri198863}. This problem does not arise in supersymmetric theories, because supersymmetry is introduced as a new symmetry between bosons and fermions. The quadratically divergent corrections $\sim \Lambda_{\mbox{\footnotesize UV}}^2$ to the Higgs mass of any supersymmetric multiplet cancel~\cite{Witten1981513,Dimopoulos1981150}.  

%SUSY breaking
Supersymmetry is broken at low energy scales, allowing the masses of the supersymmetric partners to avoid current experimental observations. Soft SUSY breaking mass terms maintain the approximate cancelation of the Higgs mass correction terms. Various supersymmetric models are known where supersymmetry is broken spontaneously. The models used to interpret the experimental results extend the minimal supersymmetric standard model by gravity-mediated breaking terms, referred to as minimal supergravity models (mSUGRA) or as con\-strained-\-MSSM (cMSSM)~\cite{PhysRevLett.49.970,Barbieri,Hall,NILLES19841}. In a second type of models the supersymmetry is broken by gauge-mediated breaking terms (GMSB)~\cite{GGMa,GGMd2,GGMd3,GGMd4,GGMd5,GGMd1,GGMd}. Anomaly-mediated supersymmetry breaking~\cite{Randall:1998uk,Giudice:1998xp} is another breaking scenario, which will not be considered in the following.

%The Hierarchy and the little hierarchy problems
The mass scale of the SUSY particles is generally undetermined by the theory, but can be constrained by the following considerations. The higher-order corrections $\sim\lambda m^2\ln(\Lambda_{\mbox{\footnotesize UV}}/m)$ to the Higgs mass $m_H^2$ depend on the particle masses but only logarithmically on the cutoff scale and thus matter only if the differences in mass of the members of a supermultiplet become too large. This is in particular relevant for the third quark generation supermultiplet of the top with the large top-quark Yukawa couplings at first-loop order, and for gluinos at second-loop order. Heavy third generation squarks and gluinos have large contributions to $\Delta m_H^2$. The non-observation of the Higgs boson near the electroweak scale at LEP was referred to as little hierarchy problem~\cite{Barbieri:2000gf}, as a large Higgs mass requires large radiative corrections and thus large stop masses between $300$\,GeV and $1$\,TeV~\cite{Papucci:2011wy} in the MSSM. 
The supersymmetry naturalness requirement is illustrated~\cite{Papucci:2011wy} by the tree-level relation in the MSSM
\begin{equation}
-\frac{m_Z^2}{2} = |\mu|^2 + m_{H^2_u}
\label{eq:naturalness}
\end{equation}
with the $Z$-boson mass $m_Z$, where the $\mu$-term is directly linked to the higgsino masses, and $m_{H_u}$ to the gluino and stop masses as discussed above. A natural supersymmetry spectrum without fine-tuning requires light stop, gluino, and higgsino masses. A natural supersymmetry scenario should be accessible at the LHC and motivates in particular searches for low-mass superpartners of the top-quark.

\subsection{Supersymmetry models and simplified scenarios}

%R-Parity and dark matter
Another motivation for supersymmetry is, that it can deliver a candidate particle for the dark matter, if the $R$-parity~\cite{FARRAR1978575} is conserved. The baryon number $B$ and lepton number $L$ that are conserved in the SM become in supersymmetry the quantum number $R$-parity, which is defined as $R=(-1)^{3B+L+2S}$ where $S$ is the spin quantum number. $R$ equals $+1$ for \SM and $-1$ for SUSY particles. $R$-parity violating (RPV) trilinear and bilinear terms exists in the superpotential~\cite{Rparity}:
\begin{equation}
\hspace{-8mm}W_{{R}_p\hspace{-2.2mm}\slash} = \frac{1}{2}\lambda_{ijk}L_iL_jE_k^c + \lambda'_{ijk}L_iQ_jD_k^c + \frac{1}{2}\lambda''_{ijk}U_i^cD_j^cD_k^c + \mu_iH_mL_i\quad
\end{equation}
where $L$ and $Q$ are the lepton and quark SU(2) doublet superfields, $E$, $U$, $D$ the singlet superfields, and $i,j,k$ are the family indices. The gauge indices are not shown. The coupling strengths are given by the Yukawa constants $\lambda$, $\lambda'$, and $\lambda''$. The bilinear term allows the mixing of the lepton and Higgs superfields. RPV implies lepton- or baryon number violation and sufficiently large couplings allow the decay of the lightest supersymmetric particle, the LSP. If $R$-parity is conserved because all RPV couplings vanish or at least are sufficiently small, then supersymmetric particles are only produced in pairs, and the LSP is stable. If the RPV-couplings are so small or zero, that the LSP-lifetime is large compared to the age of the universe, then the LSP is a particle candidate for dark matter. The dark matter candidate has to be a massive, only weakly interacting particle, like the lightest neutralino \chione. In the following, RPV couplings are assumed to be zero.

%cMSSM
The minimal supergravity model or the cMSSM is a MSSM scenario with gravity-mediated SUSY breaking determined by five parameters $m_0$, $m_{1/2}$, $A_0$, $\tan\beta$, and the sign of $\mu$. The universal scalar mass $m_0$ determines the mass of the scalar sparticles, i.e. the squarks and the sleptons masses at the GUT scale $M_{\footnotesize GUT}\approx10^{16}$~GeV. The common mass of the gauginos and higgsinos at $M_{\footnotesize GUT}$ is $m_{1/2}$. $A_0$ is the universal trilinear coupling defined at $M_{\footnotesize GUT}$ and $\tan\beta$ is the ratio of the vacuum expectation values of the two Higgs doublets. The absolute value of the higgsino mass parameter $|\mu|$ is determined by the electroweak symmetry breaking, leaving only the sign of $\mu$ as discrete free cMSSM parameter. $A_0$, $\tan\beta$, and the sign of $\mu$ have generally only a small influence on the experimentally observables and are choosen such, that the predicted Higgs mass is consistent with the measurement of approximately $m_H=125$\,GeV. The $m_0$ and $m_{1/2}$ parameters determine the masses and branching ratios of the supersymmetric particles. The relation of the supersymmetric particles masses are therefore constrained by effectively only two degrees of freedom. On one hand this allows for concise comparisons of experimental results in the $m_0$ and $m_{1/2}$ parameter plane, on the other hand many different mass spectra and branching ratios viable in other SUSY models are not examined. Past  $\sqrt{s}=7$\,TeV searches for supersymmetry often used the cMSSM for interpretation of the experimental results. Recently, less constrained models like the phenomenological minimal supersymmetric standard model (pMSSM)~\cite{Djouadi:1998di,CMS-PAS-SUS-2015-10,ATLAS-SUS-2014-08} gained attention. The pMSSM is more difficult to scrutinize and the results are harder to display, because of the much larger number of $19$ free parameters. The theoretical viable phase space of the cMSSM is challenged~\cite{Fittino-2015,Buchmueller:2013rsa} by recent experimental results discussed in the following.

%GMSB
Models of gauge mediated supersymmetry breaking can guarantee flavor universality for the MSSM sfermion masses avoiding the so-called SUSY flavor-problem~\cite{Dimopoulos:1995ju}. The lightest supersymmetric particle is here the gravitino \Grav.  The \Grav is produced in decays of the next-to-lightest SUSY particle (NLSP), which in the studied GMSB scenarios is assumed to be the neutralino \chione, together with a \SM boson $\gamma,Z^0,H$. Prompt decays of the NLSP into the \Grav are assumed in all GMSB scenarios discussed in the following. Non-prompt decays would lead to displaced vertices or heavy stable charged particles, depending on the nature of the NLSP. The coupling to the gravitino is significantly weaker compared to other particles and inverse proportional to its mass. The gravitino mass is negligible in the studied scenarios, typically significantly smaller than $1$\,GeV. Depending on their mass, gravitinos are dark matter candidate particles~\cite{Steffen:2006hw,MOROI1993289}. The direct decay $\chaone\to W^\pm\Grav$ occurs only if the chargino and neutralino masses are almost mass-degenerate and the decay $\chaone\to W^\pm\chione$ is suppressed. The GMSB final state topology depends strongly on the nature of the NLSP and therefore on the neutralino mixing. The SUSY model of general gauge mediation (GGM)~\cite{GGMe,GGMf,Ruderman:2011vv,Kats:2011qh,Kats:2012ym} is used in the following to interpret the experimental results of GMSB inspired searches.

% SMS, Fittino & the end of the cMSSM
In order to characterize the wide range of possible signal scenarios in terms of masses of supersymmetric particles, production channels, and decay modes, i.e. in terms of directly experimentally accessible parameters, so-called simplified model spectra (SMS)~\cite{ArkaniHamed:2007fw,Alwall:2008ag,Alves:2011wf} have been developed. The current experimental results at $\sqrt{s}=8$\,TeV are commonly interpreted using these simplified scenarios~\cite{SMS_ATLAS,SMS_CMS}. The SMS are effective-Lagrangian descriptions of single processes involving just a small number of new particles, to which the analyses have direct sensitivity. These simplified models feature a clear final state topology: All supersymmetry particles which do not directly enter the production and decay chain are effectively decoupled and at high mass scales, in sharp contrast to full models of supersymmetry, that have characteristic complex compositions of processes and final states. The relation of the relevant SUSY masses for the studied process can be chosen freely, which overcomes limitations of full models with few free parameters, like the constrained-MSSM. While the decomposition of a supersymmetry model scenario into different SMS is easily possible, interpreting a combination of different SMS results in any SUSY model is more difficult~\cite{SModelS}. For example, analyses strongly dependent on data-driven background estimation methods can be subject to signal contamination in the control regions, which is usually irrelevant for an individual simplified model scenarios of a single process, but can significantly lower the sensitivity to full model scenarios of many supersymmetry processes.

No excess in the data recorded by \ATLAS or \CMS incompatible with statistical fluctuations is observed. The derived limits obtained by searches for supersymmetry are summarized in this article with the help of simplified models. Exclusion contours derived by different analyses and sometimes by different model assumptions are compared in the summary figures, if the free parameters of the studied models are compatible. The shown analyses and results illustrate the general sensitivity to the most relevant SUSY parameter space. More information and interpretations are available in the quoted references. 
Physics beyond the \SM scenarios other than supersymmetry are discussed for example in~\cite{Golling:2016thc}.

\subsection*{Organization of this article}
%Organisation of this article/paper
The article is organized as follows:  The LHC experiments \ATLAS and \CMS and different experimental aspects relevant for searches for supersymmetry are introduced in section~\ref{sec:detector}.
The spectrum of the various searches for supersymmetry are discussed, from very inclusive to very specific search strategies. In section \ref{sec:inclusive} inclusive search strategies in the all-hadronic final state and with leptons are discussed, together with elementary standard procedures to model the SM background using the data. The results are discussed with respect to the simplified pair production of gluinos and light-flavour squarks and in the constrained MSSM. In section~\ref{sec:gluinomediated} gluino-mediated third generation squark production is discussed, while the direct production of third generation squarks is covered in section~\ref{sec:third}. Electroweak production of supersymmetric particles is presented in section~\ref{sec:weak}, and gauge-mediated supersymmetry breaking scenarios are presented in section~\ref{sec:gauge}. Resonances and signals with kinematic edges are discussed in section~\ref{sec:edge}, before the experimental status of the searches for supersymmetry is summarized in the conclusion in section~\ref{sec:conclusion}.

\section{Experimental methods}
\label{sec:detector}
The analysis strategies differ in the way the \SM background estimation is approached and in the choice of the kinematic variables used to enrich the signal-to-background ratio for the statistical interpretation. In the following the experiments are introduced and then different experimental methods are discussed. Data-driven background estimation techniques will be summarized, which are necessary for the  precise estimation of the \SM background.

\subsection{The ATLAS and CMS experiments at the LHC}

\ATLAS~\cite{ATLAS} and \CMS~\cite{CMS} are multipurpose experiments located at the proton-proton collider LHC. The LHC delivered collision data at a center-of-mass energy of up to $\sqrt{s}=8$\,TeV, corresponding to about $20$\,fb$^{-1}$ of integrated luminosity for each of the experiments. \ATLAS and \CMS have comparable sensitivity and discovery potential, though the detector designs differ in detail. 

The \ATLAS detector includes a silicon pixel detector, a silicon microstrip detector, and a straw-tube tracker that can also provide transition radiation measurements for electron identification. The inner tracking detectors are enclosed in a superconducting solenoid producing a magnetic field of $2$\,T, that allows for precise tracking up to pseudo-rapidities $|\eta|<2.5$. A high-granularity liquid-argon (LAr) sampling calorimeter with lead absorber is used as electromagnetic calorimeter, hadronic showers are measured by an iron/scintillator tile calorimeter and by a LAr calorimeter in the end-caps. The very large muon spectrometer in the magnetic field, provided by three air-core toroidal magnets, consists of a set of fast trigger chambers and high resolution muon chambers for the precise measurement of muon momenta, making use of the large leverage arm, due to the size of the \ATLAS detector.

\CMS in contrast is a rather ``compact'' detector but much heavier compared to \ATLAS, because of the iron return yoke for the magnetic field, which is produced by a single solenoid magnet. The \CMS muon system consisting of high resolution muon drift chambers and fast responsive resistive plate chambers used for triggering is integrated in the iron yoke. The superconducting solenoid magnet delivers a magnetic field of $3.8$\,T enclosing the calorimeter and the tracking system, consisting out of a large silicon strip and the innermost silicon pixel detector. The inner part of the calorimeter system is a lead-tungstate crystal electromagnetic calorimeter, the outer part is a brass-scintillator sampling calorimeter. The inferior energy resolution of the stand-alone \CMS hadronic calorimeter for hadronic jets compared to \ATLAS is compensated by a particle-flow-algorithm~\cite{CMS-PAS-PFT-09-001}, which aims at the best possible use of all detector components, in order to reconstruct the momenta of all identified particles.

\subsection{Kinematic variables for signal selection and background rejection}

%%MET
The classic kinematic variable to search for signals of supersymmetry is the missing transverse energy, defined as the absolute value $\MET=|\METvec|$ of the momentum imbalance of all reconstructed objects in the event:
\begin{equation}
\METvec = -\left(\sum_i^{\mbox{\footnotesize jets}} \ptvec^{~i} + \sum_i^{\mbox{\footnotesize leptons}} \ptvec^{~i} + \sum_i^{\mbox{\footnotesize photons}} \ptvec^{~i} \right).
\label{eq:MET}
\end{equation}
The \MET offers a good separation power between signal and background events, as in supersymmetry the lightest stable particles, e.g. the \chione or the \Grav, can carry away a large amount of energy. The energy of the only electroweakly interacting LSP cannot be detected, resulting in missing transverse energy which is typically larger than the \MET produced in \SM background processes. An excess of events in the high energy tail of a kinematic variable like \MET is a signal expected for supersymmetry, in particular for models that contain a dark matter candidate particle.

Though \MET is a very signal sensitive variable, i.e. the variable is able to separate signal and \SM backgrounds with good power, other kinematic variables have been studied and used in analyses in addition to, or instead of \MET. The precise prediction of the \SM background in the high energy tail of the search variable is of crucial importance for a good signal sensitivity of the search. This is particularly difficult for \MET, because the variable is directly affected by all other objects in the event, as illustrated by Eq.~(\ref{eq:MET}). \MET is  sensitive to multiple effects like detector noise, multiple $pp$-interactions (pile-up), energy depositions not clustered in jets, and the energy resolution of all reconstructed objects most relevantly of jets. The $M_{T2}$, \alphat, Razor, and \meff variables introduced in the following are less sensitive to these effects, but maintain signal sensitivity and signal to background discrimination power, by exploiting different properties of the signal or the \SM background.

%meff
The variable $\meff^{\mbox{\tiny incl}}$ is defined as the scalar sum of the transverse momentum of all jets or leptons in the event and \MET:
\begin{equation}
\meff^{\mbox{\tiny incl}} = \sum_{\mbox{\footnotesize leptons}} p_T^l + \sum_{\mbox{\footnotesize jets}} p_T^j + \MET.
\label{eq:meffincl}
\end{equation}
In the all-hadronic state \meff is equivalent to the variable \St used in other experiments, defined as $\St = \Ht + \MET$.
\Ht is the scalar sum of hadronic energy clustered in jets exceeding a transverse momentum threshold, typically of the order of $\pt>30$\,GeV. The \meff is correlated with the overall mass scale of the hard scattering. The \MET/\meff ratio is useful to remove events, where the \MET is largely due to poorly measured jets.

%MT2
The kinematic variable $M_{T2}$ aims at the mass reconstruction of pair-produced particles that each decay into visible and one invisible particle. The $M_{T2}$ is defined~\cite{mt2_lester,mt2_barr} analogously to the transverse mass \mt, which is used for example for $W$-mass measurements in $W\to l\nu$ events at hadron colliders: 
\begin{equation}
\mt = \sqrt{2 p_T^l\MET\left(1-\cos \Delta\phi(\vec{p}_T^{\ l},\METvec)\right)}.
\label{eq:mt}
\end{equation}
The \MET corresponds in $W\to l\nu$ events to the not reconstructed transverse energy of the neutrino.
Searches for supersymmetry with the $M_{T2}$ variable assume, that the signal events are due to the pair-production of two SUSY particles of the same mass (e.g. $\tilde{g}\tilde{g}$) and each undergo cascade decays into at least one or more jets and the LSP (e.g. a neutralino $\tilde{\chi}^0_1$), as shown in Fig.~\ref{fig:feyn_incl}. The visible objects, such as jets, are clustered into two pseudo-jets using hemisphere algorithms~\cite{hemisphere}.  $M_{T2}$ is defined as the larger of the two transverse masses $M_T^{(1)}$  and $M_T^{(2)}$, where the transverse mass $M_T^{(i)}$ is calculated analog to Eq.~(\ref{eq:mt}), where the $i^{\mbox{\footnotesize th}}$ pseudo-jet replaces the lepton and the corresponding transverse momentum of \chione replaces \MET.
\begin{equation}
M_{T2} = \mbox{min}_{\ptvec^{\chione(1)} + \ptvec^{\chione(2)} = \METvec } \left[ \mbox{max}(M_T^{(1)}, M_T^{(2)}) \right].
\label{eq:mt2}
\end{equation}
The momenta of the invisible particles $\pt^{\chione(1)}$ and $\pt^{\chione(2)}$ are inaccessible, only the sum of \pt is constrained by \MET. The ambiguity is resolved by minimizing $M_{T2}$ over all possible momenta of the undetected particles that fulfill the constraint.

%alpha_T
The \alphat variable~\cite{alphat_randall} aims at the best possible suppression of QCD multijet events, that are characterized by non-genuine \MET. The \MET in the QCD multijet background events is created by jet-resolution effects or non-prompt neutrino production in the hadronization, but not by undetectable particles created in the hard interaction.  In a dijet event the dimensionless variable \alphat is defined as
\begin{equation}
\alphat = \frac{E_T^{\mbox{\footnotesize jet}_2}}{m_T^{\mbox{\footnotesize dijet}}}.
\label{eq:alphat}
\end{equation}
where $E_T^{\mbox{\footnotesize jet}_2}$ is the transverse energy of the less energetic \second jet and $m_T$ is the transverse mass of the dijet system defined analog to Eq.~(\ref{eq:mt}), where the transverse momentum vectors of both jets replace the lepton and the neutrino i.e. the \MET. 
%\begin{equation}
%M_T = \sqrt{ \left(\sum_{i=1}^2 E_T^{\mbox{\footnotesize jet}_i}\right)^2 - \left(\sum_{i=1}^2 p_x^{\mbox{\footnotesize jet}_i}\right)^2 - \left(\sum_{i=1}^2 p_y^{\mbox{\footnotesize jet}_i}\right)^2}.
%\label{eq:mtdijet}
%\end{equation}
In contrast to the missing transverse energy, \alphat is less sensitive to jet energy mismeasurements and therefore to beam conditions and the detector performance. In events with genuine \MET as expected for the SUSY signal the \alphat variable has a long tail with values larger than $0.5$. For QCD dijet events \alphat is constrained to values $\leq 0.5$; the two jets in a perfectly measured event are balanced in $E_T$ and thus $\mt=2E_T$ and $\alphat=0.5$. Any jet energy mismeasurement reduces $E_T^{\mbox{\footnotesize jet}_2}$, while the jet direction in the transverse plane is not and \mt is only minimally affected, leading to $\alphat\leq 0.5$ for QCD dijet events. Multijet events are combined into a pseudo-dijet system using hemisphere algorithms. 

%Razor 
The ``Razor''~\cite{Razor} variables $R^2$ and $M_R$ are motivated by the pair production of two heavy particles such as squarks or gluinos, each decaying to an undetected particle and jets. Multiple jets in each event are combined by hemisphere algorithms into two pseudo-jets. The razor variables are defined as:
\begin{eqnarray}
M_R   &=& \sqrt{ (p^{j_1} + p^{j_2})^2 - (p_z^{j_1} + p_z^{j_2})^2 },\label{eq:mr} \\ 
M_T^R &=& \sqrt{ \frac{\MET(p_T^{j_1} + p_T^{j_2}) - \METvec(\vec{p}^{j_1} + \vec{p}^{j_2}) }{2} },\\
R     &=& \frac{M_T^R }{M_R}. \label{eq:razor}
\end{eqnarray}
$M_T^R$ quantifies the transverse momentum imbalance of the event, while $M_R$ is proportional to the mass-scale of the produced particles in the event. The shape of the \SM backgrounds can be estimated in data sideband regions of $M_R$ and the dimensionless variable $R^2$, where the potential signal contribution is negligible.  The name ``Razor'' refers to the modeling of the SM background, which efficiently enables the separation of signal and the \SM background, and might also be inspired by the first set of $\sqrt{s}=7$\,TeV SUSY searches at CMS nicknamed ``reference analyses'', abbreviated e.g. Ra1.

\subsection{Background estimation techniques using data events\label{sec:inclBKGest}}

The key to searches for supersymmetry is the precise understanding of the various \SM background contributions, especially in the high energy tail of distributions like \MET and at high jet multiplicities. An excess over the \SM expectation in the observed data would indicate a signal of ``New Physics'', such as supersymmetry.

The tails of signal-sensitive distributions like \MET are hard to simulate by Monte Carlo methods for the known \SM backgrounds and therefore carry large or even unknown uncertainties. Rare or unknown detector and reconstruction effects  must be modeled sufficiently correct, like detector responses and noise, efficiencies and resolutions. The common way to avoid or at least reduce the influence of these effects and the corresponding systematic uncertainties, is to make use of the data itself. Data sideband regions, similarly affected by detector and reconstruction effects but depleted of signal can, appropriately weighted, be used instead of simulated samples, such that the systematic effects cancel out. Depending on the variable separating the signal and control regions, the weights for the control events can also be derived from the data: This is usually possible if the variable is uncorrelated to the kinematics of the event, as for example to a certain extent the particle identification criteria like lepton isolation or the $b$-tag probability. If a variable correlated with the kinematics is used to separate the regions, like the jet multiplicity or the hadronic activity \Ht in the event, MC simulations are usually required to derive the normalization of the control events. Depending on the details of the reweighting and the available MC statistics, this is comparable to using Monte Carlo simulation for the modeling of a background in the signal region, that was ``tuned'' by reweighting or validating it in a data sideband region. Exemplary data-driven background estimation methods for the all-hadronic final state are discussed in the following in more detail. 

%gerneral QCD for allhadronic searches
%QCD: R+S
\subsubsection*{The ``rebalance + smear'' method to estimate the QCD multijet background}
The QCD multijet background in the all-hadronic final state is particularly difficult to model with Monte Carlo simulations because of the large production cross section and the multiple but very rare effects that lead to \MET in the reconstructed events. Even though multijet events have no intrinsic \MET, jet energy mismeasurements and misreconstructions summarized as jet energy resolution can lead to a sizable missing transverse energy tail. Typically, high-statistics data sideband regions are employed to model the QCD multijet background. These sidebands can be obtained by loosening a selection such as the \MET requirement, a lepton identification, or for example a b-tagging qualifier. 
Alternatively, jet-resolution measurements derived from $\gamma$+jets and dijets events can be used to create a high-statistics generated ``Pseudo-Monte Carlo'' QCD sample in a data-driven fashion: In a first step data events are selected similar to the signal selection criteria, except for the \MET-cut. The reconstructed jets in every event are then varied in energy according to the jet-resolution probability density function (PDF), such that the \MET in each event vanishes. In a second step, the jets are smeared again according to random numbers drawn from the jet resolution PDFs. Each seed event is used of the order of $100$ times with different random numbers to minimize the statistical uncertainty of the sample. Signal contamination as well as contamination from other \SM backgrounds with genuine \MET to the sample of proxy events is negligible because of the large QCD cross section and because the first rebalancing step transforms the contaminating events into QCD-like events. This ``rebalance + smear''-method predicts the QCD multijet and other fully hadronic \SM backgrounds without genuine \MET with a good precision. 

%ttbar+W: lost-l
\subsubsection*{The ``lost-lepton'' and the ``embedding'' methods to estimate the top and $W$ backgrounds}
An important background to the hadronic final state can arise from \SM backgrounds with single leptons, i.e. electrons or muons including $e$ or $\mu$ from leptonic tau decays. The background composition is dominated by $W$+jets and $t\bar{t}$ events, where the lepton is not identified and the event therefore passes the signal selection including the lepton-veto cut. This happens if the lepton is close to a jet or not-isolated, not properly reconstructed, or out of the geometrical acceptance region.  These ``lost-lepton'' backgrounds are modeled using a data-control sample with exactly one isolated muon. The control sample can be enlarged using isolated electrons in addition to reduce statistical uncertainties at the price of larger systematics. To limit possible signal contamination to the control sample that could lead to an over-prediction, the control events are required to have small $m_T(l,\MET)$, typically less than $100$\,GeV consistent with $W\to l\nu$ from SM $W$+jets or $t\bar{t}$ backgrounds. The control events are weighted according to the lepton \mbox{(in-)efficiencies} and kinematic acceptance factors, which can be obtained from the MC simulation and validated in data on the $Z\to l^+l^-$ peak with a ``tag \& probe'' method. The tag \& probe method utilizes that $Z\to ll$ events can be selected in the data with high purity (``tag''), without applying a specific selection, for which the efficiency should be calculated, on one of the leptons (the ``probe''). The relative precision of the ``lost-lepton'' background estimation is of the order of $10\%$ and up to $40\%$ in large jet multiplicity signal regions.

%ttbar,W hadronic tau: embedding
If the background contains a hadronically decaying tau lepton the event also has a large probability to pass the lepton veto. The same data control sample with one isolated muon is used for the prediction of hadronically decaying taus $\tau_h$ from SM processes like $W$+jets and $t\bar{t}$. The ``embedding method'' corrects the control events for the muon efficiencies and replaces the muon in each event by a simulated $\tau_h$ jet, whose \pt value is randomly sampled from an $\tau_h$ response PDF. For each seed event the tau-jet response function is sampled of the order of one hundred times. \MET, \Ht, and the jet multiplicity are recalculated for all events. 

For supersymmetry searches in the final state with one isolated lepton the lost-lepton and hadronic-tau backgrounds are equally relevant, but here the events originate from the dilepton \SM processes dominated by $t\bar{t}$, $Z$, and diboson production. Accordingly, a dilepton data control sample can be used to model the background.

%Znunu: gamma jets, W->lnu, Z->mumu
\subsubsection*{The ``$Z\to\nu\bar{\nu}$'' estimation method}
An irreducible background for search analyses in the all-\-ha\-dro\-nic final state originates from $Z\to\nu\bar{\nu}$+jets. Three different data control regions can be used to estimate the background: $Z\to l^+l^-$ events offer a straightforward possibility with small systematics. However, since the branching ratio of the $Z$ to neutrinos is roughly three times larger than to electrons and muons, the prediction suffers from large statistical uncertainties especially in the high energy tails of the search variables, where the signal is expected. Better statistical precision is expected from $W\to l\nu$ events, which can be used as an alternative control sample. But in this case a clean selection without contamination is more challenging, because the background from other \SM processes or possible signal events is larger than for dileptons in the narrow invariant mass window around the $Z$ mass. Commonly used is therefore a $\gamma$+jets control sample exploiting the similarity to $Z$+jets, which is given at high boson transverse momenta. The total uncertainties are dominated by the theoretical uncertainties on the $\gamma$+jets/$Z$+jets cross section ratio.  The uncertainties can be constraint using $Z\to\mu\mu$ data. The relative prediction precision is of the order of $25\%$ at low and $45\%$ at high jet multiplicities.

\subsection{Limit calculation}
%\subsection{Monte Carlo simulation}
%Using the production cross section calculated to next-to-leading-order plus next-to-leading-logarithm accuracy, and

Upper limits on the amount of signal events, that could be present in the observed data under the assumption that the data are statistically distributed according to the \SM background-only expectation, are derived using the CLs method \cite{Junk,Read}. The CLs method, often called modified frequentist approach, is designed to avoid the exclusion of possible signals, to which the analysis is not really sensitive to. A statistical under-fluctuation in the data compared to the background-only expectation $b$ could lead to an exclusion of zero or even a negative amount of signal events $s$ at a certain confidence level. To avoid this unwanted behavior, the $\mbox{CL}_s$ is defined as
\begin{equation}
\mbox{CL}_s = \frac{\mbox{CL}_{s+b}}{\mbox{CL}_b}.
\label{eq:cls}
\end{equation}
The $\mbox{CL}_{s+b}$ is the probability to observe a test statistic value $Q^s$ at least as signal-like as the one observed $Q_{\mbox{\footnotesize obs}}^s$ under the signal+background test hypothesis $H_1$ for a certain signal expectation $s$. $\mbox{CL}_b$ is analogously the probability to observe $Q^0\geq Q_{\mbox{\footnotesize obs}}^0$ under the background-only null-hypothesis $H_0$. For the test statistics $Q$ typically a likelihood-ratio is chosen, which allows for the most efficient separation of both hypotheses~\cite{NeymanPearson}. Ignoring systematic uncertainties, the test statistic can be defined as the ratio of two Poisson probability functions
\begin{equation}
Q = \frac{\mbox{Poisson}(x,s+b)}{\mbox{Poisson}(x,b)}
\label{eq:q}
\end{equation}
where the measurement $x$ can be the observed data or the outcome of a pseudo-experiment. The test statistics of multiple exclusive signal search regions can be combined to a single function, as likelihoods are multiplicative. Different ways to incorporate systematic uncertainty nuisance parameters are used by the discussed analyses. The standard limit calculation procedure procedure is defined by the \ATLAS and \CMS Higgs boson search combination~\cite{CMS-NOTE-2011-005}. 

A signal of at least $s$ events is excluded at a confidence level (CL) of $1-\alpha$, if the $\mbox{CL}_s\leq \alpha$. The resulting limits and exclusion contours presented in the following have been derived at $95\%$~CL. The expected limit is defined by the $50\%$ quantile of calculated limits, where the observed data event yield is replaced repeatedly by random pseudo-data event yields distributed according to the background-only expectation. 

For a single-channel counting experiment the limit is generally independent of any signal hypothesis and can be translated into a cross section limit using the signal acceptance and the data luminosity, i.e. a limit with respect to a signal cross section expectation for a specific point of a given supersymmetry model, defined by a set of model parameters. Multi-channel counting experiments implicitly depend on the relative signal acceptance in the different search bins and depend therefore on the signal model. Exclusion contours in a plane of the supersymmetry model parameters are derived, by comparing the calculated limits to the signal expectations per point. In this article, only limits with respect to the supersymmetric particle masses in specific models or signal scenarios are discussed. The cross section limits are not shown, but can be found in the quoted references.

\section{Inclusive searches for strong production of gluinos and first or second generation squarks}
\label{sec:inclusive}
Generally, inclusive searches for supersymmetry have the largest sensitivity in the studied supersymmetric models and with respect to the wide range of supersymmetric particle masses and cross sections~\cite{Beenakker:1996ch,Kramer:2012bx}. However, more specialized searches extend the reach significantly in relevant corners of the phase space, as will be discussed in the following sections. Inclusive searches make few model assumptions, typically only missing transverse energy \MET or a similar correlated quantity is required in the selected events. 

\begin{figure*}[tb]
%\begin{center}
\hspace*{\fill}%
\subfigure[$\tilde{g}\tilde{g}\to q\bar{q}q\bar{q}\chione\chione$]{
\includegraphics[scale=1.0]{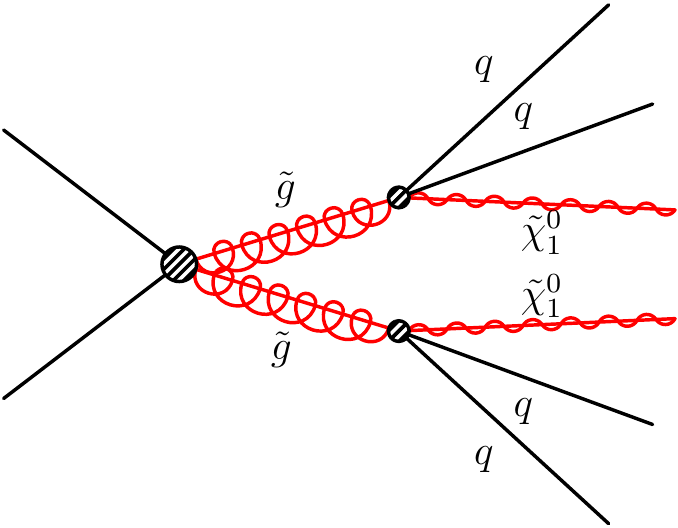}%incorrect filename
\label{fig:feyn_glgl_qqqqchichi}
}
\hfill
\subfigure[$\tilde{q}\tilde{q}\to q\bar{q}\chione\chione$]{
\includegraphics[scale=1.0]{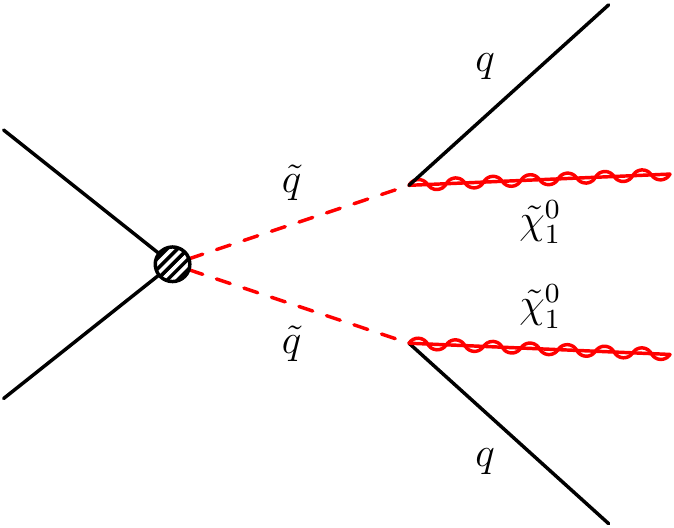}%incorrect filename
\label{fig:feyn_sqsq_qqchichi}
}
\hspace*{\fill}%
\caption{Effective Feynman diagrams for the simplified pair production of gluinos (a) and first- or second generation squarks (b).}
 \label{fig:feyn_incl}	
%\end{center}
\end{figure*}

The \MET requirement, or any other similar kinematic property exploiting the non-observation of two LSPs with high energy,  implicitly constrains the analysis sensitivity to R-parity conserving models with a stable and electrically neutral, thus only weakly interacting lightest supersymmetric particle. However, also $R$-parity violating SUSY scenarios can be probed, that lead either through neutrinos from non-zero $LL\bar{E}$ $\lambda_{ijk}$ or $LQ\bar{D}$ $\lambda'_{ijk}$ couplings to \MET in the final state, or lead to \MET because the RPV couplings are sufficiently small so that the LSP decays unobserved outside the detector. Squarks and gluinos are generally produced at higher energy scales compared to \SM processes because of the larger particle masses due to the existing limits~\cite{Feng:2009te}. This typically leads to either long decay chains with many SM particles as for example for $\tilde{g}\tilde{g}$ production as shown in Fig.~\ref{fig:feyn_glgl_qqqqchichi} or to few high energy particles and therefore also large \MET as for $\tilde{q}\tilde{q}$ production as shown in Fig.~\ref{fig:feyn_sqsq_qqchichi}. Because of the large hadronic branching ratio, the final state of strongly-produced supersymmetry events consists generally of many jets leading to good sensitivities for searches in the all-hadronic final state, which comes at the price of a more difficult to estimate \SM background. 

\subsection{All-hadronic final state\label{sec:inclQCD}}

%\cite{ATLAS-SUS-2013-02}% Strong / Inclusive
% 0 leptons + 2-6 jets + Etmiss [Incl. squarks &amp; gluinos],  05/2014
% https://atlas.web.cern.ch/Atlas/GROUPS/PHYSICS/PAPERS/SUSY-2013-02
\ATLAS~\cite{ATLAS-SUS-2013-02,ATLAS-SUS-2013-04} as well as \CMS~\cite{CMS-SUS-2013-12} have searched for SUSY in the fully hadronic final state with the classical \MET variable, no leptons, and at least two jets. 

The $\MET+\Ht$ search~\cite{CMS-SUS-2013-12} at \CMS relies on the ``rebalance + smear''-method to predict the QCD multijet background in the \MET tails. A precision of $10-90\%$ depending on the signal region can be achieved. The method has been validated in QCD enriched data control regions as well as on Monte Carlo simulation. The backgrounds from $t\bar{t}$ and $W$+jets, where at least one lepton fails identification are estimated using the ``lost lepton'' method. The method is validated with Monte Carlo simulation and the precision of the prediction is typically $10-20\%$, depending on the jet multiplicity. The ``embedding method'' is used to estimate the contribution from hadronically decaying taus. For the irreducible $Z\to\nu\bar{\nu}$ process $\gamma$+jets events are used to model the background. The \MET, \Ht, and the jet multiplicity variables are used to define $36$ exclusive signal regions, in order to optimize the sensitivity to the different regions of the signal parameter space for gluino or squark pair or associated production. High jet multiplicity bins are for example more sensitive to $\tilde{g}\tilde{g}$ production, while $\tilde{q}\tilde{q}$ processes produce a smaller particle multiplicity, but more \MET. The individual counting experiments per signal region are combined into one test statistic. The characteristic feature of the analysis is the careful estimation of all \SM backgrounds, using only data-driven methods that are validated in data control regions as well as with Monte Carlo simulation.

The \MET and $2-6$ jets \ATLAS analysis~\cite{ATLAS-SUS-2013-02} follows a slightly different strategy compared to the 
previously discussed search, where the statistically exclusive signal regions were combined in a final likelihood fit. Here, the $15$ signal regions are defined inclusively, i.e. containing at least 2,3,4,5, or 6 jets and are further divided into inclusive loose, medium, tight, or $W$-candidate selections according to the variables $\Delta\Phi_{\mbox{\footnotesize min}}$,\linebreak $\MET/\sqrt{\Ht}$, \meff, and \MET/\meff. Each region is optimized to target a specific scenario depending on the production scenario and the particles masses m($\tilde{g}$), m($\tilde{q}$). The \SM backgrounds for each signal region are estimated from data sideband regions scaled by transfer factors estimated from Monte Carlo simulation except for the QCD multijet background. The QCD background contribution to the signal regions is small, but the total uncertainties on the prediction are still relevant. The QCD control region is obtained by inverting the $\Delta\Phi_{\mbox{\footnotesize min}}>0.2-0.4$ cut, which depends on the signal region, and the requirement on $\MET/\sqrt{\Ht}$ or \MET/\meff. The transfer factors are determined by a data-driven technique applying a jet energy resolution function to estimate the impact of a jet mismeasurement. The systematic uncertainties of the QCD background prediction are dominated by the uncertainties on the estimated transfer factors and are smaller than $8\%$ for all signal regions. The total uncertainty of the combined background estimation is between $8\%$ and $61\%$. The results from the $15$ signal regions are combined, by using the result from the signal region with the most stringent expected sensitivity to set a limit on a given signal parameter point. 

%\cite{ATLAS-SUS-2013-04}% Strong / Inclusive, 0l 7-10 jets
% 0 leptons + &gt;=7-10 jets + Etmiss [Incl. squarks &amp; gluinos],  08/2013
% https://atlas.web.cern.ch/Atlas/GROUPS/PHYSICS/PAPERS/SUSY-2013-04
The \ATLAS search using 7-10 jets~\cite{ATLAS-SUS-2013-04} and up to two $b$-tags aims at longer supersymmetry decay chains and complements the inclusive analysis with $2-6$ jets~\cite{ATLAS-SUS-2013-02}. The analysis targets gluino mediated squark production and has also good sensitivity to gluino mediated stop production as discussed in the section~\ref{sec:gluinomediated} and to gluino mediated chargino \chaone or neutralino \chitwo production as discussed in the following section~\ref{sec:incl_leptons}, that all lead to many jets in the final state. The sensitivity of the analysis is enhanced by the sum of masses $M^\Sigma_{\mbox{\footnotesize jets}}$ of large radius jets in the event. These large radius jets are clustered by an anti-$k_t$ algorithm with a distance parameter $R=1$ from the four-momenta of $R=0.4$ anti-$k_t$ jets with $\pt>20$\,GeV. The jet multiplicity, the $b$-tagged jet multiplicity, $M^\Sigma_{\mbox{\footnotesize jets}}$, as well as the $\MET/\sqrt{\Ht}$ variables are used to define $19$~partially overlapping signal regions. For the QCD multijet background estimation the observation that the \MET resolution is proportional to $\sqrt{\Ht}$ independently of the jet multiplicity is exploited: The $\MET/\sqrt{\Ht}$ shape is extracted from data sidebands in the jet multiplicity, for each bin of $b$-tag multiplicity as the shape depends on the number of $b$-tags in the event. Soft unclustered energy in the event distorts the proportionality and leads to the largest systematic uncertainty on the QCD background estimation.

SUSY searches relying on the kinematic variables $M_{T2}$, \alphat, and the Razor variable $R^2$ and $M_R$ have been pursued. The general advantage of these analyses is the reduced dependency on pileup, unclustered energy, and other detector effects that influence the \MET in an event. Especially QCD multijet background events can acquire \MET through these effects and are difficult to estimate. The alternative kinematic variables therefore offer a better handle to reduce or to estimate QCD multijet events. Disadvantages arise from the more complicated definition and used assumptions compared to the \MET variable which is clearly defined for all signals and \SM backgrounds. 

% QCD for MT2
The $M_{T2}$ variable explored at ATLAS and CMS~\cite{CMS-SUS-2013-19,ATLAS-SUS-2014-07} allows to control the QCD multijet contribution, which is small at large values of $M_{T2}$. The amount of QCD background is further reduced by requiring $\Delta\Phi_{\mbox{\footnotesize min}}>0.3$, where $\Delta\Phi_{\mbox{\footnotesize min}}$ is the minimal azimuthal angle between the \METvec and any of the four leading jets in \pt, again exploiting the origin of \MET from jet energy mismeasurements. The surviving QCD background is estimated from the $\Delta\Phi_{\mbox{\footnotesize min}}<0.3$ control region, scaled according to a transfer factor depending on $M_{T2}$. Two analysis strategies are pursued for the all-hadronic final state at CMS~\cite{CMS-SUS-2013-19}: An inclusive search for supersymmetry spans the signal region by the $M_{T2}$ variable, the jet and the $b$-tagged jet multiplicity, and by \Ht. The second approach makes use of $M_{T2}$ and the invariant dijet mass of $b$-tagged jets $m(b\bar{b})$ aiming at the reconstruction of a light Higgs boson produced in supersymmetry cascade decays.

The \alphat search for supersymmetry~\cite{CMS-SUS-2012-28} uses a different kinematic variable to identify possible signal events and to efficiently suppress the QCD multijet background. QCD events do not have genuine \MET, so that the measured \MET is mostly due to jet resolution effects, resulting in $\alphat\lesssim0.5$ for QCD multijet. Detector effects do not significantly influence \alphat in an event, so that the QCD background can be efficiently removed by a $\alphat>0.55$ selection requirement. The signal region is further categorized by the \Ht, the jet multiplicity, and the $b$-tagged jet multiplicity to maximize the analysis sensitivity to wide range of possible signals of supersymmetry like $\tilde{g}\tilde{g}$, $\tilde{g}\tilde{q}$, and $\tilde{q}\tilde{q}$ production and in particular also those with third generation squarks, that lead to $b$-jets. The analysis determines the sum of the remaining \SM background from QCD multijet, $t\bar{t}$, $W$, and $Z$ processes by a binned simultaneous likelihood fit to event yields in the signal regions and in $\mu+$jets, $\mu\mu+$jets, and $\gamma+$jets control regions. Data collected in the first half of the $8$~TeV data taking period corresponding to an integrated luminosity of $11.7$~fb$^{-1}$ was used.

Searches for new physics relying on the shape of the ``Razor''~\cite{Razor} variables $R^2$ and $M_R$ have been carried out at the \ATLAS~\cite{ATLAS_7TEV_Razor,ATLAS-SUS-2013-20} and the CMS experiments~\cite{CMS-SUS-2013-04,CMS-PAS-SUS-2014-11}. The final state with at least one  $b$-tagged jet and with and without leptons has been analyzed at CMS using the full $8$~TeV data set as reported in~\cite{CMS-SUS-2013-04}. The \ATLAS inclusive search with at least one lepton~\cite{ATLAS-SUS-2013-20}, using also the Razor variables is discussed below in the following section~\ref{sec:incl_leptons}.

%Discuss gl-chi and sq-chi exclusion contours: 8 degen. quarks, mention other contours. Discuss cMSSM, quote Fittino best fit, and other interpretations. Mention interpretations also for third gen searches.
\begin{figure*}[tb]
\subfigure[]{\includegraphics[width=0.5\textwidth]{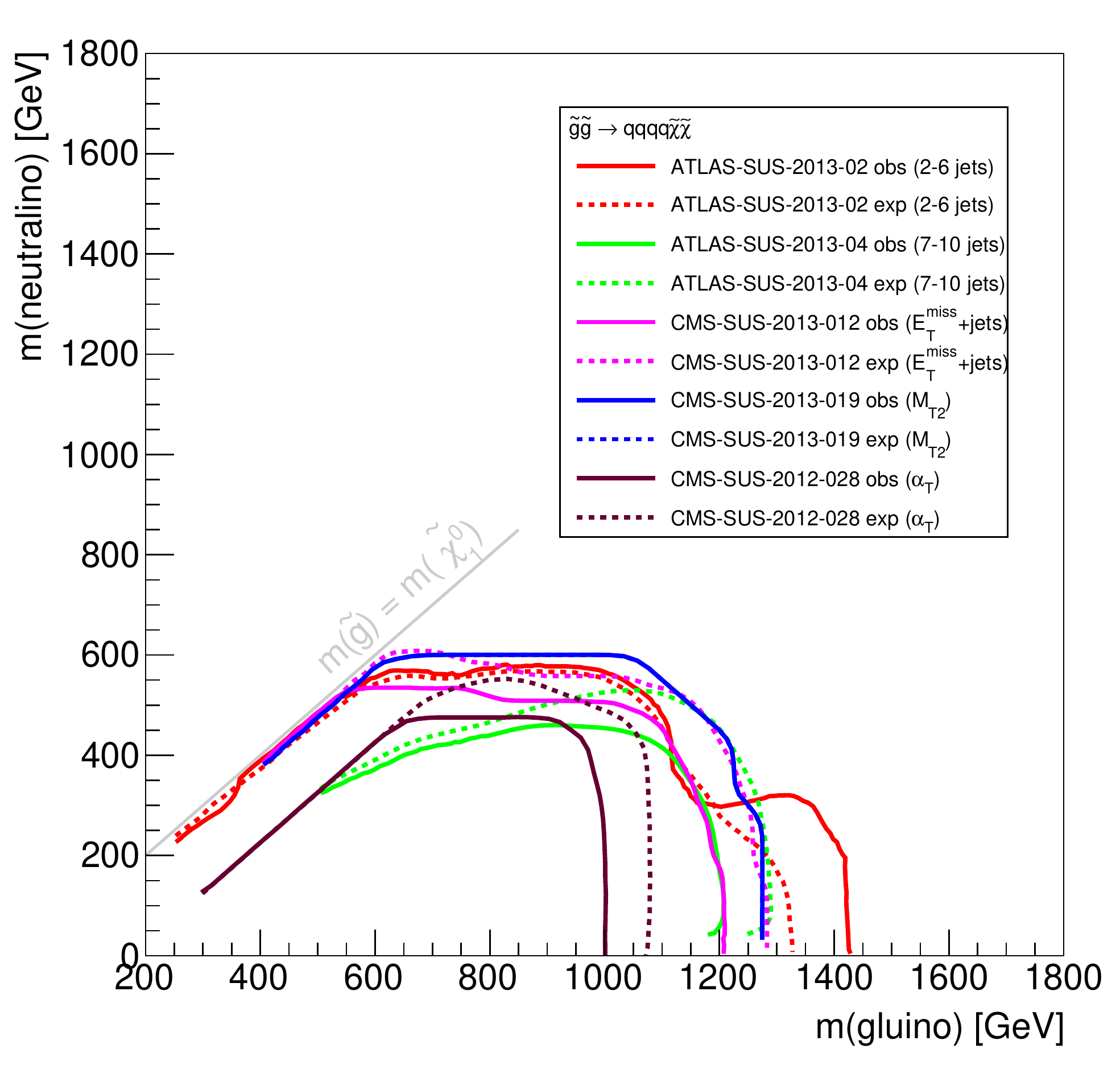}}
\subfigure[]{\includegraphics[width=0.5\textwidth]{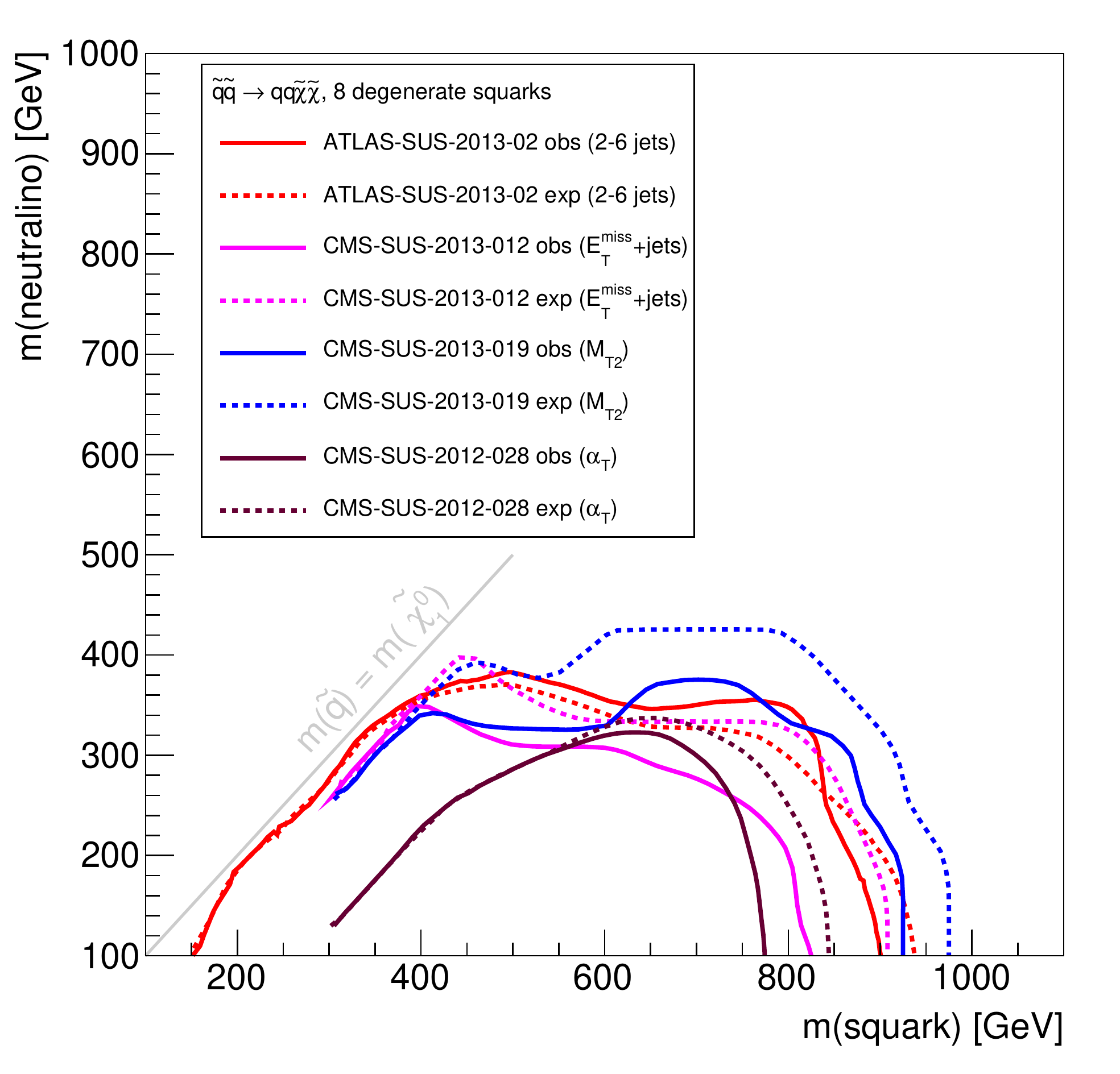}}
  \caption{Exclusion contours at $95\%$~CL for the inclusive searches for strong production of supersymmetry signal events of \ATLAS and \CMS.  Gluino pair production is shown in (a), where the gluino decay through off-shell squarks as $\tilde{g}\to qq\chione$. Squark pair production asssuming eight mass degenerate light flavor squarks is shown in (b), where the squarks decay as $\tilde{q}\to q\chione$. The analysis labels refer to Table~\ref{tab:overview1}. }
 \label{fig:inclT1T2}	
\end{figure*}

The inclusive searches for supersymmetry in the all-hadronic final state are interpreted in the simplified models of squark- or gluino-pair production as shown in Fig.\ref{fig:feyn_incl}. For the on-shell squark-pair production case eight mass-degenerated light squarks $\tilde{u}_{L,R}$, $\tilde{d}_{L,R}$, $\tilde{c}_{L,R}$, $\tilde{s}_{L,R}$ are assumed, that each decay into a quark and the neutralino LSP. The squark mass and the neutralino mass are varied for each point of the generated signal Monte Carlo simulation scan. In the case of the gluino-pair production scenario all squark masses are assumed to be decoupled at high energy scales. Each gluino undergoes an effective three-body decay into two quarks and the neutralino LSP. Again, two parameters define all experimental observables; the gluino mass and the neutralino mass. All hadronic inclusive searches follow different search and background estimation strategies with different strengths, setting the strongest limits in the $\tilde{g}\tilde{g}$ and the light-flavor $\tilde{q}\tilde{q}$ scenarios. The \MET and $2-6$ jets analysis as well as the \alphat and $M_{T2}$ analyses are expected to perform better for squark-pair production, where lower jet multiplicities but more energetic jets are produced, leading to an optimal performance of the hemisphere algorithms, compared to the $7-10$ jets and the $\MET+\Ht$ analyses, that perform best for gluino-pair production with many jets in the final state. In the cMSSM this corresponds to low values of the universal scalar mass $m_0$, where squark-pair production is dominant because m($\tilde{q}$)$<$m($\tilde{g}$) and high values of $m_0$ where m($\tilde{g}$)$\ll$m($\tilde{q}$) leading to dominant $\tilde{g}\tilde{g}$ production. The resulting limits in the cMSSM are compared in the following section~\ref{sec:incl_leptons} to the sensitivity of the inclusive leptonic searches.

The results of the all-hadronic searches for supersymmetry~\cite{ATLAS-SUS-2013-02,ATLAS-SUS-2013-04,CMS-SUS-2013-12,CMS-SUS-2013-19,CMS-SUS-2012-28} are translated into cross section limits in the $\tilde{g}\tilde{g}$ and light-flavor $\tilde{q}\tilde{q}$ pair-production.  Model-dependent exclusion contours, as shown in Fig.~\ref{fig:inclT1T2} for the simplified pair-production are derived, by comparing the cross section limits to the signal cross section prediction in the corresponding model. For the simplified model of gluino-pair production with $\tilde{g}\to qq\chione$ gluinos masses up to $1.4$\,TeV and neutralino masses up to $600$\,GeV are probed. Similarly, for the simplified model of squark-pair production with $\tilde{q}\to q\chione$ squark masses up to $900$\,GeV and neutralino masses up to $350$\,GeV are probed. The squark mass limits are weaker compared to the gluino mass limits. In the diagonal region of Fig.~\ref{fig:inclT1T2}, where the gluino or squark masses become almost mass-degenerate with the neutralino LSP mass, only little hadronic energy in the form of soft hadronic jets is created in the gluino or squark decays. The analyses acceptance drops, depending on the details of jet energy and multiplicity requirements. Uncertainties from initial state radiation and parton density functions for \SM background and signal Monte Carlo simulations are relevant in this region of phase space.

\subsection{Inclusive searches with leptons\label{sec:incl_leptons}}

\begin{figure*}[tb]
\hspace*{\fill}%
\subfigure[$\tilde{g}\tilde{g}\to q\bar{q}q\bar{q}WWZ\chione\chione$]{
\includegraphics[scale=1.0]{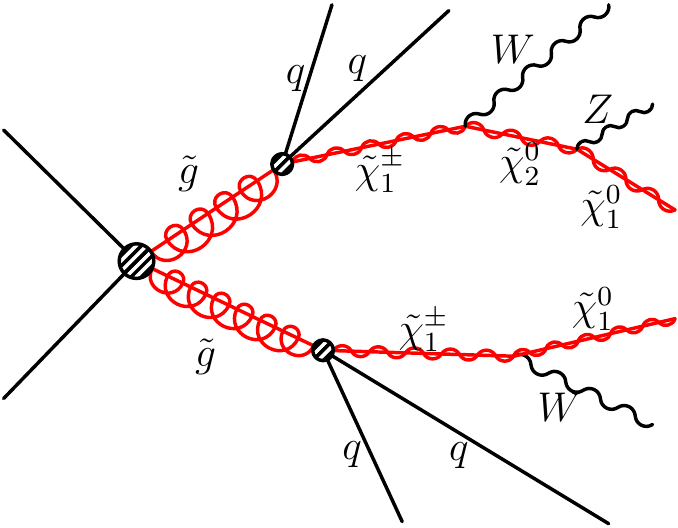}
}\hfill
\subfigure[$\tilde{q}\tilde{q}\to q\bar{q} q\bar{q}llll\chione\chione$]{
\includegraphics[scale=1.0]{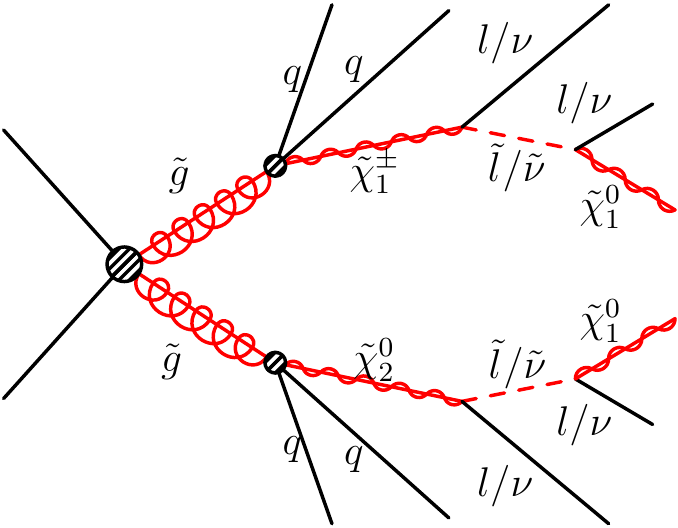}
}
\hspace*{\fill}%
  \caption{Effective Feynman diagrams for gluino mediated production of gauginos leading to quarks and $W^\pm$, $Z^0$ bosons in the final state (a) and gluino decays through gauginos and sleptons leading to leptons and/or neutrinos in the final state (b).}
 \label{fig:feyn_inclWZ}	
\end{figure*}

The additional requirement of leptons in the final state restricts the analyses to signal scenarios with longer decay chains, where leptons are created through sleptons $\tilde{l}^\pm$, $\tilde{\nu}$ or through vector boson $Z^0$, $W^\pm$ decays, as shown in Fig.~\ref{fig:feyn_inclWZ}. Also third generation squark production discussed in the following section can lead to lepton final states through top-quark decays. Direct electroweak slepton production offers usually no competitive sensitivity on slepton or gaugino masses compared to the indirect production through gluino- or squark decays, if the gluino or a squark mass is sufficiently small, or the slepton masses are not too light~\cite{Beenakker:1996ch,Kramer:2012bx}.  If the signal contains leptons, additional sensitivity compared to the inclusive all-hadronic analyses can be gained by the explicit selection of leptons. The branching fraction of neutralinos and charginos to leptons is usually smaller than to jets, but this is also true for the \SM background and a lepton in the final state simplifies the background estimation, because more possibilities for kinematically identical data sideband regions for the modeling of a background or for validation are offered.

Combining all-hadronic analyses and searches with leptons in the final state from the same experiment is, unfortunately, often difficult if the analyses were not designed with a combination in mind from the beginning. The reason for this is the non-trivial correlation of systematic uncertainties and most importantly the partially overlapping signal and control regions. Nevertheless, several \ATLAS and \CMS analyses have combined search channels and have been reinterpreted in several models of new physics beyond the \SM, e.g.~\cite{ATLAS-SUS-2014-06,CMS-SUS-2013-04}.

%% 1l %% vvvvvvvvvvvvvvvvvvvvvvvvvvvvvvvvvvvvvvvvvvvvvvvvvvvvvvvvvvvvvvvvvvvvvvvvvvvvvvvvvvvvvvvvvvvvvvvvvvvvvvvvvvvvvvv

%\cite{ATLAS-SUS-2013-20}% Strong Inclusive, 1-2 leptons
% 1/2 leptons + jets + Etmiss [incl. squarks &amp; gluinos, mUED], 1/2015
% https://atlas.web.cern.ch/Atlas/GROUPS/PHYSICS/PAPERS/SUSY-2013-20/

A search for supersymmetry with at least one isolated electron or muon, jets, and \MET~\cite{ATLAS-SUS-2013-20} has been carried out by the \ATLAS experiment on a dataset of $20$\,fb$^{-1}$. Different signal regions with optimal sensitivity to different signal scenarios have been defined: At least one low \pt or one high \pt lepton selection, $e$ or $\mu$, offer complementary sensitivity to low and large mass splittings. Dilepton selections, $ee$, $e\mu$, or $\mu\mu$, cover different signal production modes and cascade decay chains. Various further criteria are applied individually in the signal regions, optimized for the different signal models. For example, small jet multiplicities are required for squark pair-production, large multiplicities for gluino production (Fig.~\ref{fig:feyn_incl}). For gluino pair production dileptons are expected through decays via sleptons and sneutrinos or via intermediate gaugino decays as shown in Fig.~\ref{fig:feyn_inclWZ}.  The dominant background originates from $t\bar{t}$ production. In the dilepton channels the SM $t\bar{t}$ is suppressed by applying a veto on $b$-tagged jets.
The transverse mass \mt, defined in Eq.~(\ref{eq:mt}) is used in all single-lepton regions to reject the sub-dominant background from $W\to l\nu$. Similarly, the invariant dilepton mass $m_{ll}$ is used to reject $Z\to l^+l^-$ in dilepton regions, where no on-shell $Z^0$-bosons are expected in the signal.
The signal regions are defined orthogonal, except for the specifically designed inclusive single-lepton and low-\pt dilepton regions. 
The binned variables $\meff^{\mbox{\tiny incl}}$, \MET, $\MET/\meff$, and the Razor variable $M_R$, defined in Eq.~(\ref{eq:mr}), are used to exploit the expected signal shape in order to optimize model-dependent limits. 
The dominant \SM backgrounds from $t\bar{t}$, $W$+jets, and $Z$+jets are modeled by Monte Carlo simulation, which are normalized for the numerous signal regions in kinematically similar control regions, obtained e.g. by inverting the $b$-tag requirement, the $m_T$, or the $m_{ll}$ selection. The normalization factors are constrained by a simultaneous fit based on the profile likelihood method to all control regions per signal region and are checked in multiple validation regions.
Another non-negligible background to signal-regions with leptons originates from QCD-multijet and $Z(\to\nu\nu)$+jets events, where the lepton requirement is fulfilled by non-prompt leptons or misidentified jets. This ``fake-lepton'' background can be estimated from the data, by measuring the lepton fake-rate $f_{j\to e}$ in events with loosely identified leptons and applying it to an appropriately weighted data control sample with the same requirements than the signal selection, except for the lepton identification. Since the fake-rate is usually orders of magnitude smaller than the lepton identification efficiency, the data-driven background estimation has small uncertainties of statistical origin due to the much larger control sample. The total uncertainties of the estimation is dominated by the sytematical uncertainty of $f_{j\to e}$ and on the quality of the kinematic similarity of the signal and the loose-lepton control region.

%% >2l %% vvvvvvvvvvvvvvvvvvvvvvvvvvvvvvvvvvvvvvvvvvvvvvvvvvvvvvvvvvvvvvvvvvvvvvvvvvvvvvvvvvvvvvvvvvvvvvvvvvvvvvvvvvvvvvv
%\cite{CMS-SUS-2013-13}%Strong Inclusive, 2l same sign, Third
%Search for new physics in events with same-sign dileptons and jets in pp collisions at 8 TeV, https://twiki.cern.ch/twiki/bin/view/CMSPublic/PhysicsResultsSUS13013

A slightly different approach is used by the like-sign (LS) dilepton analysis~\cite{CMS-SUS-2013-13} at the CMS experiment. Like-sign lepton final states are very rare in the \SM, but occur naturally in supersymmetry since, for example, the two decays chains of pair produced gluinos are not correlated. The dominant SM background therefore arises from misidentified ``fake leptons'', which are estimated from the data with loose lepton identification criteria. Rare \SM processes yielding like-sign dileptons like diboson production, $t\bar{t}V$, and $HV$, where the $V$ denotes a vector boson $Z^0$ or $W^\pm$, are estimated from Monte Carlo simulation. Four variables, that are able to discriminate signal against \SM backgrounds, are used to define several exclusive signal regions: \MET, the scalar sum of hadronic transverse energy clustered in jets \Ht, the jet multiplicity, and the $b$-tagged jet multiplicity. For each variable two sets of regions are defined; one loose selection with depleted signal contribution in order to validate the SM background description, and one with tighter requirements where exclusive signal bins are defined. The analysis is carried out with two different \pt thresholds for the two leptons of either $10$\,GeV each, or with thresholds of $20$\,GeV for both leptons. The low-\pt selection increases the sensitivity to SUSY models with off-shell $W$-bosons. The \Ht threshold in this case is increased from $200$\,GeV to $250$\,GeV to ensure the full efficiency of the trigger. The high-\pt selection is partially overlapping but has complementary sensitivity to scenarios with on-shell bosons. In addition to the results interpreted in the signal scenarios discussed in this section, the analysis uses the higher $b$-tag multiplicity regions and a low \MET signal region to set limits on third generation squark production and $R$-parity violation models, respectively.

\begin{figure*}[bt]
\subfigure[]{\includegraphics[width=0.5\textwidth]{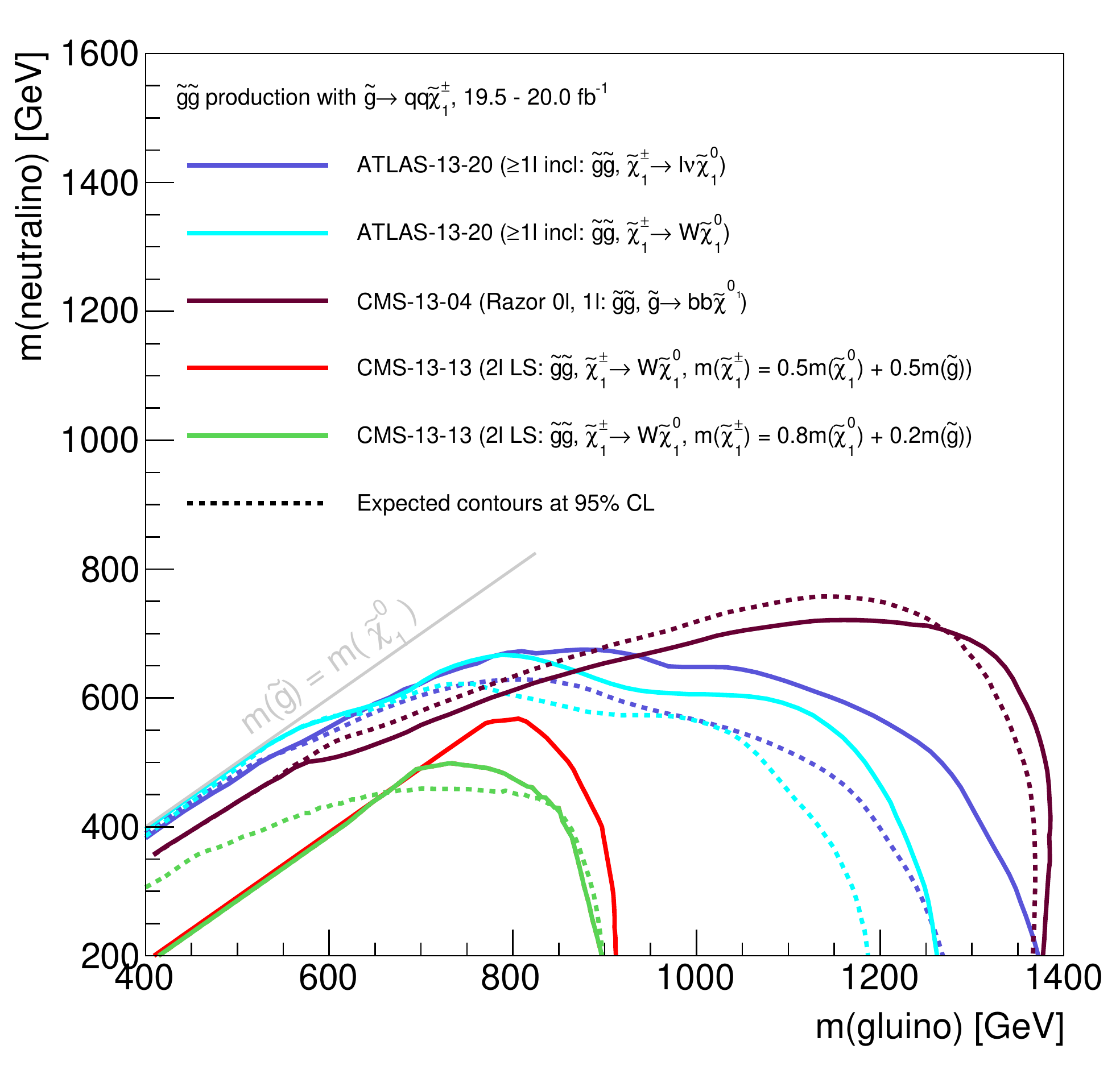}\label{fig:gl_chi_glgl-leptons}}
\subfigure[]{\includegraphics[width=0.5\textwidth]{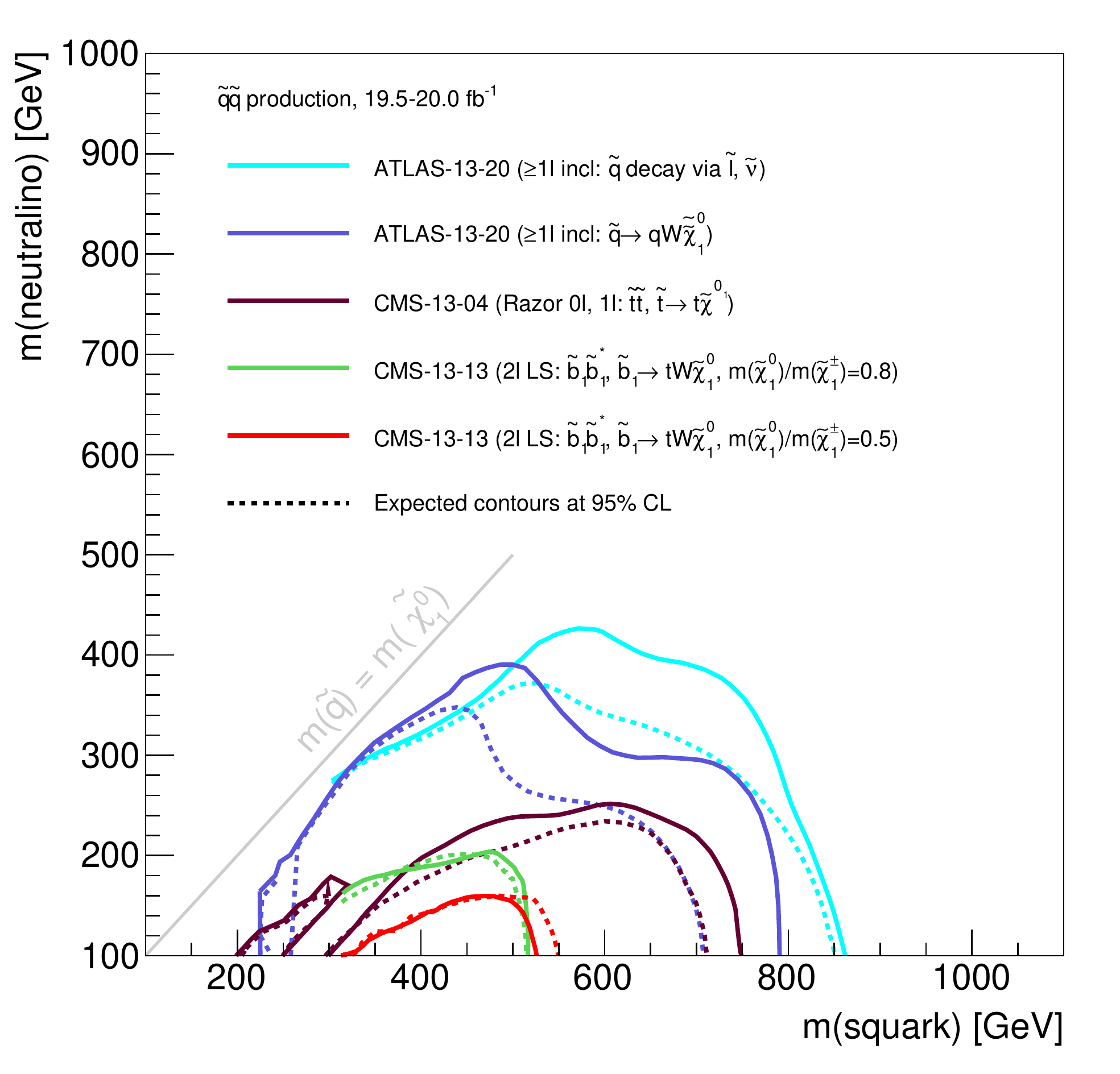}\label{fig:sq_chi_sqsq-leptons}}
  \caption{Exclusion contours at $95\%$~CL of the inclusive searches with leptons for strong production of supersymmetry signal events at \ATLAS and \CMS. Result for $\tilde{g}\tilde{g}$ production is shown in (a), for different gluino decay topologies. The results for $\tilde{q}\tilde{q}$ production is shown in (b), where either light-flavor squarks or single $\tilde{b}$ or $\tilde{t}$ squarks are assumed to be pair produced. The different squark decay modes assumed by the analyses are specified in the legend. The analyses use different analyses strategies as discussed in the text. The analysis labels refer to Table~\ref{tab:overview1}. }
 \label{fig:inclLeptons}	
\end{figure*}
\begin{figure*}[bt]
\begin{center}
\includegraphics[width=0.8\textwidth]{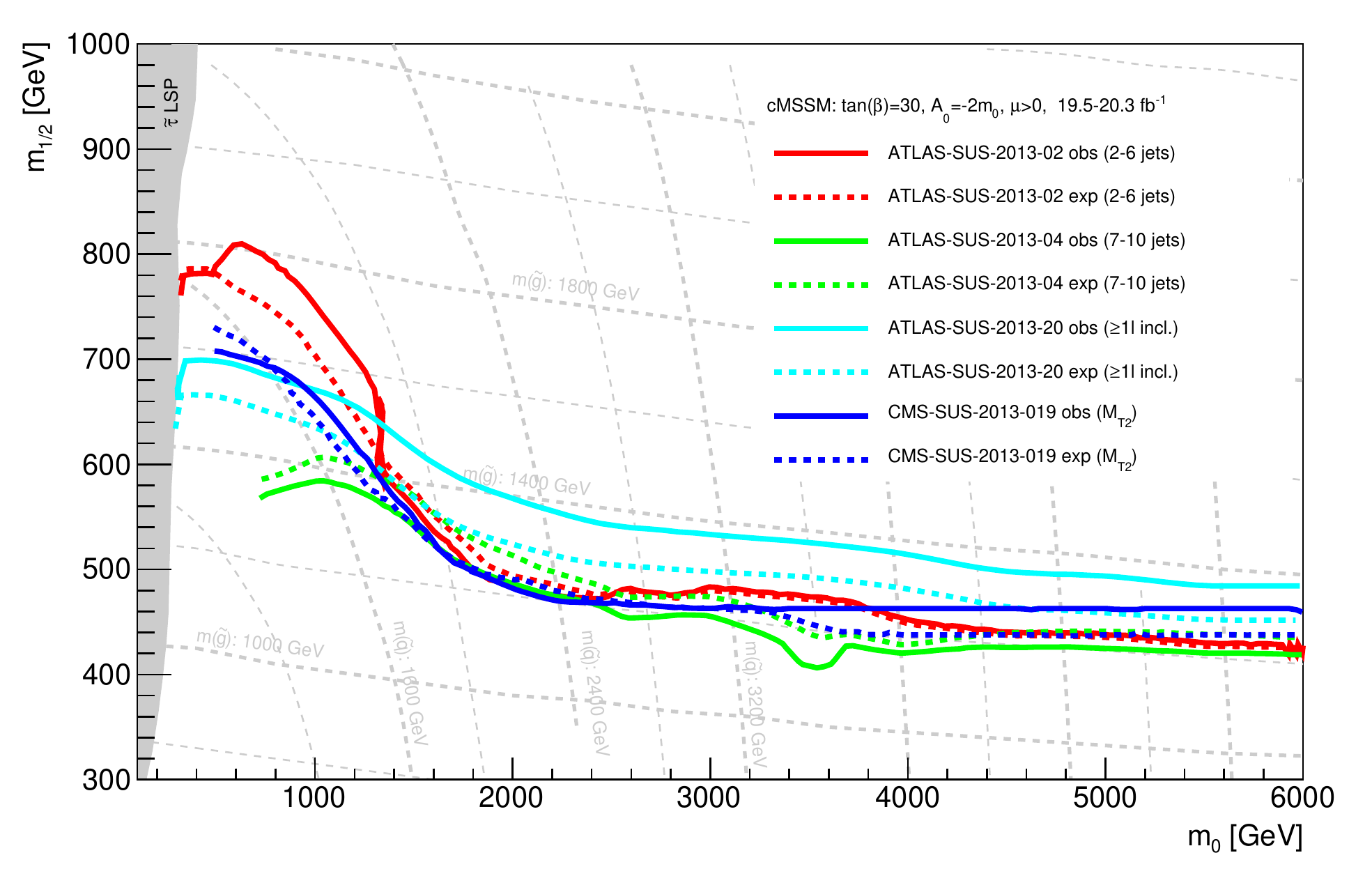}
\end{center}
  \caption{Exclusion contours at $95\%$~CL in the constrained MSSM/mSUGRA with $\tan\beta=30$, $\mu>0$, and $A_0=-2\cdot m_0$, such that the Higgs mass is approximately $m_H=125$\,GeV in the studied parameter space. The CMS $M_{T2}$ analysis used a cMSSM scan with $A_0=-2\cdot\mbox{max}(m_0,m_{1/2})$. The analysis labels refer to Table~\ref{tab:overview1}.}
 \label{fig:cMSSM}	
\end{figure*}

The results of the different inclusive searches with leptons in the final state are shown in Fig.~\ref{fig:inclLeptons} for the simplified gluino-pair and for squark-pair production. The analyses assume a variety of different production and decay topologies, that make direct comparisons of the different analyses sensitivities problematic. Interpretations in the same signal model are not available. The shown exclusion contours give however a good overview of the experimental sensitivity for a variety of scenarios. Typically long decay chains are studied, compared to the previously discussed results displayed in Fig.~\ref{fig:inclT1T2}. 

For $\tilde{g}\tilde{g}$ production shown in Fig.~\ref{fig:gl_chi_glgl-leptons}, the gluino effectively decays into two quarks and the lighter chargino $\tilde{g}\to qq\chaone$, the \chaone decays further into the \chione LSP as identified in the figure legend. The chargino mass is fixed in between the gluino or squark mass and the neutralino \chione with equal mass difference, unless otherwise specified in the figure legend. 
In the signal scenarios, where intermediate sleptons or sneutrinos are produced in the final state, the gluinos or squarks decay with equal probability via either the lightest chargino \chaone or the next-to-lightest neutralino \chitwo. These subsequently decay via left-handed sleptons (or sneutrinos) into a lepton (or neutrino) and the lightest neutralino \chione. The masses of the intermediate charginos/neutralinos are set to be equal, while the slepton and sneutrino masses (all three lepton flavors are mass degenerate in this model) are fixed to lie in the middle of m(\chaone/\chitwo) and m(\chione). 

For $\tilde{q}\tilde{q}$ production shown in Fig.~\ref{fig:sq_chi_sqsq-leptons}, different production and different decay topologies are assumed by the shown analyses. The inclusive search with at least one lepton~\cite{ATLAS-SUS-2013-20} assumes pair production of first- or second-generation squarks that decay either through sleptons or SM $W$-bosons. The Razor analysis~\cite{CMS-SUS-2013-04} with one $b$-tagged jet and combined no-lepton and one-lepton final state is targeted at stop-pair production, where the stop decays into a SM top-quark and the \chione. More analyses specialized for this decay are discussed in Sec.~\ref{sec:third}. The results of the like-sign dilepton search~\cite{CMS-SUS-2013-13} are interpreted with respect to bottom squark pair production, where the sbottom decays via charginos as $\tilde{b}_1\to t W \chione$. Two scenarios for chargino masses $m(\chaone)=1.25m(\chione)$ and $m(\chaone)=2m(\chione)$ are shown. 

The free parameters of the simplified model are the gluino (squark) mass and the neutralino \chione mass, which span the plot plane. Depending on the model of supersymmetry, gluino masses up to $1.32$\,TeV, squark masses up to $840$\,GeV, and neutralino masses up to $650$\,GeV can be excluded. 

Exclusion contours for the cMSSM for $\tan\beta = 30$, $A_0 = −2m_0$ and $\mu > 0$ in the plane of the universal scalar mass $m_0$ and the common mass of the gauginos and higgsinos $m_{1/2}$ are derived, as shown in Fig.~\ref{fig:cMSSM}. Models like the cMSSM are helpful to compare different analysis search results on equal footings. However,  only few analyses have produced exclusion contours in the cMSSM, because of the limitations of the model that significantly reduce the possible relations between supersymmetric masses or branching fractions. The cMSSM exclusion contours of the inclusive analyses discussed above, where available, are shown in Fig.~\ref{fig:cMSSM}. All light-flavour squark masses below $1.6$\,TeV and gluino masses below $1.35$\,TeV masses can be excluded at $95\%$~CL. The exclusion contours reach $m_{1/2}<800$\,GeV at small values of $m_0$, and $m_{1/2}<500$\,GeV at large values of $m_{0}$. 

The  \MET and $2-6$ jets analysis~\cite{ATLAS-SUS-2013-02} targeted at low jet multiplicities is most sensitive at low $m_0$ where squark masses are light and $\tilde{q}\tilde{q}$ production is largest. At large values of $m_0$, where squarks are heavy and gluino mediated production leads to long decay chains, the inclusive search with leptons~\cite{ATLAS-SUS-2013-20} sets the best exclusion.

The Fittino and Mastercode collaborations~\cite{Fittino-2015,Buchmueller:2013rsa} have investigated the remaining allowed parameter space with a global fit, in order to identify the set of cMSSM parameters best compatible with \SM precision measurements, astro\-phy\-sics, and direct LHC searches. The precision observables include for example the anomalous magnetic moment of the muon, direct dark matter detection bounds, the dark matter relic density, and the Higgs boson mass. The global minimum was found at small values of $m_0$ just outside the direct search limits from the LHC experiments with respect to $m_{1/2}$. The fit $p$-value, i.e. the consistency with the model is smaller than $10\%$. In this sense, the cMSSM can be excluded by the LHC searches at a confidence level of at least $90\%$~\cite{Fittino-2015,Buchmueller:2013rsa}.

\section{Inclusive searches for gluino mediated production of third generation squarks}
\label{sec:gluinomediated}
As discussed before, the mass difference to the \SM partners of the third generation multiplets with large Yu\-ka\-wa couplings should not be too large~\cite{MassDiff}, in order to minimize necessary fine-tuning. 
The lighter mass-eigenstate $\tilde{t}_{1}$ is typically the lightest squark. Since the $\tilde{t}_{L}$ belongs to the same weak isospin doublet as the bottom squark $\tilde{b}_L$, and is therefore controlled by the same supersymmetry-breaking mass parameter, a light $\tilde{t}_{1}$ can also imply light $\tilde{b}_{1}$. In simplified models of third generation squark production $\tilde{t}_{1}$ (or the $\tilde{b}_1$) is typically assumed to be the only accessible squark, all other squarks are assumed to be decoupled at high energy scales. Direct strong pair production of third generation quarks is possible and will be discussed in the following section. Here, the gluino mediated production, favored by naturalness arguments, i.e.~Eq.~(\ref{eq:naturalness}), shall be discussed, which is similar to the inclusive searches for gluino and light-flavor squark production discussed previously. In general, an increased signal sensitivity to stop or sbottom masses can be achieved by the usage of $b$-tagging or top-tagging information. The relevant gluino-mediated stop-quark production channels are shown in Fig.~\ref{fig:feyn_inclThird}. Assuming sufficiently heavy gluinos, top-squark decays can lead to final states with up to four real top-quarks and two neutralino LSPs in the final state, e.g. $\tilde{g}\to t\tilde{t}\to t t \chione$, depending on m($\tilde{t}_1$) and m(\chione).

\subsection{Searches in the all-hadronic final state}

\begin{figure*}[tb]
\hspace*{\fill}%
\subfigure[$\tilde{g}\tilde{g}\to t\bar{t}t\bar{t}\chione\chione$]{
\includegraphics[scale=1.0]{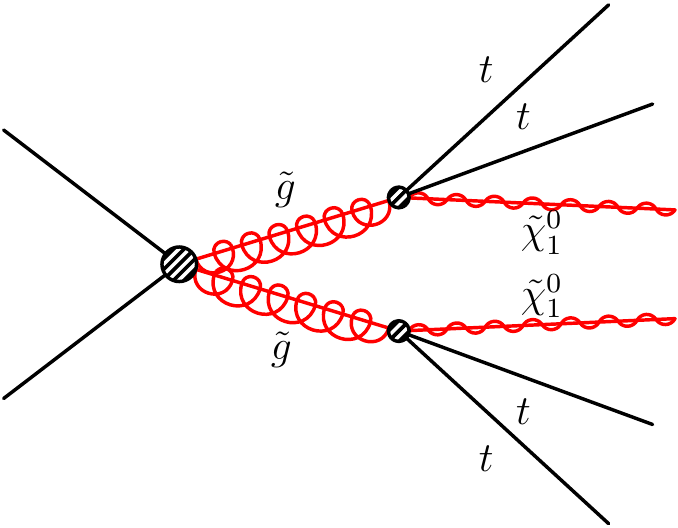}
}\hfill
\subfigure[$\tilde{g}\tilde{g}\to t\bar{c}\bar{t}c\chione\chione$]{
\includegraphics[scale=1.0]{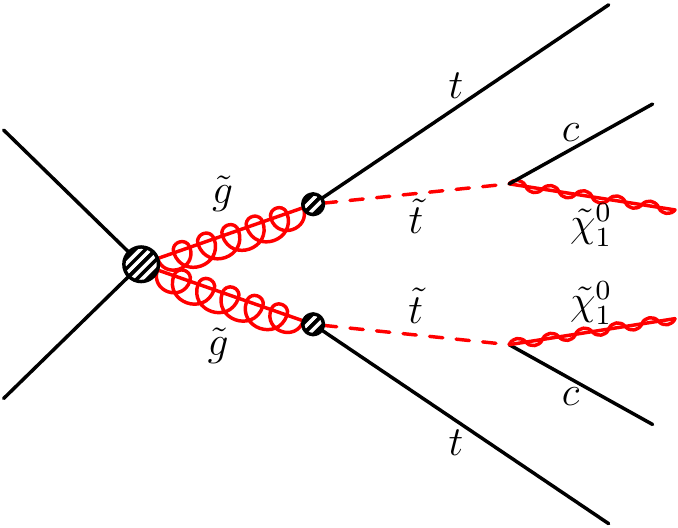}
}
\hspace*{\fill}%
  \caption{Effective Feynman diagrams for the simplified gluino mediated top production and the effective decay through an off-shell stop $\tilde{t}_1$ (a) and gluino mediated stop production and the successive decay $\tilde{t}_1\to c\tilde{\chi}^0_1$ (b) relevant if the stop and the neutralino LSP are almost mass degenerate.}
 \label{fig:feyn_inclThird}	
\end{figure*}

%\cite{CMS-SUS-2012-28}%Strong Inclusive & GLUINO MED. & Third, alphat
%Search for supersymmetry in final states with missing transverse energy and 0, 1, 2, 3, or ≥ 4 b jets in 8 TeV pp collisions, https://twiki.cern.ch/twiki/bin/view/CMSPublic/PhysicsResultsSUS12028

Analyses~\cite{CMS-SUS-2012-28,CMS-SUS-2013-19} based on the kinematic variables $m_{T2}$ and \alphat introduced in Eq.~(\ref{eq:mt2}) and Eq.~(\ref{eq:alphat}) were carried out at CMS. The variables provide powerful discrimination against the QCD multijet background. The $b$-tag multiplicity as well as the total number of jets are used to define exclusive signal regions, maximizing the analysis sensitivity to different signal scenarios like gluino- or squark pair production, and in particular also to direct and gluino mediated production of third generation quarks. In addition to $\tilde{g}\tilde{g}, \tilde{g}\to qq\chione$ and light-flavor $\tilde{q}\tilde{q}$ production discussed in the last section, the analyses results are interpreted for gluino-pair production, where the gluino decays to two top-quarks and the neutralino LSP, as shown in Fig.~\ref{fig:feyn_inclThird}(a).

%\cite{CMS-SUS-2012-24}%Strong Inclusive, Third
%Search for supersymmetry using the shape of the HT and MET, and b-jet multiplicity distributions, https://twiki.cern.ch/twiki/bin/view/CMSPublic/PhysicsResultsSUS12024
The search analysis~\cite{CMS-SUS-2012-24} uses the shape of the conventional \MET and \Ht variables to search for supersymmetry. By dividing \MET and \Ht into four bins each and in addition to the $b$-tag multiplicity $N_{\mbox{\footnotesize $b$-tags}}=1,2,\geq 3$ bins, 176 mutually exclusive signal regions are defined, in order to target gluino-mediated top- or bottom-squark production. The dominant \SM backgrounds originate from QCD multijet production, from ``lost leptons'' of $t\bar{t}$+jets and $W$+jets processes and from $Z\to\nu\bar{\nu}$. The shapes of these backgrounds are obtained from data-sidebands: For the QCD background, control events are selected with low $\Delta\Phi_{\mbox{\footnotesize min}}$, lost-lepton proxy event samples require exactly one electron or one muon with a small transverse mass $m_T<100$\,GeV as defined by Eq.~(\ref{eq:mt}) to limit possible signal contamination, and for the $Z$ to invisible background $Z\to e^+e^-$ and $Z\to\mu^+\mu^-$ events are selected. The shape of these three backgrounds is fitted to the three-dimensional signal region in \MET, \Ht, and the number of $b$-tags, except for the $Z\to\nu\bar{\nu}$ background, where only \MET and \Ht are fitted to solve the low-statistics problem of the $Z\to l^+l^-$ control events.
In the limit of a massless lightest supersymmetric particle, gluinos with masses below $1020$~GeV are excluded for gluino mediated stop production.

\subsection{Top squark searches with leptons in the final state}

Extending the examined final states to include leptons can increase the sensitivity to gluino mediated stop production $\tilde{g}\to t\tilde{t}\to tt\chione$, because up to four leptons from the $W$-bosons from the top-quark decays can be produced. The disadvantage of the small branching ratio can be compensated by a better handle on the separation and the description of the remaining \SM backgrounds.

\begin{figure*}[tb]
\subfigure[]{\includegraphics[width=0.5\textwidth]{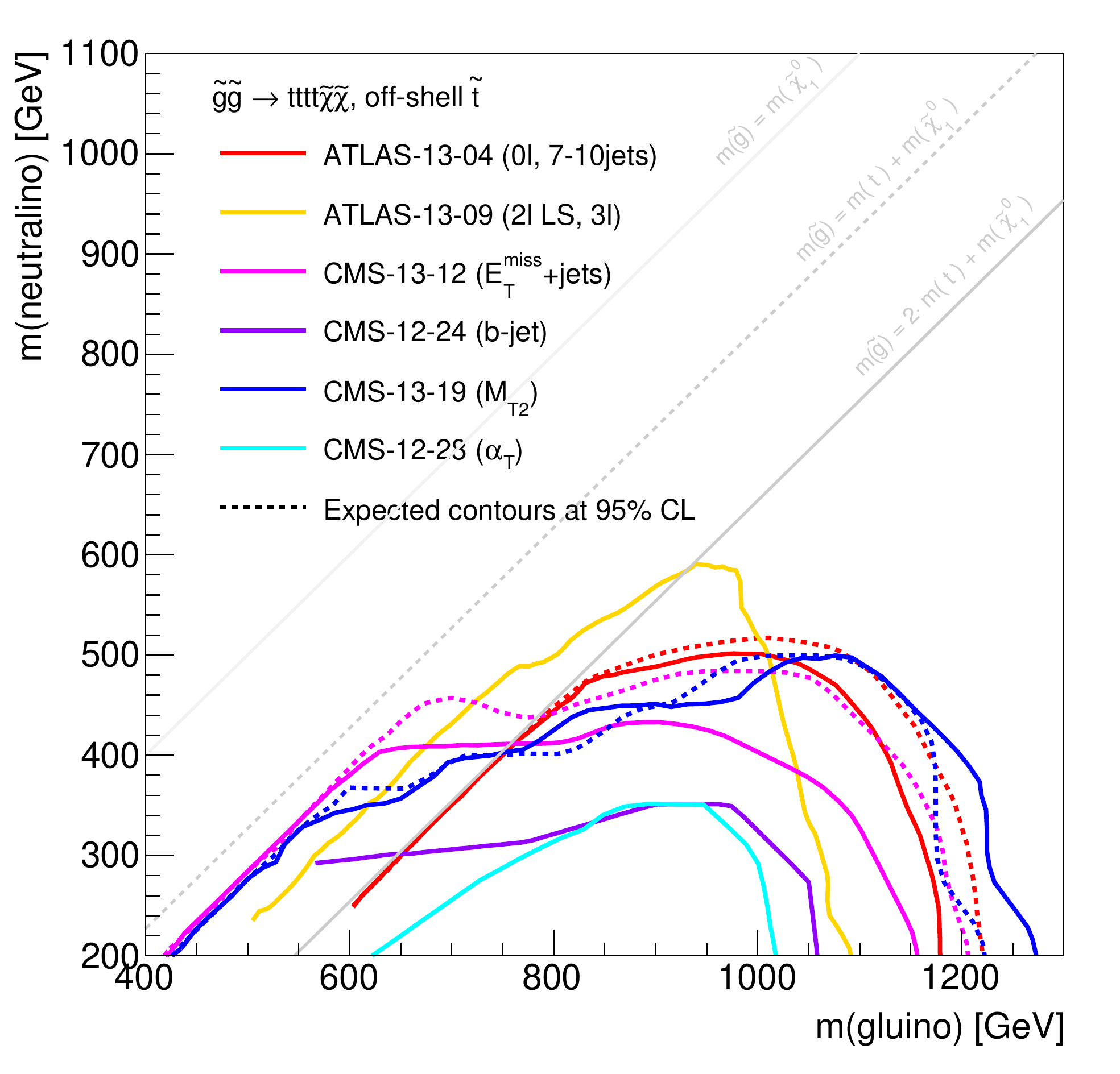}}
\subfigure[]{\includegraphics[width=0.5\textwidth]{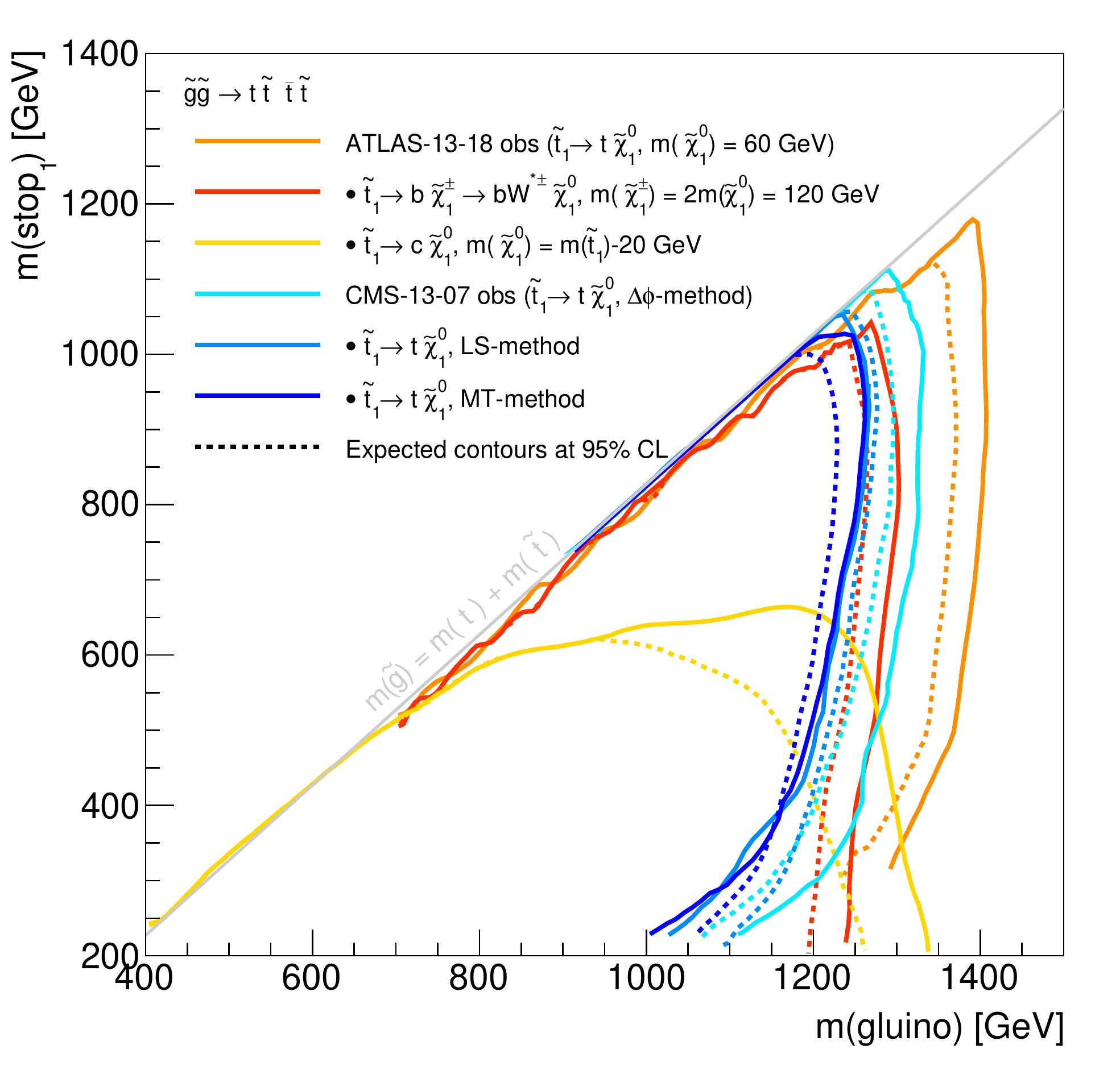}}
  \caption{Exclusion contours at $95\%$~CL for the inclusive searches for strong gluino pair-production and the subsequent decay of both gluinos $\tilde{g}\to t\bar{t}\tilde{\chi}$ in the gluino - neutralino mass plane (a) and the m($\tilde{g}$) - m($\tilde{t}_1$) plane (b). The analysis labels refer to Table~\ref{tab:overview1}.}
 \label{fig:glmedstop}	
\end{figure*}

%\cite{CMS-SUS-2013-07}%Strong - not so inclusive, 1 l
%Search for supersymmetry using events with a single lepton, multiple jets, and b-tags, https://twiki.cern.ch/twiki/bin/view/CMSPublic/PhysicsResultsSUS13007
The search for gluino-mediated top-squark production in $19.3$\,fb$^{-1}$ reported by CMS~\cite{CMS-SUS-2013-07} is based on events with a single isolated lepton (electron or muon) and at least six high-\pt jets, at least two of which are identified as $b$-jets. Two scenarios with on- or off-shell top squarks $\tilde{g}\to t\tilde{t}^{(*)}$ are considered. In both scenarios the stop decays into a top quark and the neutralino LSP $\tilde{t}^{(*)}\to t \chione$. The final state contains four \SM top quarks and \MET, which means that four jets originating from $b$-quarks and can be $b$-tagged. The probability that in at least one of the four $W$-bosons decays one charged lepton $e$ or $\mu$ is created is $40\%$, well motivating the single lepton requirement. The SM background is dominated by $t\bar{t}$ processes, where large \MET stems from the leptonic decay of a heavily boosted $W$-boson and $b$-tag multiplicities larger than two can originate from gluon splitting $g\to b\bar{b}$ or mistagged jets. Contributions from $W$+jets and other rare diboson processes are suppressed by the high jet multiplicity requirement. Three complementary analysis strategies with respect to the chosen kinematic variables and data-sets are pursued: The first two methods rely on the evaluation of the \MET distribution in bins of the amount of hadronic energy in the event \Ht. The \MET shape of the remaining SM backgrounds is estimated in two independent ways. The missing transverse momentum template method (MT) determines a parametrized description of the \MET distribution by fits to control regions at low-\Ht. The lepton spectrum method (LS) exploits the correlation of the direction of the lepton momentum and the \METvec originating from a neutrino from the same $W\to l\nu$ decay in the \SM background. Another independent method is based on the azimuthal angle $\Delta\Phi(W,l)$ between the reconstructed $W$-direction and the lepton to separate a signal-depleted background dominated region at small $\Delta\Phi(W,l)\leq1$. The $\Delta\Phi(W,l)$ variable is comparable to $\mt(l,\MET)$ with respect to the separation power of signal and \SM $t\bar{t}$ and $W$ background, but has superior resolution in this analysis. The signal region is binned in $S_T^{\mbox{\footnotesize lep}}=\MET+\sum_{\mbox{\footnotesize leptons}}\pt$, which includes also signal events with low \MET but high lepton momenta. The three different data-driven analysis strategies MT, LS, and $\Delta\Phi(W,l)$ increase the robustness of the search, while avoiding uncertainties due to potential mismodelling of SM simulation.
Upper limits are set for gluino pair production with $\tilde{g}\to t\tilde{t}^{(*)}$, where each of the two top squarks decays into a top quark and the lightest supersymmetric particle, as shown in Fig.~\ref{fig:glmedstop} in the gluino - neutralino LSP mass plane and in the gluino - stop mass plane.

%\cite{ATLAS-SUS-2013-09}% Strong / Inclusive, 2l same sign, Third or 3l
% 2 same-sign / 3 -leptons + 0-3 b-jets + Etmiss [Incl. squarks &amp; gluinos],04/2014
% https://atlas.web.cern.ch/Atlas/GROUPS/PHYSICS/PAPERS/SUSY-2013-09

Two isolated leptons ($e$ or $\mu$) with the same electric charge (like-sign, LS), or at least three isolated leptons are required in the search for strongly produced supersymmetric particles~\cite{ATLAS-SUS-2013-09} at the \ATLAS experiment. In addition to requiring multiple energetic jets, also the $b$-tag multiplicity is taken into account to increase the sensitivity in particular to gluino mediated stop production. The analysis strategy and the SM background composition is comparable to its CMS counterpart~\cite{CMS-SUS-2013-13} discussed in the previous section.

%\cite{ATLAS-SUS-2014-06}%ATLAS: Inclusive squark/gluino searches [Summary]
%\cite{ATLAS-SUS-2013-18}%ATLAS: 0/1 leptons, 3 bjets
The final state with and without at least one high \pt lepton $e$ or $\mu$ was analyzed simultaneously by the \ATLAS analysis~\cite{ATLAS-SUS-2013-18}, combining the limits from different signal regions, based on their expected sensitivity. The analysis was designed specifically for gluino mediated production of third generation squarks and requires at least three $b$-tagged jets. The \SM backgrounds are divided into reducible backgrounds, where at least one $b$-tagged jet is mistagged, and irreducible backgrounds dominated by $t\bar{t}+b/b\bar{b}$ production with genuine $b$-jets. The reducible backgrounds are modeled using data-sidebands weighted according to the $b$-misidentification rate. The rate was measured using MC simulation and validated in $t\bar{t}$ enriched data control regions. The irreducible background is described by Monte Carlo simulation, where the dominant contribution from $t\bar{t}+b/b\bar{b}$ is normalized in a control region with two isolated leptons and relaxed \MET requirements. The sensitivity to different signal scenarios is optimized by binning the $0$-lepton signal region with respect to \meff, \MET, \Ht, $\Delta\Phi_{\mbox{\footnotesize min}}$, and $\MET/\sqrt{\Ht}$. The $1$-lepton signal region is binned in \meff and the transverse mass $m_T(l,\MET)$. Due to the simultaneous analysis of the $0$- and $1$-lepton final states and the combination of multiple signal regions very compatible sensitivities to gluino mediated $3^{\mbox{\footnotesize rd}}$~generation squarks production are obtained.

\ATLAS searches for gluinos and first- and second-generation squarks in final states containing jets and missing transverse momentum, with or without leptons or $b$-jets are summarized in Ref.~\cite{ATLAS-SUS-2014-06}. The combination is further extended by a new search for squarks and gluinos in inclusive final states with \MET, high-\pt jets, with and without leptons and $b$-tags, thus improving the sensitivity to a wide range of supersymmetry models and in particular to gluino-mediated third generation squark production. The resulting limits on gluino mediated stop production are shown in Fig.~\ref{fig:glmedstop} for the discussed analyses. In the case of decays via virtual top squarks and for light LSPs, gluino masses below $1.26$\,TeV are excluded. Viable scenarios remain possible with light $\tilde{t}$ and $\tilde{g}$ masses below approximately $0.5$ and $1.5$\,TeV, respectively, in particular if the mass difference between the gluino and the neutralino LSP is similar to or smaller than the \SM top mass.

%\todo{gl vs m(stop) plot. Consider sbottom?}

\section{Direct production of top squarks}
\label{sec:third}

Naturalness arguments prefer light top squark masses $m(\tilde{t}_1)$, as previously discussed, leading to a $\tilde{t}\tilde{t}$ production cross-sec\-tion~\cite{StopXsec} interesting for direct searches in this topology. The produced top squarks decay into the neutralino and further \SM particles, depending on the mass difference between stop and neutralino $\Delta m(\tilde{t},\chione)$. Scenarios where the $\tilde{t}_1$ is the lightest supersymmetric particle eliminate the possibility of a supersymmetric candidate particle to explain the dark matter problem of the universe. These scenarios are not considered here, because astrophysical observations place stringent constrains on strongly and/or electromagnetically interacting LSPs \cite{PhysRevLett.42.1117}, or require sufficiently large $R$-parity violating couplings to explain the non-observation of exotic atoms or nuclei~\cite{Rparity}.

\begin{figure*}[tbp]
\begin{center}
\includegraphics[width=0.7\textwidth]{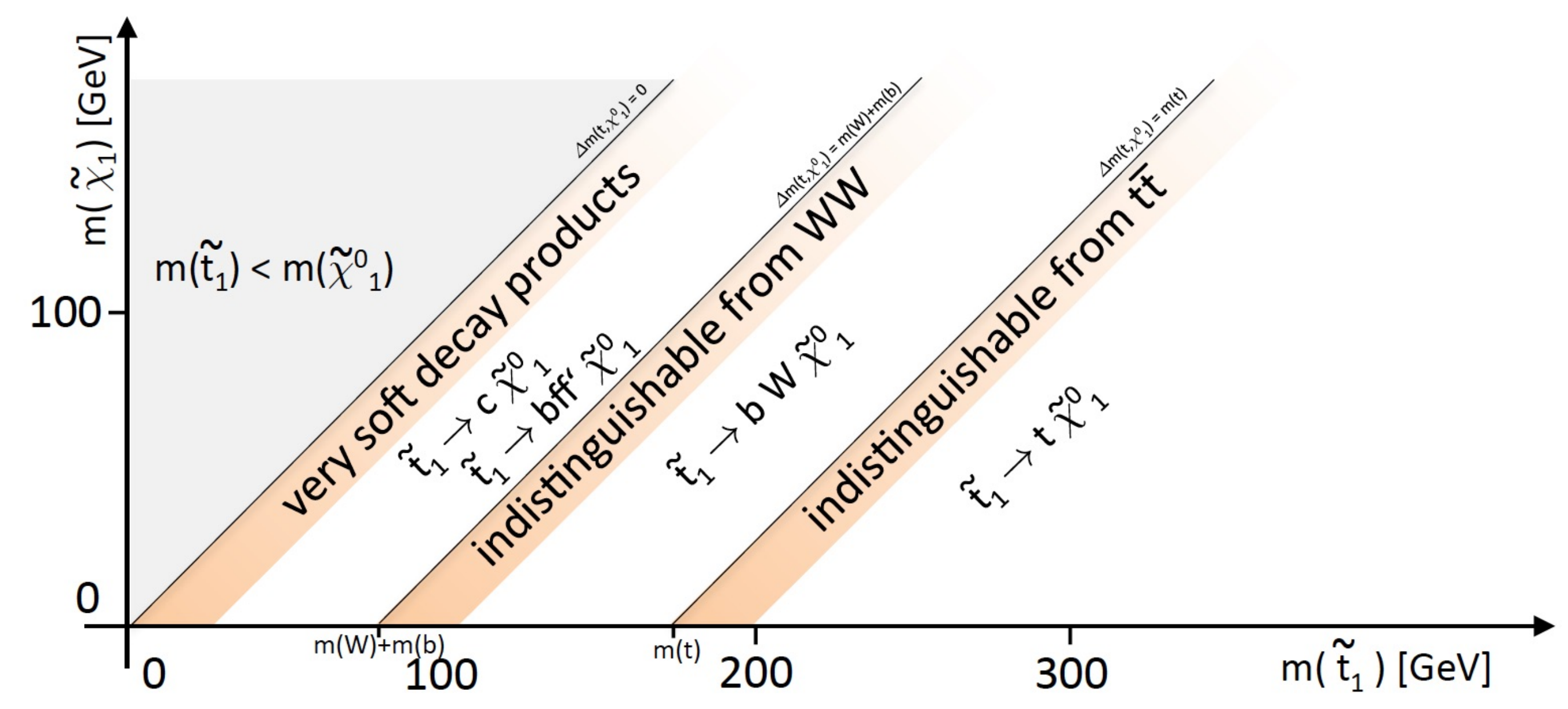}
\caption{Top squark decay topologies depending on the mass difference $\Delta m(\tilde{t}_1,\chione)$ between the lightest stop and the neutralino LSP.}
 \label{fig:stop_chi_scheme}	
\end{center}
\end{figure*}

\begin{figure*}[tb]
\hspace*{\fill}%
\subfigure[$\tilde{t}\tilde{t}, \tilde{t}$~four-body decay]{
\includegraphics[scale=1.0]{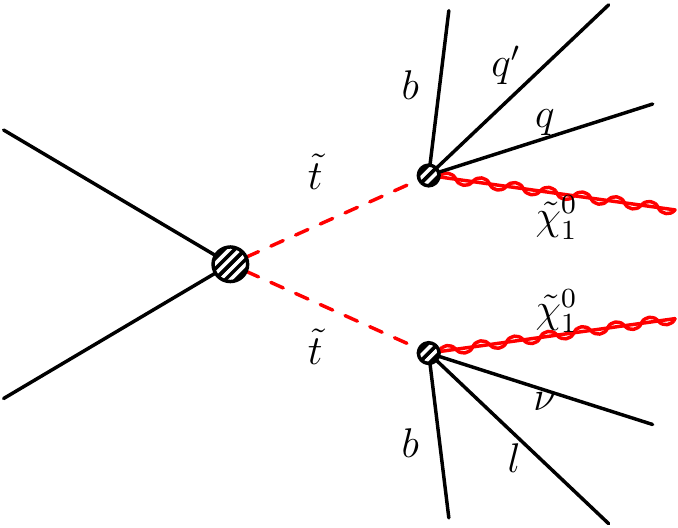}
\label{fig:feyn_stst_4body}
}\hfill
\subfigure[$\tilde{t}\tilde{t}\to bb WW \chione\chione$]{
\includegraphics[scale=1.0]{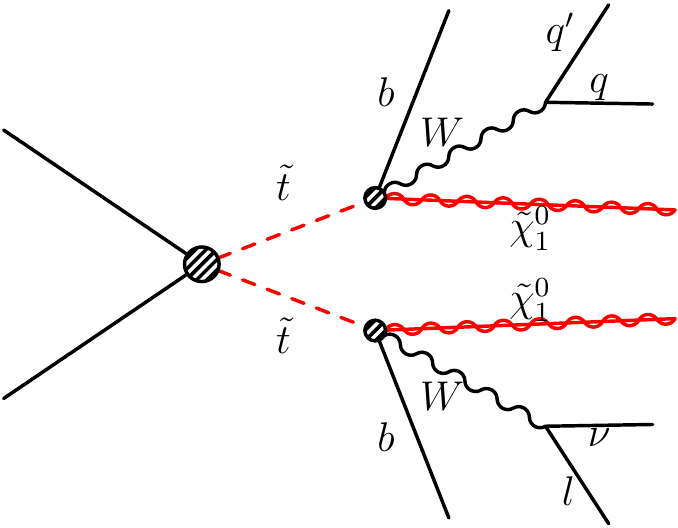}
\label{fig:feyn_stst_blnuchi1}
}
\hspace*{\fill}%
\\

\hspace*{\fill}%
\subfigure[$\tilde{t}\tilde{t}\to tt\chione\chione$]{
\includegraphics[scale=1.0]{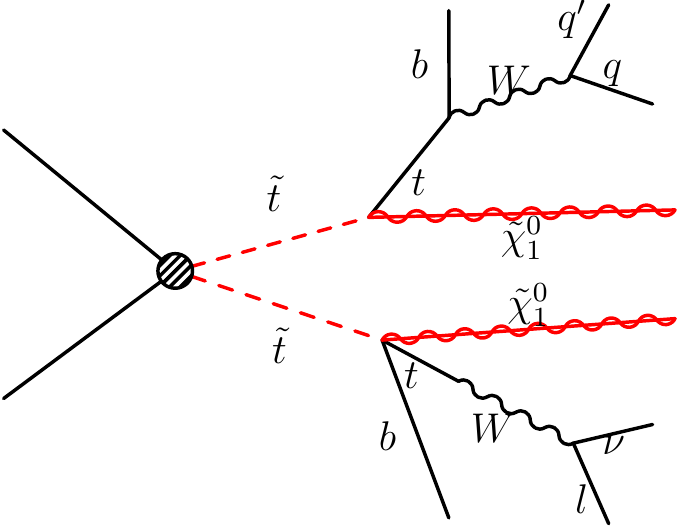}
\label{fig:feyn_stst_ttchi1chi_detailed}
}\hfill
\subfigure[$\tilde{t}\tilde{t}\to b\bar{b}\chaone\chaone$]{
\includegraphics[scale=1.0]{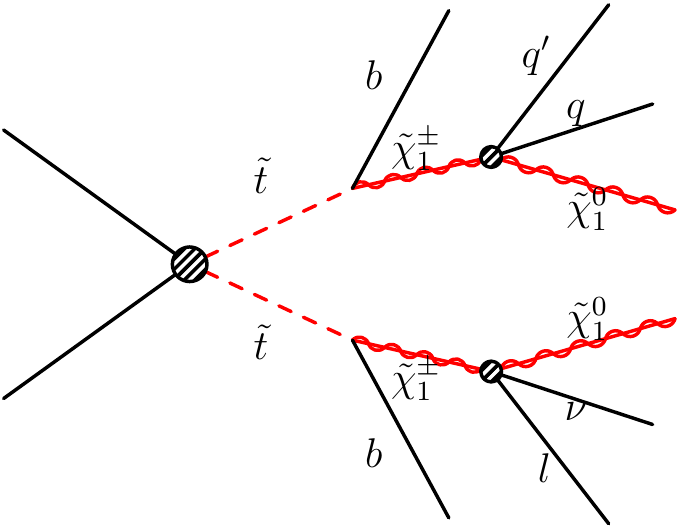}
\label{fig:feyn_stst_bbchacha}
}
\hspace*{\fill}%
  \caption{Effective Feynman diagrams for direct stop production and the following stop decay to the neutralino LSP and 
a stop four-body decay (a),  a stop three body decay into the neutralino, a $b$-quark and an on-shell $W$-boson (b),
  an on-shell top (c), or through trough a chargino $\tilde{t}\to b\chaone$ followed by a three-body decay of the \chaone (d).}
 \label{fig:feyn_third}	
\end{figure*}

In the following, pair production of the lightest stop quark $\tilde{t}\tilde{t}$ is considered, where the details of the $\tilde{t}$ decay topology depend on $\Delta m(\tilde{t},\chione)$ as demonstrated by Fig.~\ref{fig:stop_chi_scheme}. The four different decay regimes will be discussed with respect to the most sensitive analyses. The relevant Feynman diagrams are shown in Fig.~\ref{fig:feyn_third} and the results of the \ATLAS and \CMS experiments are summarized in Fig.~\ref{fig:Third}. Pair production of sbottom quarks is not discussed here, as the scenarios and the analyses are very comparable to the pair production of light-flavor quarks, except for the $b$-jets, that can be tagged. Results for $\tilde{b}\tilde{b}$ can be found in the quoted references.

%\subsection[Stop-decay to soft decay products and $\chione$ for $\Delta m(\tilde{t}_1,\chione) < m(W) + m(b)$]{Stop decay to soft decay products and $\mathbf \chione$ for $\mathbf{ \Delta m(\tilde{t}_1,\chione) < m(W) + m(b)}$ }
\subsection[Stop-decay to soft decay products and $\chione$ for $\Delta m(\tilde{t}_1,\chione) < m(W) + m(b)$]{Stop-decay to soft decay products and $\chione$ for \linebreak $\Delta m(\tilde{t}_1,\chione) < m(W) + m(b)$}

At very small mass differences $\Delta m(\tilde{t}_1,\chione)$, i.e. at almost mass-degenerate stop and neutralino masses  $\tilde{t}_1 - m(\chione)  \gtrsim m(c)$ the available energy in the stop decay is neither sufficient to produce on-shell top quarks $\tilde{t}_1\to t\chione$ nor to produce on-shell $W$-bosons in the succeeding decay $t^*\to W b$. The stop therefore undergoes a four-body decay $\tilde{t}_1\to b f f' + \chione$ into one $b$-quark, one up- and down-type fermion, and the LSP. This includes the stop decay involving charginos $\tilde{t}_1\to b\chaone$ that subsequently decay into an off-shell $W$-boson and the neutralino LSP. The four-body decay competes with the flavor-changing $\tilde{t}_1\to c \chione$ loop-suppressed decay that leads to two $c$-jets and \MET in the detector. The transverse momenta of the $c$-jets depend on the transversal boost of the stop quarks and on $\Delta m(\tilde{t}_1,\chione)$, therefore the \pt($c$-jet) is often so small, that these jets cannot be reconstructed and identified. The events are essentially invisible, unless initial state radiation (ISR), balancing the $\tilde{t}\tilde{t}$ system in the transversal plane, leads to additional jets. 

The analysis strategy with good sensitivity to scenarios with very small mass splittings $\Delta m(\tilde{t}_1,\chione)$ is therefore a search for monojet signatures~\cite{ATLAS-SUS-2013-21,CMS-SUS-2014-01,ATLAS-EXO-2013-13}. The single jet originates from initial state radiation and is balanced by \MET, because the stop-stop system decays into soft decay products, i.e. $c$-jets below the reconstruction threshold, and invisible neutralino LSPs. Few \SM processes have genuine monojet signatures, e.g. $Z$+jet where the $Z$ decays into neutrinos. Hence, the largest part of the background originates from instrumental background, where objects are not reconstructed or identified, leaving missing energy in the detector and a jet. Events with large \MET and one high \pt jet are selected. More soft jets in the event are tolerated, as more than one ISR jet can be produced in the signal process, jets from other simultaneous $pp$-collisions (pileup) may be present, or one of the $c$-jets is reconstructed. Events with reconstructed leptons above $10$\,GeV are rejected.
The \ATLAS analysis~\cite{ATLAS-SUS-2013-21} requires $\MET>120$\,GeV and \pt($1^{\mbox{\footnotesize rst}}$ jet)$>120$\,GeV and rejects events with \pt($4^{\mbox{\footnotesize th}}$ jet)$>30$\,GeV or $\Delta\Phi(\mbox{jet},\METvec)\leq0.4$. The CMS analysis~\cite{CMS-SUS-2014-01} requires $\MET>250$\,GeV, \pt($1^{\mbox{\footnotesize rst}}$ jet)$>110$\,GeV and rejects events if \pt($3^{\mbox{\footnotesize rd}}$ jet)$>60$\,GeV or $\Delta\Phi(1^{\mbox{\footnotesize rst}} \mbox{jet}, 2^{\mbox{\footnotesize nd}} \mbox{jet})>2.5$. 

For sufficiently large mass-splittings $\Delta m(\tilde{t}_1,\chione)\gtrsim20$\,GeV the $c$-jets can receive enough transversal boost to be reconstructed. The monojet analysis~\cite{ATLAS-SUS-2013-21,ATLAS-SUS-2014-03} is expanded by an orthogonal search region explicitly requiring the identification of one of the $c$-jets originating from $\tilde{t}_1\to c\chione$. A $c$-tagging algorithm is applied on the first four leading jets, as the $c$-jets are often softer compared to jets from initial state radiation or pileup. The algorithm employs multivariate methods to combine the information from the impact parameters of displaced tracks and topological properties of decay vertices reconstructed within the jet to produce a likelihood ratio for charm-jets, light-flavor or gluon jets, and $b$-jets. The medium working point has a $c$-tagging efficiency of about $20\%$ and rejection factors of about $8$ and $200$ for $b$-jets and light-flavor or gluon jets, respectively.

The competing four-body decay $\tilde{t}_1\to bff'\chione$, where the $ff'$ are the up- and down-type fermions as shown in Fig.~\ref{fig:feyn_stst_4body}, offers a complementary final state. Searches requiring one or two leptons~\cite{ATLAS-SUS-2013-15,ATLAS-SUS-2014-07,CMS-SUS-2014-21} attempt also the reconstruction of the soft stop decay products, i.e. reconstructing soft electrons or muons with transverse momenta down to $5$\,GeV or $7$\,GeV. Different signal regions with respect to the lepton and $b$-tagged jet multiplicity are defined. One high-\pt jet consistent with ISR is required, this does not reduce the signal acceptance too much, because the $\tilde{t}\tilde{t}$ system has in the selected events an equivalent transverse boost, ensuring sufficiently energetic decay products of the top squarks to pass the reconstruction thresholds. 

Using monojet-topologies, top squarks of masses up to about $250$\,GeV can be excluded at $95\%$~CL for scenarios where the top squark and the lightest neutralino are nearly degenerate in mass. In compressed scenarios of top squark pair production, analyses of the decay channel  $\tilde{t}_1\to b f f'$ lead to similar exclusion contours up to stop masses of about $320$\,GeV for mass splittings $\Delta m(\tilde{t}_1,\chione)<80$\,GeV, as shown in Fig.~\ref{fig:Third}.

%\subsection[Stop-decay to $W$, $b$-jet and $\chione$ for $m(W) + m(b) \leq \Delta m(\tilde{t}_1,\chione) < m(t)$]{Stop decay to $\mathbf W$, $\mathbf b$-jet and $\mathbf \chione$ for $\mathbf{ m(W) + m(b) \leq \Delta m(\tilde{t}_1,\chione) < m(t)}$ }
\subsection[Stop-decay to $W$, $b$-jet and $\chione$ for $m(W) + m(b) \leq \Delta m(\tilde{t}_1,\chione) < m(t)$]{Stop-decay to $W$, $b$-jet and $\chione$ for \linebreak $m(W) + m(b) \leq \Delta m(\tilde{t}_1,\chione) < m(t)$}

The relevant stop decay mode, if $\Delta m(\tilde{t}_1,\chione)$ is in between the $W+b$  mass and the top-quark mass, is the three-body decay $\tilde{t}_1\to Wb\chione$ through an off-shell top-quark into an on-shell $W$-boson and a $b$-quark, as shown in Fig.~\ref{fig:feyn_stst_blnuchi1}. If the mass-gap $\Delta m(\tilde{t}_1,\chione)$ is just enough to produce the on-shell $W$ and $b$, the transverse momentum of the $b$-jet is only determined by the stop-\pt and therefore too soft in order to be reconstructed by the detectors. In this case, only the decay products of the two $W$-bosons can be identified and the experimental challenge is to distinguish the signal from \SM $WW$ production. The precise estimation of the SM $WW$ production is crucial. Monte Carlo simulation validated in data control regions is used to model the $WW$ production. The precision of the \SM $WW$-production cross section is an important source of systematic uncertainty and has been measured by \ATLAS~\cite{AtlasWW8} and CMS~\cite{CMSWW8} and compared to theory predictions~\cite{Kim:2014eva,Curtin:2014zua}. 

Searches are carried out in the one-lepton and jets final state \cite{ATLAS-SUS-2013-15,CMS-SUS-2013-11,CMS-SUS-2014-15} or with two leptons \cite{ATLAS-SUS-2013-19,CMS-SUS-2014-15} using multiple signal selection regions and different kinematic variables, optimized for different mass-splittings.

The exclusion contours for the simplified model of $\tilde{t}_1\tilde{t}_1$ production assuming a $100\%$ branching fraction into the LSP and a virtual top quark  $\tilde{t}_1\to t^*\chione$ reaches up to $m(\tilde{t}_1)<300$\,GeV and $m(\chione)<160$\,GeV. The limit deteriorates, as $\Delta m(\tilde{t}_1,\chione)$ approaches the diagonals defined by $\Delta m=m(W) + m(b)$ and $\Delta m=m(t)$. Near the low-mass diagonal $\Delta m\geq m(W) + m(b)$ the signal acceptance drops, because the $b$-jet momenta is insufficient to pass the identification thresholds, and the remaining signal looks like \SM $WW$. When $\Delta m(\tilde{t}_1,\chione)$ drops below the diagonal, the previously discussed four-body decay channel opens. The high-mass diagonal $\Delta m\geq m(t)$ is discussed in the following.

%\subsection[Stop-decay to top-quark and $\chione$ for $\Delta m(\tilde{t}_1,\chione) = m(t)$ ]{Stop decay to top-quark and $\mathbf \chione$ for $\mathbf{ \Delta m(\tilde{t}_1,\chione) = m(t)}$ \label{sec:third_ontop}}
\subsection{Stop-decay to top-quark and $\chione$ for $\Delta m(\tilde{t}_1,\chione) = m(t)$\label{sec:third_ontop}}

In the case where the stop - neutralino mass difference approximately matches the top-quark mass, two on-shell top-quarks are created in these events at the kinematic threshold. The transverse momenta of the top-quarks and the neutralinos are determined only by the transverse momenta of their top-squark decay parents. The \MET, i.e. the vectorial sum of the neutralino transverse momenta can be very small, so that the visible part of the events of supersymmetry $\tilde{t}_1\tilde{t}_1$ production is very similar to the \SM $t\bar{t}$-production process. As before, the precise knowledge of the \SM cross-section, in this case of $t\bar{t}$-production, is crucial in this  stop - neutralino mass region.

Precision measurements of the top-production cross section at \ATLAS and \CMS~\cite{ATLAS-TOP-2013-04,Khachatryan:2016mqs,Khachatryan:2015fwh} in the various $t\bar{t}$ decay channels allow through comparisons to the calculated cross-section from theory the setting of limits on the supersymmetric $\tilde{t}\tilde{t}$ process. Interpreting the result in a simplified $\tilde{t}\to t \chione$ model allows to exclude top squark masses between approximately $150$\,GeV and $177$\,GeV~\cite{ATLAS-TOP-2013-04,ATLAS-SUS-2014-07} for a massless neutralino LSP. The limit deteriorates fast for heavier neutralinos.

The measurement of the $t\bar{t}$ spin-spin correlation~\cite{ATLAS-TOP-2014-07,CMS-TOP-2014-23} is another example of \SM precision measurements with sensitivity to signals of new physics. Probing the correlation of the top spins expected for \SM pair-production of top quarks, offers a handle on the region $\Delta m(\tilde{t}_1,\chione) = m(t)$, without relying on the precise calculation of the SM $t\bar{t}$ cross-section. The \SM top-pair production process is dominated at low energy scales by the fusion of gluon pairs with the same helicities, leading to top quarks with antiparallel spins in the center-of-mass system. At large energy scales the gluons have opposite helicities, resulting in a top-pair with parallel spins~\cite{Mahlon:2010gw,Bernreuther:2013aga}. In supersymmetry the spins of the top quarks, i.e. from the spin-$0$ stop decays $\tilde{t}\to t \chione$, are not correlated. The angular distributions of the charged leptons from the top and antitop decays, e.g. $t\to b l^+\nu$, are powerful tools to analyze the top-spins. The azimuthal angle $\Delta\phi$ between both leptons is a sensitive variable to probe the $t\bar{t}$ spin-correlations. Final states with electrons and muons ($ee$, $\mu\mu$, $e\mu$) are examined. The CMS analysis~\cite{CMS-TOP-2014-23} analyzes the $\Delta\phi$-shape and is limited by the theoretical uncertainty on $\Delta\phi$. The corresponding \ATLAS analysis~\cite{ATLAS-TOP-2014-07} uses the $\Delta\phi$-distribution in combination with the SM $t\bar{t}$-production cross-section prediction to extract a limit on the $\tilde{t}_1\tilde{t}_1$-production cross-section of approximately $20$\,pb, excluding $\tilde{t}_1$-masses between the top-mass and $191$\,GeV for a simplified scenario with a branching fraction of $100\%$ for $\tilde{t}_1\to t\chione$ and m$(\chione)=1$\,GeV.

Alternatively, if the $\tilde{t}_1\tilde{t}_1$ production is too similar to the \SM top production, searching for the production of the heavier stop mass eigenstate $\tilde{t}_2\tilde{t}_2$ is viable~\cite{ATLAS-SUS-2013-08}. The heavier state can decay into the lighter state and a $Z$-boson $\tilde{t}_2\to Z\tilde{t}_1$. The analysis requires two leptons of opposite charge, same flavor and invariant mass consistent with an on-shell $Z$ decay. Additional signal regions with three leptons help to suppress the SM $t\bar{t}$ background, where the third lepton in the signal may originate from the top-decay from $\tilde{t}_1\to t\chione$, or from other decay modes of the heavier stop $\tilde{t}_2\to H \tilde{t}_1$ or $\tilde{t}_2\to t \chione$. The analysis has also sensitivity to models of gauge mediated supersymmetry breaking; the resulting exclusion contours in the plane of the $m(\tilde{t}_2)$ and $m(\chione)$ are shown in Fig.~\ref{fig:GGM_Third} in the context of GMSB results of Sec.~\ref{sec:gauge}.

%\subsection[Stop-decay to top-quark and $\chione$ for $\Delta m(\tilde{t}_1,\chione) > m(t)$ ]{Stop decay to top-quark and $\mathbf \chione$ for $\mathbf{ \Delta m(\tilde{t}_1,\chione) > m(t)}$ }
\subsection{Stop-decay to top-quark and $\chione$ for $\Delta m(\tilde{t}_1,\chione) > m(t)$}

The dominant decay in this region is $\tilde{t}_1\to t \chione$, as shown in Fig.~\ref{fig:feyn_stst_ttchi1chi_detailed}. The stop mass is sufficiently heavy at low  $\Delta m(\tilde{t}_1,\chione)$ to produce on-shell top-quarks in addition to the neutralino LSP. The reconstruction of the stop decay products is similar efficient compared to \SM $t\bar{t}$ production. For larger values of $\Delta m(\tilde{t}_1,\chione)$, the decay products become more boosted in the transverse plane, leading to more transverse momentum and more \MET. This scenario in the parameter-space of stop-pair production is almost comparable to generic strong production of supersymmetry, i.e. pair and associated production of squarks and gluinos. The difference aside from the cross section is that the visible energy in form of jets and leptons originates from two top-quarks, which offers an advantage. This provenance information can be used by specialized top-tagging search analyses to improve the signal sensitivity to $\tilde{t}\tilde{t}$-production with respect to generic inclusive searches.

The top-decay topology leads to final states with up to two leptons. Search analyses define dedicated signal regions to cover all possible scenarios. All-hadronic final states~\cite{ATLAS-SUS-2013-16,ATLAS-SUS-2013-05,CMS-SUS-2013-23,CMS-SUS-2014-01} and final states with one lepton~\cite{ATLAS-SUS-2013-15,CMS-SUS-2013-11} or up to two leptons~\cite{ATLAS-SUS-2013-19,ATLAS-SUS-2014-07,CMS-SUS-2013-04,CMS-PAS-SUS-2014-11,CMS-SUS-2014-15} are carefully studied individually with respect to exclusive $b$-tag multiplicity bins. Top-tagging algorithms are used, where the $W$- and $t$-mass constraints are explored in order to assign jets and leptons to certain decay chains. 

\begin{figure*}[tbp]
\subfigure[]{\includegraphics[width=0.5\textwidth]{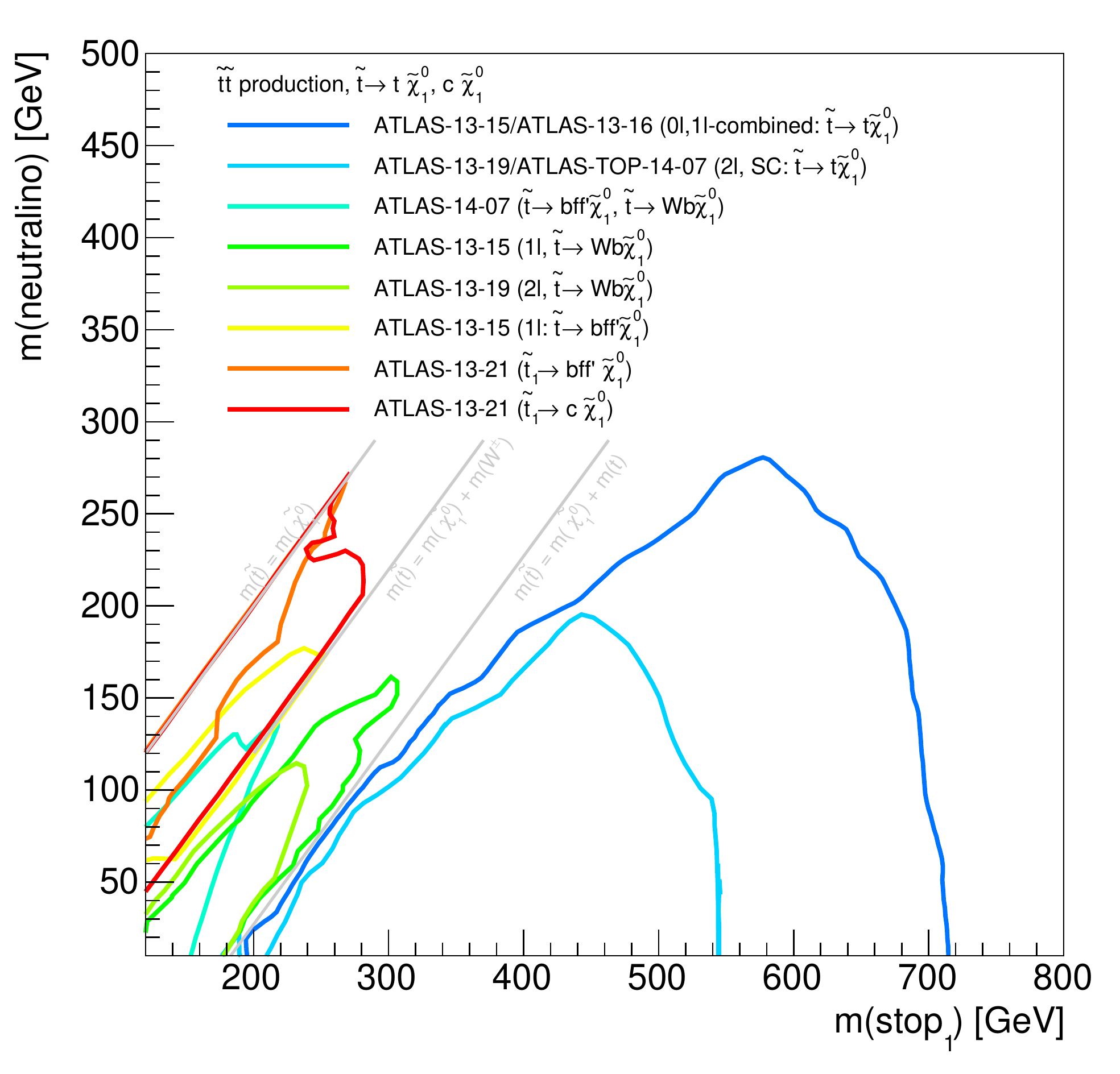}\label{fig:stop1_chi_ATLAS}}
\subfigure[]{\includegraphics[width=0.5\textwidth]{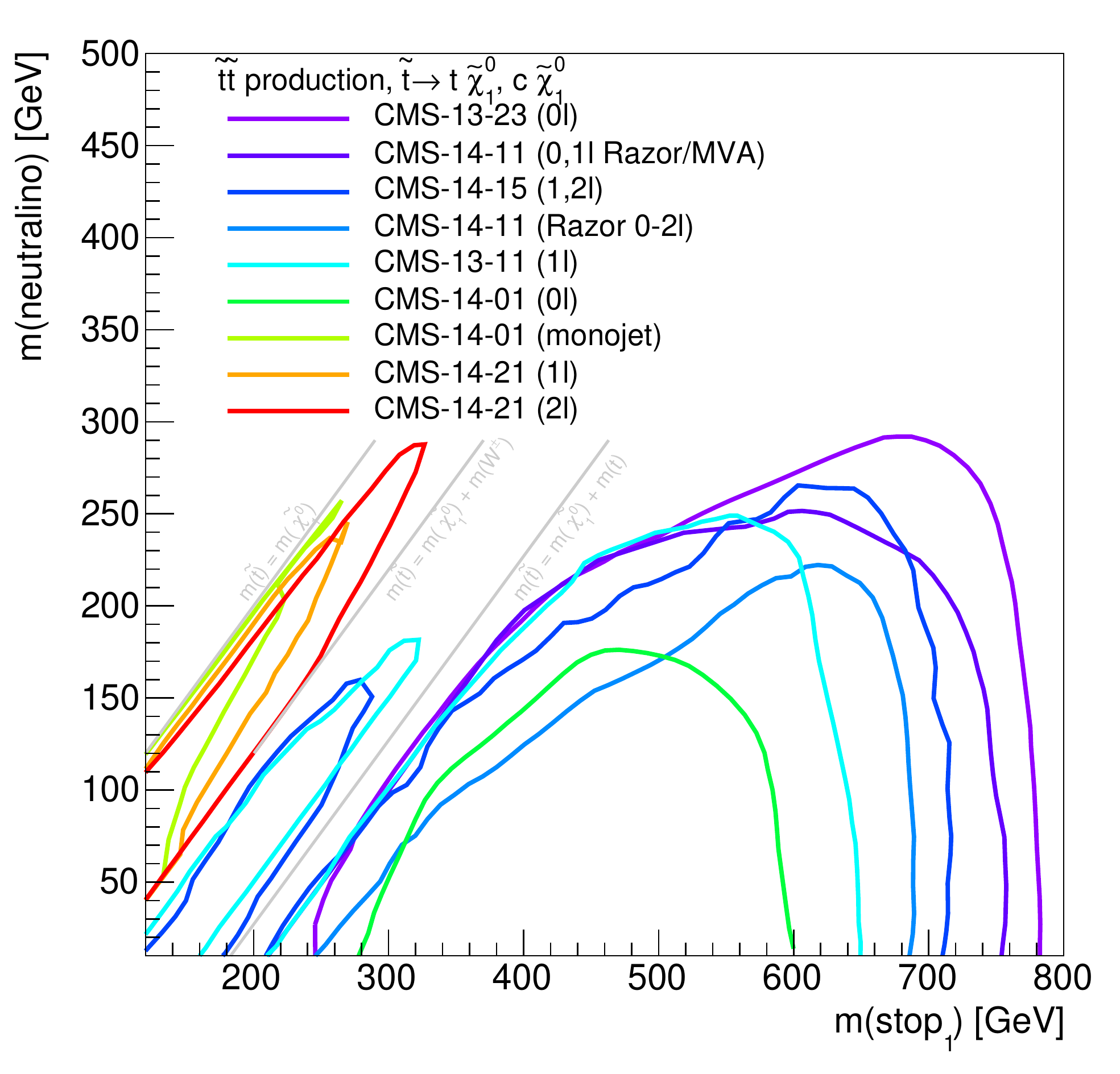}\label{fig:stop1_chi_CMS}}
  \caption{Exclusion contours at $95\%$~CL for the searches for direct pair-production of the lightest stop quark $\tilde{t}_1$ at the \ATLAS experiment (a) and at the \CMS experiment (b). The legend labels refer to Tab.~\ref{tab:overview1}.}
 \label{fig:Third}	
\end{figure*}

The obtained limits for simplified models of stop-pair production are interpreted as exclusion contours in the m$(\tilde{t})-$m$(\chione)$ plane as shown in Fig.~\ref{fig:Third}. For clarity the \ATLAS and \CMS results are shown in separate plots. Stop masses up to $800$\,GeV and neutralinos up to $300$\,GeV are examined by a wide range of different analysis strategies and high precision measurements specialized for specific regions of the parameter space. Different mass splittings $\Delta m(\tilde{t}_1,\chione)$ down to nearly mass-degenerate stops and neutralinos can be excluded. Still, some ``natural'' supersymmetry parameter space with low stop and neutralino masses remains allowed, e.g. where the $\tilde{t}\tilde{t}$-signal is hard to distinguish from \SM $WW$- or $t\bar{t}$-production. Supersymmetry can still be natural, but it can also be stealthy.

In full models of supersymmetry the variety of stop decay topologies is more complex. The details of the branching ratios depend on the specific model. More decay channels, e.g. through gauginos as shown in Fig.~\ref{fig:feyn_stst_bbchacha} exist, leading e.g. to four $W$-bosons in the final state~\cite{CMS-SUS-2014-10,ATLAS-SUS-2014-07} or to $HZ$~\cite{ATLAS-SUS-2013-08,CMS-SUS-2013-24}.

\begin{table*}[htb]
\centering
\begin{tabularx}{\textwidth}{rr>{\raggedright\arraybackslash}X>{\raggedright\arraybackslash}X}
\hline
Label & Ref. & Analysis \& final state & Model of result interpretation\\ \hline\hline
\multicolumn{4}{c}{Inclusive searches}\\
ATLAS-13-02 & \cite{ATLAS-SUS-2013-02}   & $0$l, $2-6$~jets    			         & cMSSM, \Tone, \Ttwo  \\
ATLAS-13-04 & \cite{ATLAS-SUS-2013-04}   & $0$l, $7-10$~jets, $0,1,\geq2$~$b$-jets & cMSSM, \Tone \\
CMS-13-12   & \cite{CMS-SUS-2013-12}     & \MET+jets, $0$~$b$-jets    		     & \Tone, \Ttwo, \Tonett   \\
CMS-13-19   & \cite{CMS-SUS-2013-19}     & $M_{T2}$, \Ht, $0,1,2,\geq3$~$b$-jets   & cMSSM, \Tone, \Ttwo  \\
%%            &                            & $M_{T2}$, $m(b\bar{b})$ from $H\to b\bar{b}$             & \todo{???}  \\
CMS-12-28   & \cite{CMS-SUS-2012-28}     & $\alpha_T$, $0,1,2,3,\geq4$~$b$-jets    & \Tone, \Ttwo, \Tonett, \Ttt \\ 
CMS-13-04   & \cite{CMS-SUS-2013-04}     & Razor, $0$,$1$ lepton, $b$-jets & $\tilde{g}$ decay through $\tilde{b}$, \Tonett\\
ATLAS-13-20 & \cite{ATLAS-SUS-2013-20}   & 1-2 leptons, inclusive                & $\tilde{g}$ decay through $\tilde{l}$, $\tilde{\nu}$, or \chaone \\ 
CMS-13-13   & \cite{CMS-SUS-2013-13}     & $2$l like-sign, $0,1,\geq2$~$b$-jets    & $\tilde{g}$ decay through $\tilde{l}$, $\tilde{\nu}$, or \chaone \\ \hline
\multicolumn{4}{c}{Gluino mediated production of $3^{rd}$ generation sparticles} \\
CMS-12-24  & \cite{CMS-SUS-2012-24}      & \MET+\Ht+$b$-tags 					 & \Tonett  \\
CMS-PAS-14-11&\cite{CMS-PAS-SUS-2014-11} & Razor, $\geq1$~$b$-tag, $0,\geq1$~l     & \Tonett, \Ttt\\
CMS-13-07  & \cite{CMS-SUS-2013-07}      & $1$l, $\geq2$~$b$-tag; MT, LS, $\Delta\Phi(W,l)$ & \Tonett \\
ATLAS-13-09 & \cite{ATLAS-SUS-2013-09}   & $2l$ LS, $3l$, $b$-tags				 & \Tonett \\
ATLAS-13-18 & \cite{ATLAS-SUS-2013-18}   & $0/1$ leptons combined, $3$~$b$-jets    & \Tonett \\ \hline
\multicolumn{4}{c}{Direct top squark production} \\
ATLAS-13-16     & \cite{ATLAS-SUS-2013-16}   & $0$~lepton 	& \Ttt, $\mbox{m}(t)\leq\Delta\mbox{m}$\\
ATLAS-13-15     & \cite{ATLAS-SUS-2013-15}   & $1$~lepton 	& \Ttt, $\Delta\mbox{m}<\mbox{m}(W)$, $\mbox{m}(W)\leq\Delta\mbox{m}<\mbox{m}(t)$\\
ATLAS-13-19     & \cite{ATLAS-SUS-2013-19}   & $2$~lepton 	& \Ttt, $\mbox{m}(W)\leq\Delta\mbox{m}<\mbox{m}(t)$, $\mbox{m}(t)\leq\Delta\mbox{m}$\\
ATLAS-13-21     & \cite{ATLAS-SUS-2013-21}   & compress stop decays: mono-jet, $c$-tag & \Ttt, $\Delta\mbox{m}<\mbox{m}(W)$\\
ATLAS-14-07     & \cite{ATLAS-SUS-2014-07}   & 2l, $M_{T2}$, combinations    & \Ttt, $\Delta\mbox{m}<\mbox{m}(W)$, $\mbox{m}(W)\leq\Delta\mbox{m}<\mbox{m}(t)$, $\mbox{m}(t)\leq\Delta\mbox{m}$ \\
ATLAS-TOP-14-07 & \cite{ATLAS-TOP-2014-07}   & spin-correlation measurement  & \Ttt, $\Delta\mbox{m}=\mbox{m}(t)$ \\
ATLAS-TOP-13-05 & \cite{ATLAS-TOP-2013-04}   & top cross section measurement & \Ttt, $\Delta\mbox{m}=\mbox{m}(t)$ \\ 
CMS-13-23 	    & \cite{CMS-SUS-2013-23}     & 0l  			& \Ttt, $\mbox{m}(t)\leq\Delta\mbox{m}$ \\
CMS-14-01       & \cite{CMS-SUS-2014-01}     & 0l monojet		& \Ttt, $\Delta\mbox{m}<\mbox{m}(W)$\\
CMS-13-11       & \cite{CMS-SUS-2013-11}     & 1l 			& \Ttt, $\mbox{m}(W)\leq\Delta\mbox{m}<\mbox{m}(t)$\\
CMS-14-15       & \cite{CMS-SUS-2014-15}     & 1l,2l 			& \Ttt, $\mbox{m}(W)\leq\Delta\mbox{m}<\mbox{m}(t)$, $\mbox{m}(t)\leq\Delta\mbox{m}$\\
CMS-14-21       & \cite{CMS-SUS-2014-21}     & 1l,2l 			& \Ttt, $\Delta\mbox{m}<\mbox{m}(W)$\\ 
CMS-TOP-14-23   & \cite{CMS-TOP-2014-23}     & spin-correlation measurement & \Ttt, $\Delta\mbox{m}=\mbox{m}(t)$ \\
CMS-TOP-13-04   & \cite{Khachatryan:2016mqs} & top cross section measurement ($e\mu$), \cite{Khachatryan:2015fwh} (hadronic) & \Ttt, $\Delta\mbox{m}=\mbox{m}(t)$ \\ 
\hline
\end{tabularx}
\caption{Overview table of the discussed inclusive analyses and result plots for strong and electroweak production in this article.}
\label{tab:overview1}
\end{table*}

\section{Electroweak production of supersymmetry}
\label{sec:weak}
\begin{figure*}[tb]
\hspace*{\fill}%
\subfigure[$\chaone\chitwo, \tilde{\chi}\to\tilde{l}l'$]{
\includegraphics[scale=1.0]{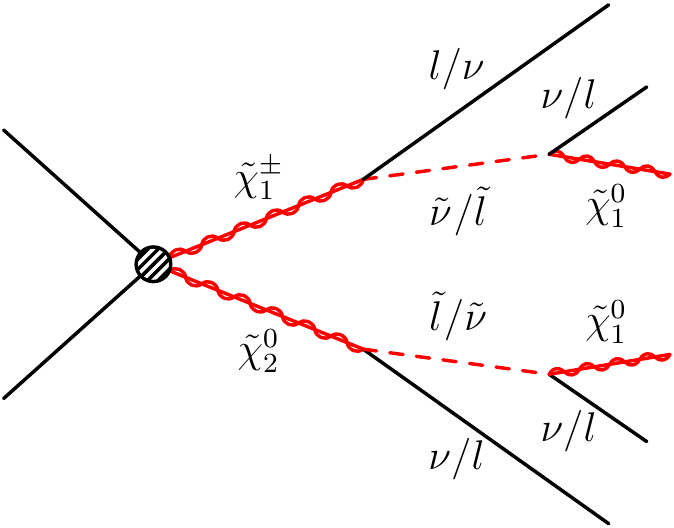}
\label{fig:feyn_cha1chi2_lllnuchichi}
}\hfill
\subfigure[$\chaone\chitwo\to W Z\chione\chione$]{
\includegraphics[scale=1.0]{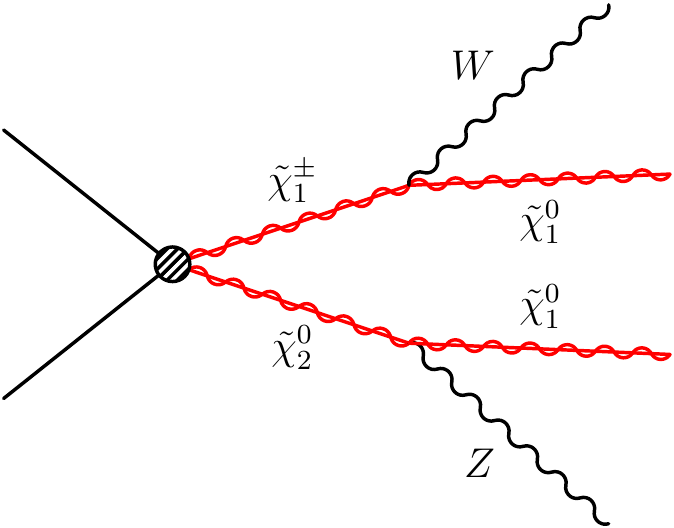}
\label{fig:feyn_cha1chi2_WZchichi}
}
\hspace*{\fill}%
\\

\hspace*{\fill}%
\subfigure[$\chaone\chitwo\to W H\chione\chione$]{
\includegraphics[scale=1.0]{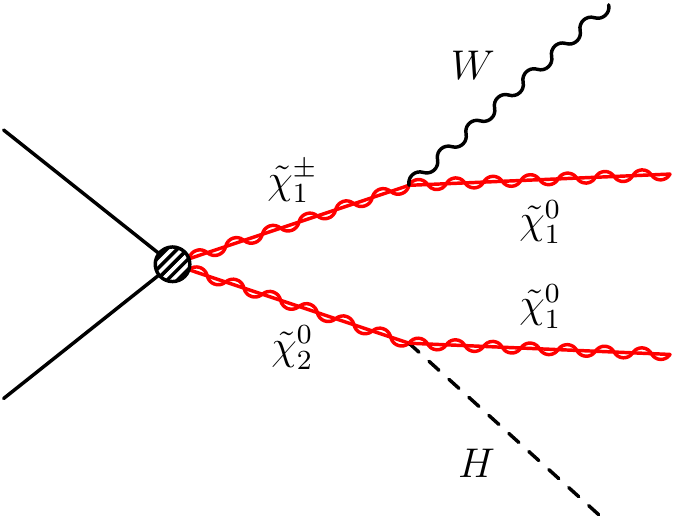}
\label{fig:feyn_cha1chi2_WHchichi}
}\hfill
\subfigure[$\chaone\chaoneT\to W W\chione\chione$]{
\includegraphics[scale=1.0]{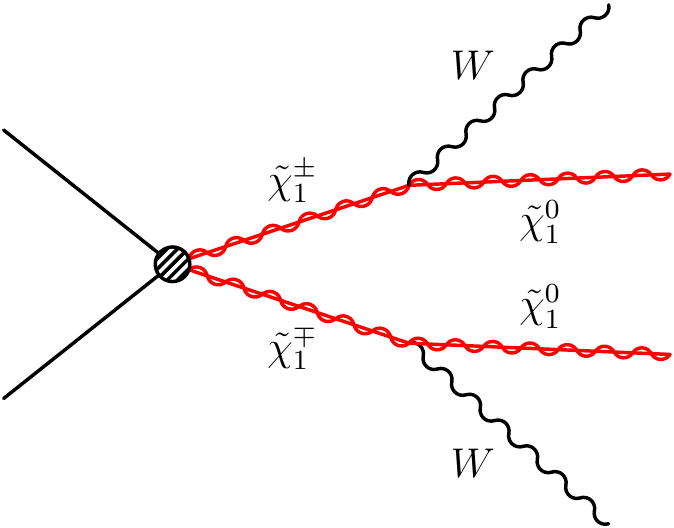}
\label{fig:feyn_cha1cha1_WHchichi}
}
\hspace*{\fill}%
  \caption{Effective Feynman diagrams for electroweak chargino and neutralino production and decay via sleptons or sneutrinos (a),  decay to the LSP and vector bosons (b), decay to LSPs, $W^\pm$ and Higgs boson (c), chargino pair-production and decay to $WW\chione\chione$.}
 \label{fig:feyn_ewk}	
\end{figure*}

Electroweak production of supersymmetric particles has significantly lower cross sections compared to strong production. The amount of energy in the final state is typically smaller compared to events with strong production, making a discovery of new physics in these channels more challenging. On the other hand, existing limits still allow rather light chargino and neutralino masses in contrast to squarks and gluinos. In the following, pair and associated production of charginos and neutralinos are considered, i.e. $\chaone\chitwo$ which generally has the largest production cross-section depending on the mixing and the subdominant $\chaone\chaone$ production, as shown in Fig.~\ref{fig:feyn_ewk}. Pair production of neutralinos is also possible, e.g. through Higgs couplings. %or through strong t-channel interactions.
Since naturalness of supersymmetry requires light higgsinos, the lightest neutralinos and charginos are often assumed to be higgsino dominated in naturalness inspired models. Simplified model scenarios are used to describe four different scenarios, depending on the masses of the involved particles and the decay topology. Charginos and neutralinos can decay hadronically as discussed previously. Decay scenarios through the superpartners of leptons and neutrinos, the sleptons $\tilde{l}$ or $\tilde{\nu}$, or via the \SM bosons $W^\pm$, $Z^0$, $H$  are discussed in detail in the following.

More searches for electroweak production of supersymmetry signal events target the pair production of sleptons $\tilde{l}\tilde{l}$, in particular of staus $\tilde{\tau}\tilde{\tau}$~\cite{ATLAS-SUS-2014-05}. Lower mass limits for staus are still very low, i.e. of the order of $100$~GeV~\cite{pdg2014}. Light staus can explain the observed dark matter relic density~\cite{relicdensity} by co-annihilation processes~\cite{coannihilation}.

\subsection{Decay through sleptons $\tilde{l}$ and neutralinos $\tilde{\nu}$}

Charged leptons or neutrinos are created if heavy gauginos \chaone or \chitwo decay via sleptons $\tilde{l}$ or sneutrinos $\tilde{\nu}$, as shown in Fig.~\ref{fig:feyn_cha1chi2_lllnuchichi}. The generation of the intermediate $\tilde{l}$ or $\tilde{\nu}$ determines the flavor of the two leptons in each decay. The invariant mass of two charged leptons originating from the same gaugino exhibits a kinematic edge. This specific type of signal is discussed in more detail in Sec.~\ref{sec:edge}. In the following, inclusive searches for electroweakly produced supersymmetry with leptons in the final state are discussed.

The simplified models simulate the pair and associated production of mass-degenerate pure wino-like charginos and neutralinos, i.e. \chaone\chitwo and \chaone\chaone, which decay through sleptons and sneutrinos into the bino-like neutralino \chione LSP. The heavy gaugino $m({\chaone})=m(\chitwo)$ and the $m(\chione)$ LSP masses have the strongest impact on the decay topology and are the parameters which are scanned. With respect to the type and the mass of the intermediate slepton or sneutrino, different scenarios are tested, as described in the following and as referred to by the legend of Fig.~\ref{fig:EWK_sl}. 

Generally, decays through $\tilde{e}_L$, $\tilde{\mu}_L$, $\tilde{\tau}_L$, $\tilde{\nu}_e$,  $\tilde{\nu}_\mu$,  $\tilde{\nu}_\tau$ with the same branching fraction and the same mass $m({\tilde{l}})=m(\tilde{\nu}) = x\cdot(m(\chaone)-m(\chione))+m(\chione)$ usually with $x=0.5$ are considered, producing up to four charged leptons in the final state.
Analysis~\cite{ATLAS-SUS-2013-11} targets this scenario requiring two oppositely charged leptons ($e$ und $\mu$) with at least $\pt^{(1)}>35$\,GeV, $\pt^{(2)}>20$\,GeV, and \MET. The dilepton invariant mass for all flavor combinations is required to be larger than $m_{ll}>20$\,GeV and the $m_{T2}$ variable as defined in Eq.~(\ref{eq:mt}) is utilized for the definition of the signal regions. These kinematic requirements achieve good sensitivity to the signal, where the mass gap between the initially produced $\chaone/\chitwo$ and the $\chione$ LSP is approximately $100$\,GeV, and therefore large enough to ensure the efficient reconstruction of the leptons. 

Dedicated analyses have been developed to probe the more challenging region at low $\chaone-\chione$ mass-splittings, as for example like-sign (LS) dilepton searches~\cite{ATLAS-SUS-2014-05,CMS-SUS-2013-06}. Like-sign dilepton states are rare in the \SM, so that less stringent selections and optimization towards compressed signal scenarios are possible.
Searches requiring three charged leptons~\cite{ATLAS-SUS-2013-12,ATLAS-SUS-2014-05,CMS-SUS-2013-06} extend the opposite-sign and like-sign dilepton searches. Because of the clean signal in the final state and the very small \SM background, also compared to the LS final states, the signal sensitivity with respect to the compressed spectra is further increased. 

Analyses with $\tau$ final states~\cite{ATLAS-SUS-2013-14,CMS-SUS-2013-06} are motivated by naturalness arguments. Accordingly, the probed signal models assume pure $\tilde{\tau}_L$ and $\tilde{\nu}_\tau$ mediated decays, where the first- and second generation sleptons and sneutrinos are too heavy to contribute to the \chaone and \chitwo decay widths. The $\tau$-leptons are difficult to identify experimentally. Leptonically decaying $\tau$-leptons create soft electrons or muons and contribute to \MET, hadronically decaying taus are reconstructed as jets and can be identified as $\tau$ using multi-variate methods, such as boosted decision trees or neural nets. The identification efficiency and the misidentification background rejection probability depend strongly on the working point and the number of charged tracks of the tau candidate jet as well as other kinematically correlated properties.   

Searches with four charged leptons~\cite{ATLAS-SUS-2013-13} complete the possible signal phase space with respect to the lepton multiplicity. Reasonable sensitivity can be achieved, which is best for scenarios with compressed spectra, where low lepton momentum thresholds are of advantage, but overall limited by the probability that all four leptons created in the $\tilde{l}$ or $\tilde{\nu}$ mediated decay of $\chaone\chitwo$ or $\chaone\chaone$ production are charged and reconstructed in the geometrical and kinematic acceptance region. Specialized signal scenarios with enhanced 4-lepton probability are studied, where pair-produced pure higgsino-like mass-degenerate heavy neutralinos $\chitwo\chithree$ decay through right-handed sleptons $\tilde{e}_R$ or  $\tilde{\mu}_R$. 

The rich decay topology with four leptons, charged or neutral, of any flavor offers a variety of final states. The searches analyzing different final states are usually defined with exclusive search regions, allowing for all possible combinations. Statistical combinations~\cite{ATLAS-SUS-2014-05,CMS-SUS-2013-06} of different final states increase the sensitivity to certain signal scenarios.

\subsection{Decay through charginos to \SM bosons $W^\pm$, $Z^0$, $H$}

Chargino and neutralino decays through $\tilde{l}$ or $\tilde{\nu}$ have been discussed above. Decays through \SM bosons are also possible, leading to typically smaller lepton multiplicities suppressed by the leptonic branching fraction of the SM bosons. If the $\chaone-\chione$ mass-splitting is sufficiently large, on-shell boson decays, in particular $Z\to l^+l^-$, offer clean signals. All-hadronic final states in spite of the largest branching fraction do not offer enough separation power to \SM processes to search for electroweak production of supersymmetry, except for final states with two Higgs bosons, that produce up to four $b$-jets. Three signal scenarios are studied: $\chaone\chitwo\to W Z\chione\chione$ as shown in Fig.~\ref{fig:feyn_cha1chi2_WZchichi},  $\chaone\chitwo\to W H\chione\chione$ as shown in Fig.~\ref{fig:feyn_cha1chi2_WHchichi}, and $\chaone\chaone\to W W\chione\chione$ as shown in Fig.~\ref{fig:feyn_cha1cha1_WHchichi}, the respective branching fractions are assumed to be $100\%$. 

\begin{figure*}[tbp]
\subfigure[]{\includegraphics[width=0.5\textwidth]{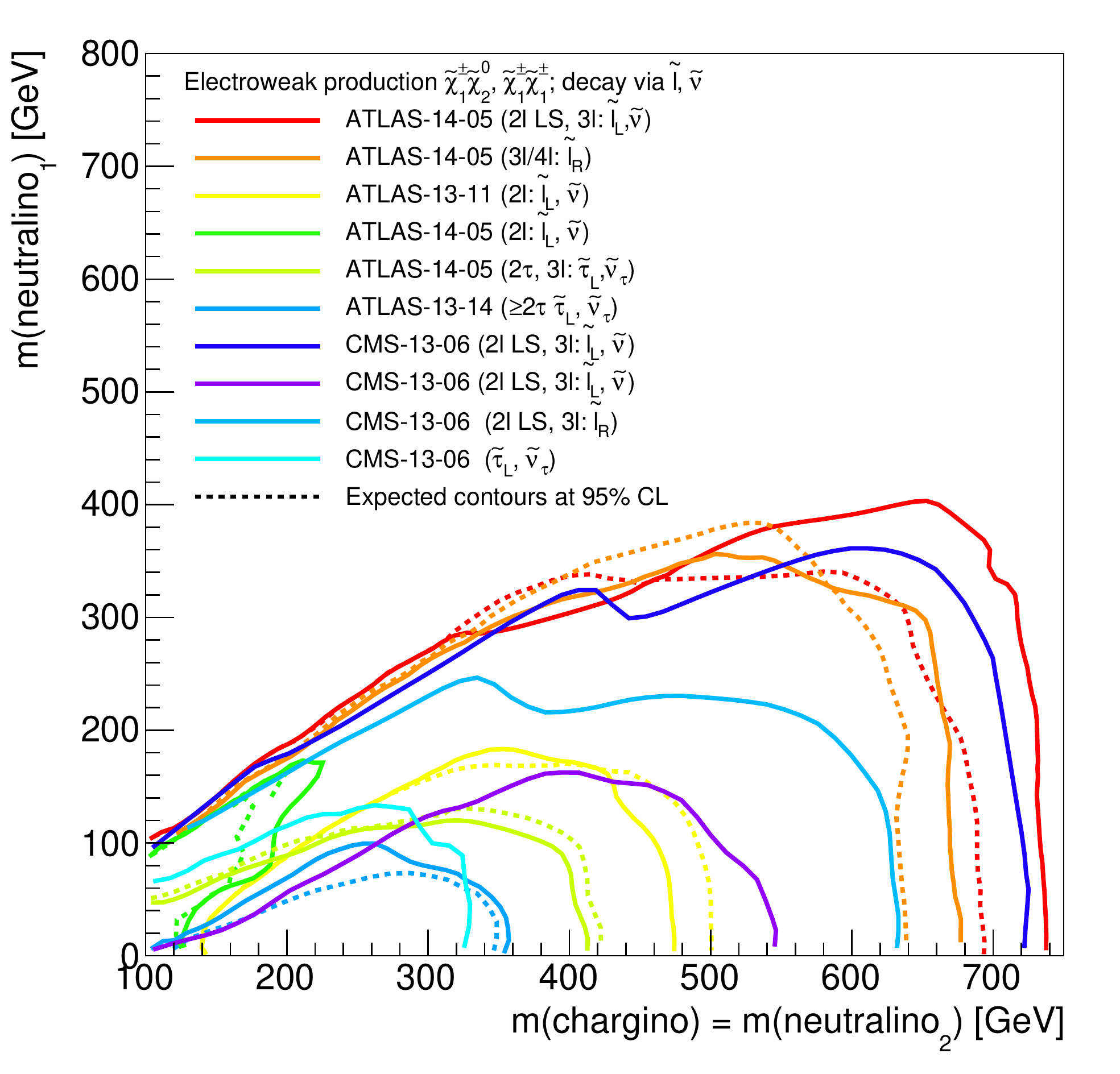}\label{fig:EWK_sl}}
\subfigure[]{\includegraphics[width=0.5\textwidth]{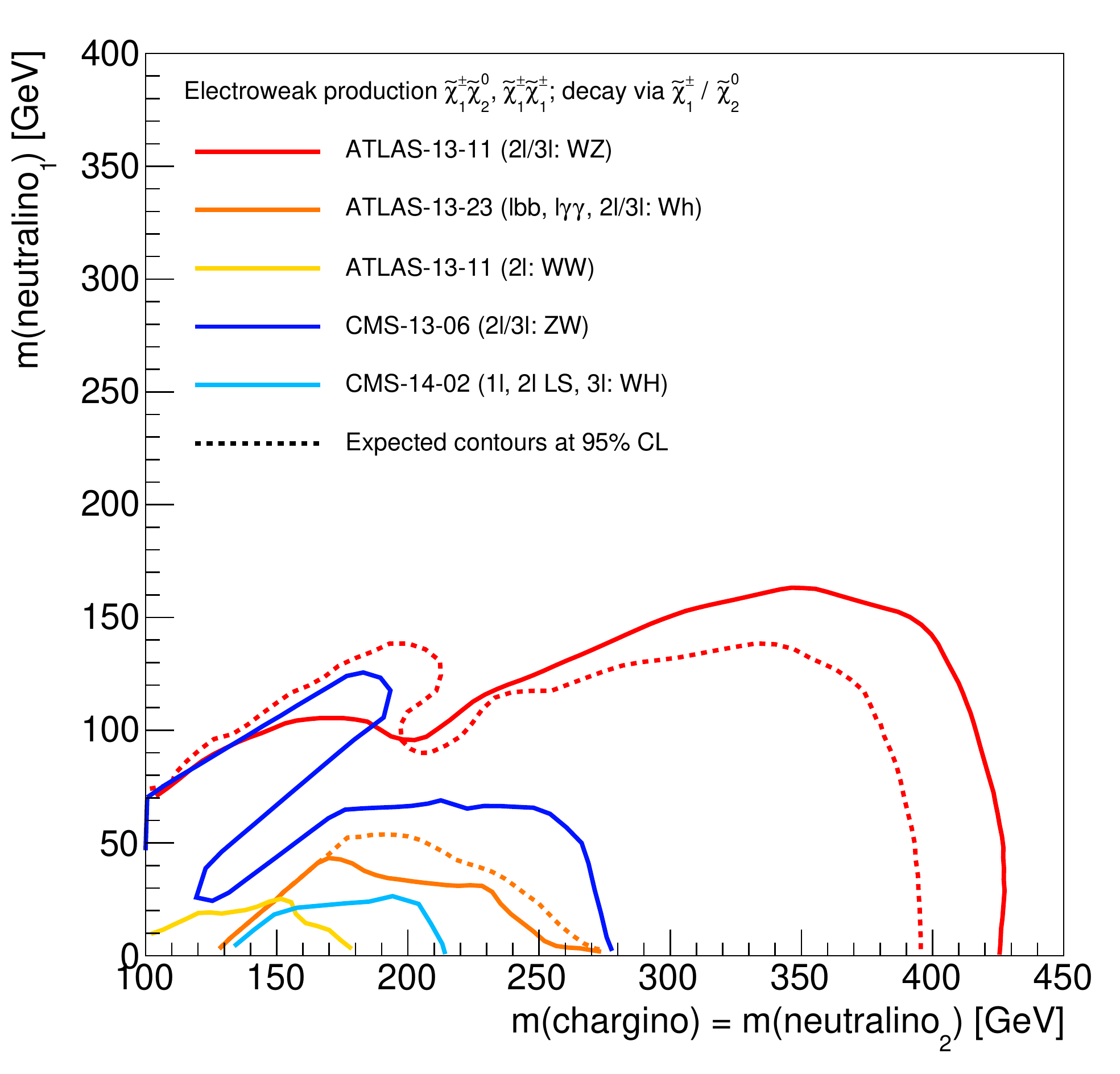}\label{fig:EWK_W}}
  \caption{Exclusion contours at $95\%$~CL for the inclusive searches for electroweak production of supersymmetry events for slepton mediated decays (a) and for decays into \SM bosons $W$, $Z$, $H$ (b). The legend labels refer to Tab.~\ref{tab:overview:ewk}.}
 \label{fig:EWK}	
\end{figure*}

Searches analyzing the $WZ$-bosons final state~\cite{ATLAS-SUS-2013-11,ATLAS-SUS-2013-12,CMS-SUS-2013-06} require two leptons including taus of the same flavor and opposite charge consistent with the $Z$-mass. With respect to the $W$-boson leptonic- as well as hadronic-decay topologies either with an additional third lepton, or two not $b$-tagged jets with an invariant mass consistent with originating from a $W$-boson, are selected. \MET is used to discriminate against the \SM background from $t\bar{t}$, $Wt$ and diboson production. The resulting exclusion contours in the plane of mass-degenerate $\chaone/\chitwo$ and \chione masses are shown in Fig.~\ref{fig:EWK_W}. The \ATLAS sensitivity is superior at large $\chaone/\chitwo$ masses, due to the combination of several individual three-lepton channels. Both contours feature a characteristic ``S''-shaped drop in significance, where the mass difference becomes $\Delta m(\chaone/\chitwo, \chione) = m(Z^0)$, i.e. where the signal events loose significant difference to the \SM $WZ$ process.

For the $WW$-scenario the search strategy~\cite{ATLAS-SUS-2013-11} is similar to the slepton-mediated analyses with two leptons of opposite charge, except that in this case the leptons from on-shell $W$-decays acquire enough energy to pass the kinematic acceptance criteria with high probability. \MET or $m_{T2}$ variables are used to separate the signal. The $\chaone\chaone$ signal cross section is smaller compared to $\chaone\chitwo$ and the signal parameter space is constrained to large mass-splittings, leading to less stringent exclusion contours compared to the $WZ$ case.

In the $WH$-scenario dedicated analyses~\cite{ATLAS-SUS-2013-23,CMS-SUS-2013-06,CMS-SUS-2014-02} require the $W$ to decay leptonically (electron or muon) and examine the different Higgs decay topologies in exclusive signal regions in combination with significant \MET. The dominant Higgs decay is $H\to b\bar{b}$, the corresponding signal region is defined by two additional $b$-tagged jets, suppressing SM backgrounds from $W$+jets and $t\bar{t}$. These backgrounds with $W\to l\nu$ are further suppressed by requiring $m_T(l,\MET)>100$\,GeV. Similarly for the rare $H\to\gamma\gamma$ decay, two photons with an invariant mass consistent within $5$\,GeV of the \SM boson Higgs mass are required. A third channel with two like-sign leptons ($e$ or $\mu$) is considered~\cite{ATLAS-SUS-2013-23}, where in addition to the lepton from $W\to l\nu$ a second lepton from $H\to WW^*$, $H\to ZZ^*$, or $H\to\tau\tau$ can be selected. The like-sign requirement significantly reduces SM backgrounds.

More combination of bosons in the final state are possible. Scenarios with two Higgs bosons or one Higgs and one $Z$-boson in the final state have been examined~\cite{CMS-SUS-2014-02}. Within models of gauge mediated supersymmetry breaking the neutralino decays into the photon and the gravitino become possible. For GMSB events final states with combinations of photons, $W^\pm$, $Z^0$ and Higgs-bosons in the final state are expected, as will be discussed in the following Sec.~\ref{sec:gauge}.

\section{Gauge mediated supersymmetry breaking}
\label{sec:gauge}

\begin{figure*}[tb]
\hspace*{\fill}%
\subfigure[$\chaone\chione\to\gamma Z(H) \Grav\Grav$+soft jets]{
\includegraphics[scale=1.0]{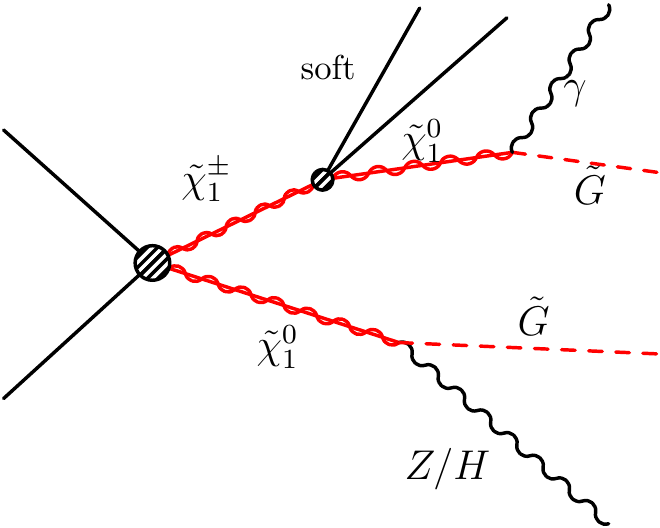}
\label{fig:feyn_cha1chi1_GammaGammaGG}
}\hfill
\subfigure[$\chaone\chaone\to\gamma  Z(H) \Grav\Grav$+soft jets]{
\includegraphics[scale=1.0]{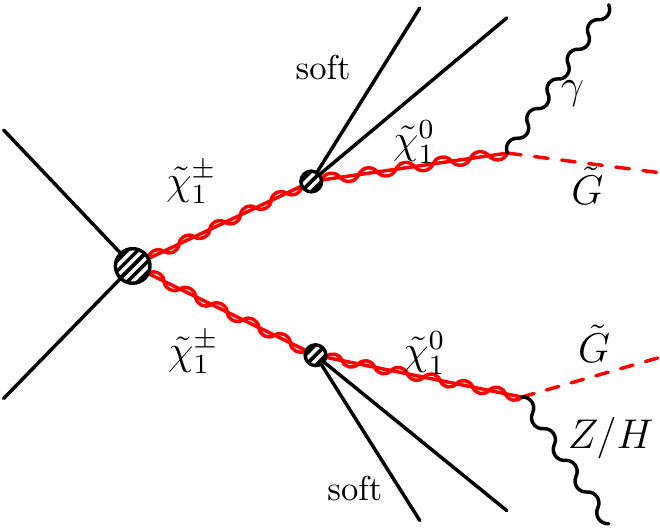}
\label{fig:feyn_cha1cha1_GammaGammaGG}
}\hspace*{\fill}%
\\

\hspace*{\fill}%
\subfigure[$\chaone\chione\to \gamma W \Grav\Grav$]{
\includegraphics[scale=1.0]{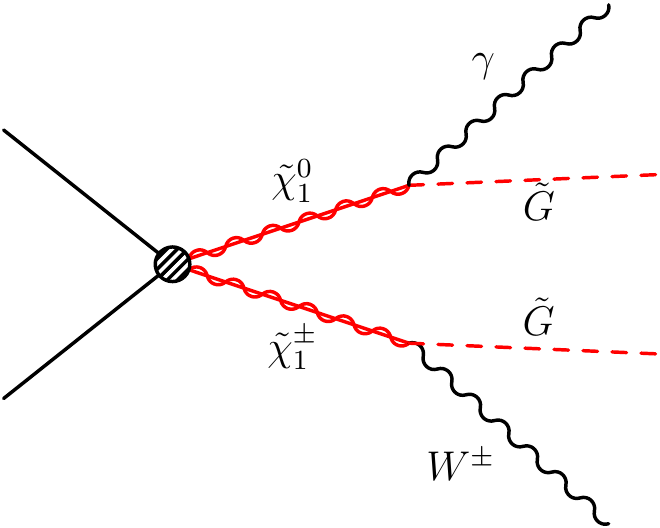}
\label{fig:feyn_cha1chi2_GammaWGG}
}\hfill
\subfigure[$\chaone\chitwo$ in the GGM]{
\includegraphics[scale=1.0]{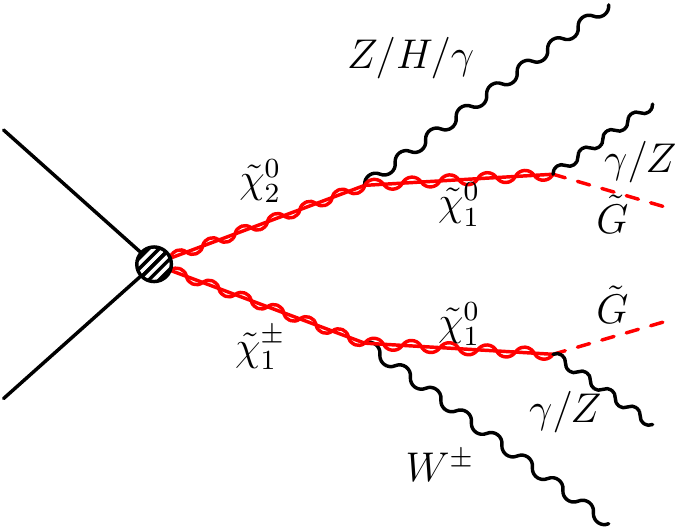}
\label{fig:feyn_cha1chi2_GGM}
}
\hspace*{\fill}%
  \caption{Effective Feynman diagrams for electroweak chargino and neutralino production in supersymmetry with gauge-mediated symmetry breaking. In the scenarios (a, b), the charginos are only slightly heavier than the neutralinos, leading to chargino to neutralino decays accompanied by soft radiation. One neutralino decays to a photon and a gravitino, while the other decays into a Z or an H boson and a gravitino with equal probability. In the Wino co-NLSP scenario (c), the neutralinos and charginos are mass-degenerate. The dominant process for electroweak GGM production is shown in (d). A small amount of hadronic energy and at least one photon and \MET are common features of all scenarios}
 \label{fig:feyn_GMSB}	
\end{figure*}

Different simplified model scenarios with fixed neutralino decays, like e.g $\chione\to\gamma\Grav$ with $100\%$ branching fraction, for strong and electroweak production are used to interpret the analysis results. Additionally, the full supersymmetry model of general gauge mediation is used, where the neutralino NLSP is assumed to be either $100\%$ bino-, wino-, or higgsino-like, which leads to a mixture of final states, in contrast to the studied SMS. The branching fraction of a bino- (or wino-) like neutralino in the GGM $\chione\to\gamma\Grav$ (or $\chione\to Z^0\Grav$) depends on $m(\chione)$ and approaches $\cos^2\theta_W$ if $m(\chione)$ is large compared to $m(Z^0)$ and similarly $\sin^2\theta_W$ for $\chione\to Z^0\Grav$ (or $\chione\to\gamma\Grav$)~\cite{Ruderman:2011vv}, where $\theta_W$ is the electroweak mixing angle. The different scenarios are shown in Fig.~\ref{fig:feyn_GMSB} and are discussed in detail in the following.

\subsection{Strong production and bino- or wino-like \chione NLSP}

Supersymmetric particles are produced through gluino or squark pair- or associated production, as discussed in Sec.~\ref{sec:inclusive}, but unlike to e.g. models with gravity mediated supersymmetry breaking, the cascade decays do not stop at the neutralino, but at the gravitino LSP. From the decays of the two neutralinos at least two \SM bosons are created, their type depends on the model scenario and in particular on the neutralino mixing. Two exemplary inclusive searches for the strong production of GMSB supersymmetry are presented, that require photons in the final state~\cite{CMS-SUS-2014-04,ATLAS-SUS-2014-01,ATLAS-EXO-2014-06} and therefore implicitly bino- or wino-like neutralino-mixings.

Two different search strategies are pursued in the analysis~\cite{CMS-SUS-2014-04}. %GMSB
%Search for supersymmetry with photons in pp collisions at sqrt(s) = 8 TeV , https://twiki.cern.ch/twiki/bin/view/CMSPublic/PhysicsResultsSUS14004
Either final states with at least one single photon and at least two jets are required, or at least two photons and one jet. While final states with single photons are ideal for wino-like neutralino mixing scenarios in the GGM, where photons in each \chione decay are suppressed by $\sin^2\theta_W$ for large $m(\chione)$, they also exhibit good sensitivity to bino-like scenarios, where diphoton final states occur with about $64\%$, and single-photon final states with $32\%$ probability. The two search strategies are also complementary with respect to the signal to SM background discriminating search variables: The single photon search uses bins in \MET to search for a signal, while the diphoton analysis makes use of the Razor variables introduced in Eq.~(\ref{eq:razor}). The \SM background are dominated by $\gamma\gamma$, $\gamma$+jet and QCD-multijet production, where jets can be misreconstructed as photons. This background is estimated using the data sideband with loosely isolated photons or by a fit to the Razor variables shapes, respectively for $1\gamma$ and $2\gamma$. A subdominant background for the single-photon analysis arises from electrons misidentified as photons, when their track is not reconstructed or properly assigned to the candidate particle. This background is modeled using an electron dataset, weighted according to the misreconstruction probability.

\begin{figure*}[tb]
\subfigure[]{\includegraphics[width=0.5\textwidth]{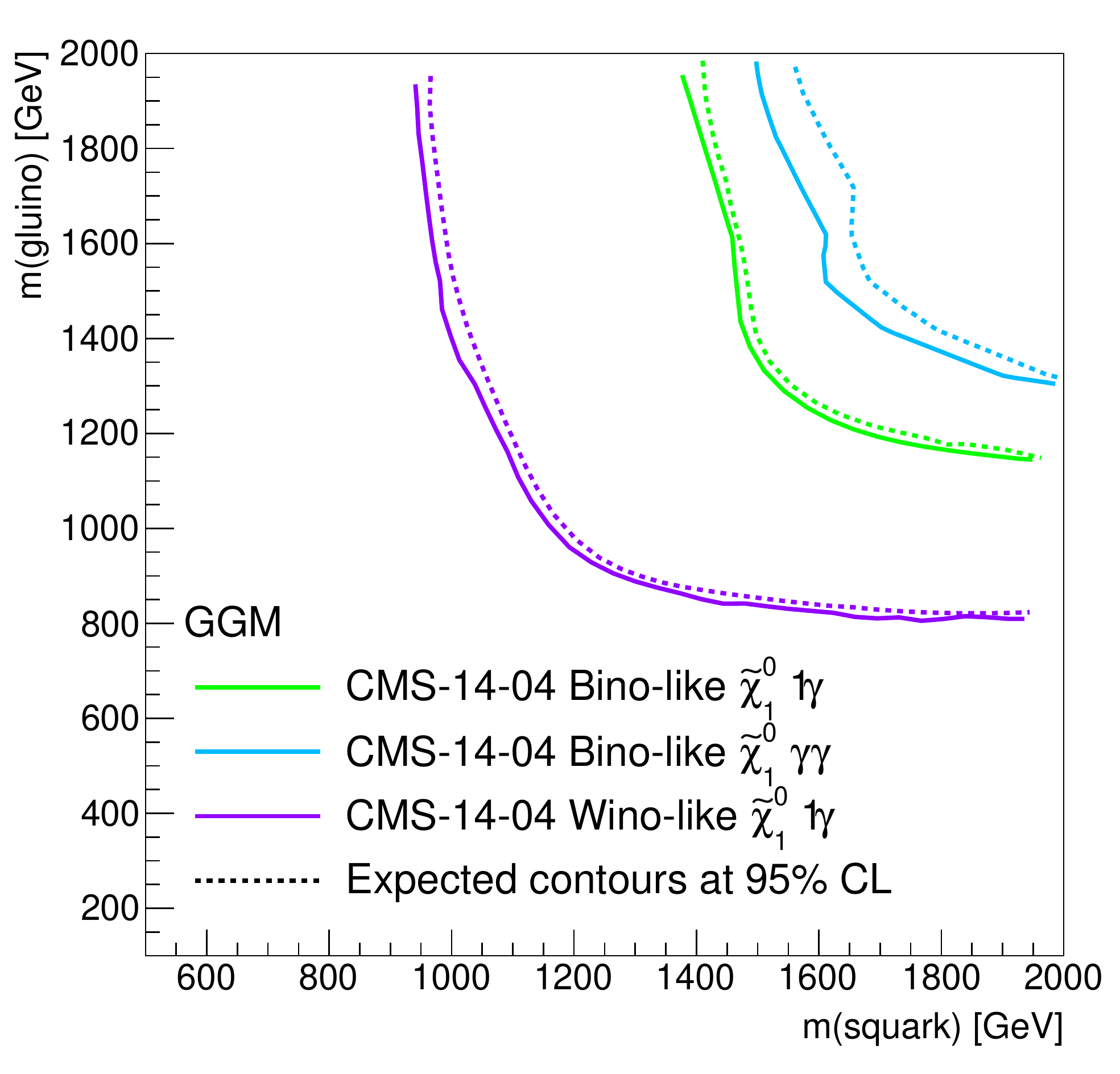}\label{fig:GGM_sq_gl}}
\subfigure[]{\includegraphics[width=0.5\textwidth]{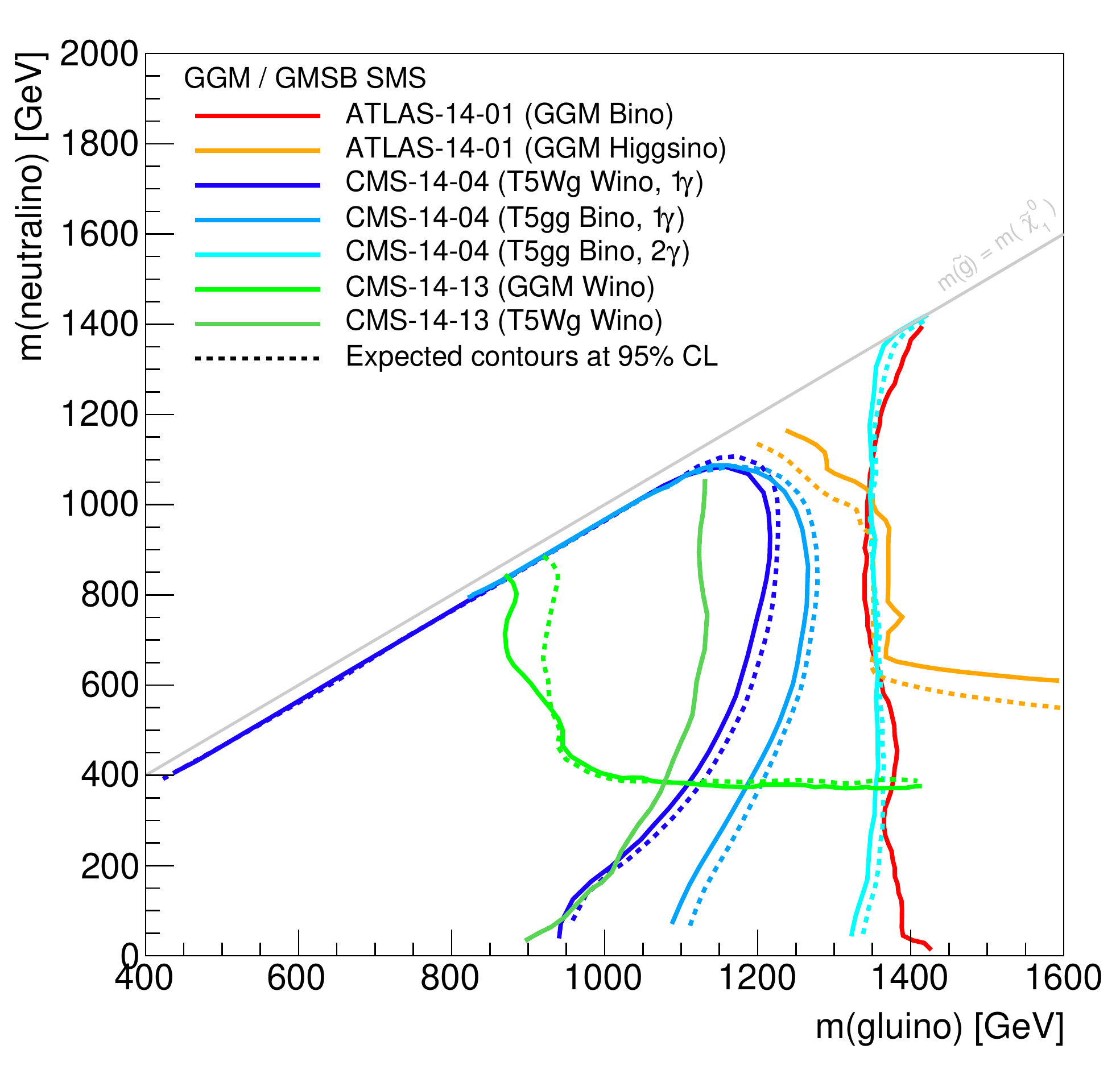}\label{fig:GGM_gl_chi}}
  \caption{Exclusion contours at $95\%$~CL for the inclusive searches for gauge mediated supersymmetry breaking signal events in the model of General Gauge Mediation (GGM) and simplified model scenarios for gluino-pair production with two different decay modes to two photons or one photon and $W$-boson in addition to \MET, as discussed in the text, in the squark - gluino mass plane (a) and in the m($\tilde{g}$) - m(\chione) plane (b). The legend labels refer to Tab.~\ref{tab:overview2}.}
 \label{fig:GGM_strong}	
\end{figure*}

Upper limits are set at $95\%$~CL on the production cross sections and translated to exclusion contours for general gauge-mediation (GGM) models as shown in Fig.~\ref{fig:GGM_sq_gl} for the squark - gluino mass plane. Similar final states are examined by~\cite{ATLAS-SUS-2014-01}. The results obtained in the GGM are shown in Fig.~\ref{fig:GGM_gl_chi}, and are compared to the sensitivity of the discussed analysis~\cite{CMS-SUS-2014-04} on simplified models of gluino-pair production. Two different simplified scenarios are considered. In the first scenario, both gluinos undergo a three-body decay $\tilde{g}\to qq\chione$, followed by the decays of both neutralinos to $\chione\to\gamma\Grav$ with a branching fraction of $100\%$. In the second scenario the gluinos decay as  $\tilde{g}\to qq\chione$  and $\tilde{g}\to qq'\chaone$ to a neutralino and a chargino degenerate in mass, followed by $\chione\to\gamma\Grav$ and $\chaone\to W^\pm\Grav$ with $100\%$ branching fraction.

%\cite{ATLAS-SUS-2013-10}% GMSB
% 1-2 taus + 0-1 leptons + jets + Etmiss [GMSB], 07/2014
% https://atlas.web.cern.ch/Atlas/GROUPS/PHYSICS/PAPERS/SUSY-2013-10

% 3 leptons (e,mu,tau) + Etmiss [chargino/neutralino], 02/2014
% https://atlas.web.cern.ch/Atlas/GROUPS/PHYSICS/PAPERS/SUSY-2013-12
GMSB scenarios where the NLSP is either a stau $\tilde{\tau}$, a smuon $\tilde{\mu}$ or a sneutrino $\tilde{\nu}$ have also been studied~\cite{ATLAS-SUS-2013-12,ATLAS-SUS-2013-10}. In strong production of gluinos and squarks, followed by cascade decays into the NLSP, in addition to jets taus or leptons and \MET are created in the NLSP decay. For scenarios where the $\tilde{\tau}$ is the NLSP, gluino mass of up to $1090$\,GeV are excluded independently of the stau mass.

\subsection{Third generation squarks and higgsino-like \chione NLSP}
%\cite{ATLAS-SUS-2013-08}% GMSB (st-chi plot)
% Z + b-jet + jets + Etmiss [Stop in GMSB, stop2], 03/2014
% https://atlas.web.cern.ch/Atlas/GROUPS/PHYSICS/PAPERS/SUSY-2013-08/

%\cite{CMS-SUS-2013-14}%GMSB, Third & Higgs
%Search for SUSY Partners of Top and Higgs Using Diphoton Higgs Decays in pp collisions at 8 TeV, https://twiki.cern.ch/twiki/bin/view/CMSPublic/PhysicsResultsSUS13014

\begin{figure*}[tb]
\hspace*{\fill}%
\subfigure[]{
\includegraphics[scale=1.0]{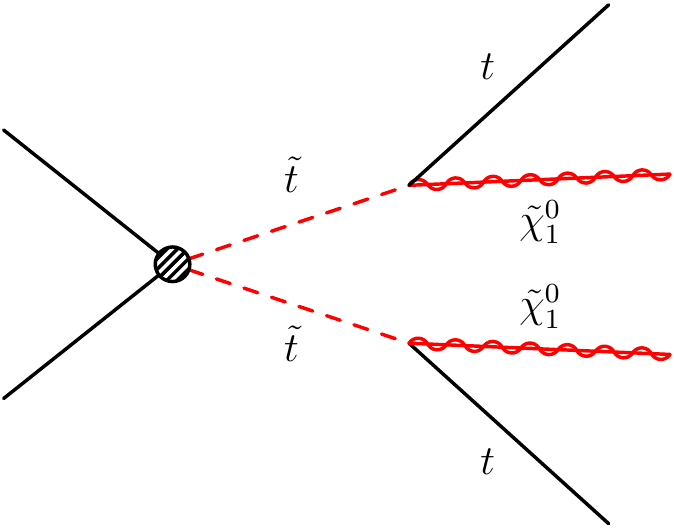}
 \label{fig:GGM_stst}	
}\hfill
\subfigure[]{
\includegraphics[width=0.5\textwidth]{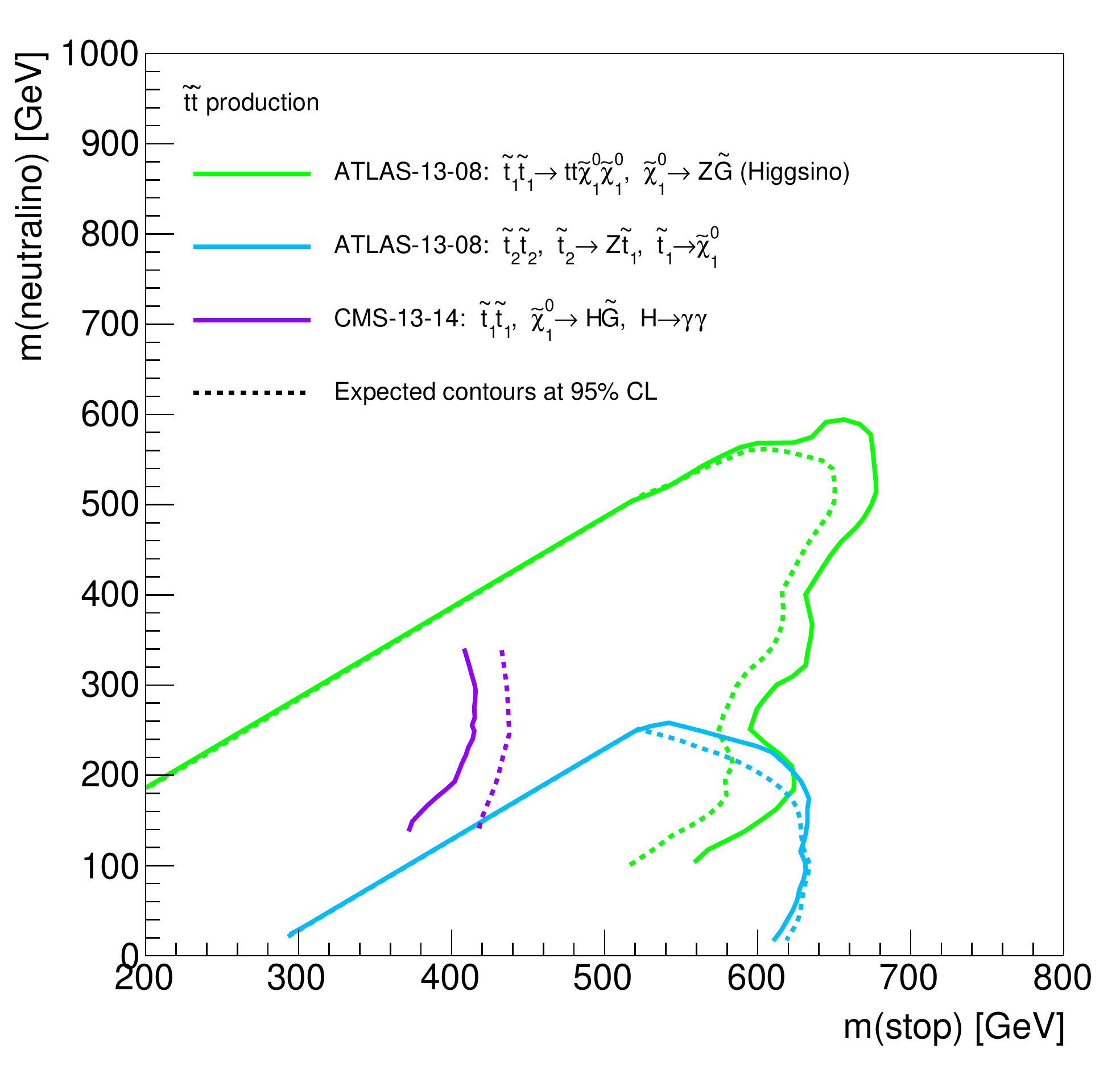}
 \label{fig:GGM_Third_lim}	
}
\hspace*{\fill}%
  \caption{Direct stop pair production (a) and exclusion contours at $95\%$~CL for top squark production in the framework of supersymmetry models of gauge mediated supersymmetry breaking (b). The legend labels refer to Tab.~\ref{tab:overview2}.}
 \label{fig:GGM_Third}	
\end{figure*}

As argued before, gluinos, third generation squarks and higgs\-ino-like neutralinos are connected by the SUSY naturalness arguments and might be the only directly accessible supersymmetry particles. Analyses~\cite{ATLAS-SUS-2013-08, CMS-SUS-2013-14} have searched for direct stop-production (Fig.~\ref{fig:GGM_stst}) in the context of gauge-mediated supersymmetry models. The \ATLAS search~\cite{ATLAS-SUS-2013-08} for direct stop production with two leptons consistent with originating from a $Z\to l^+l^-$ decay has been discussed in Sec.~\ref{sec:third_ontop}. This search has sensitivity to GMSB supersymmetry with higgsino-like neutralinos, which decay like $\chione\to H\Grav$ and $\chione\to Z\Grav$. 

The stop-higgsino search~\cite{CMS-SUS-2013-14} assumes that only the supersymmetric partners of the top-quark and the Higgs boson (higgsino-like neutralino) are accessible, thus motivating direct stop pair production. In the stop decays \SM $b$-quarks via $\tilde{t}_1\to b\chaone$ and Higgs bosons $H$ via $\chione\to H\tilde{G}$ are created. Additional very soft quarks or leptons originate from the transitions between the nearly mass-degenerate higgsinos $\chaone\to f f'\chione$. Events with two photons consistent with originating from the decay of one \SM Higgs $H\to\gamma\gamma$ and at least two $b$-quark jets are selected by the analysis. The only recently discovered Higgs boson and its measured mass value are used in this analysis to identify the signal and to suppress diphoton backgrounds. The \SM background is dominated by $t\bar{t}$ production and is estimated using the sidebands of the diphoton invariant mass. Background from SM Higgs production was found to be negligible due to the \MET requirement. The analysis sensitivity suffers from the small branching fraction $H\to\gamma\gamma$ compared to bino or wino neutralino mixing scenarios with $\gamma$, $Z$ or $W^\pm$ bosons in the final state.

The sensitivities on GMSB $\tilde{t}_1\tilde{t}_1$ production are shown in Fig.~\ref{fig:GGM_Third} and are compared to the simplified production scenario $\tilde{t}_2\tilde{t}_2$ with a neutralino LSP, as discussed in Sec.~\ref{sec:third_ontop}.

%\cite{ATLAS-CONF-2015-01}%ATLAS: Higgs to photon +MET (VBF) [PAS, GMSB, NMSSM] 	02/2015 

\subsection{Electroweak production in GMSB}

Pair and associated production of charginos and neutralinos in models of gauge mediated supersymmetry breaking as shown in Fig.~\ref{fig:feyn_cha1chi2_GGM} have the advantage of two additional bosons from the \chione decays that populate the final state, compared to the previously discussed EWK production in models with \chione LSPs. \MET is generated by the escaping gravitinos and depends on the mass scale of the produced charginos and neutralinos. Depending on the mass gap between the initially produced gauginos and the neutralino NLSP $\Delta m(\chaone/\chitwo, \chione)$, not much additional energy is present in the events. If the mass gap is small, then generally soft jets or leptons are created as shown in Fig.~\ref{fig:feyn_GMSB}(a,b). If $\Delta m$ is sufficiently large, additional on-shell \SM bosons can be created, as shown in Fig.~\ref{fig:feyn_cha1chi2_GGM}.

The \ATLAS analysis~\cite{ATLAS-SUS-2014-01} incorporates four independent searches for GMSB supersymmetry in final states with diphotons, $\gamma+b$-jet, $\gamma$+jets, or $\gamma+$lepton final states. All of these have sensitivity to inclusive strong production, but the $\gamma+b$-jet analysis is specialized on third generation quark production and $\gamma+$lepton final states have best sensitivity in electroweak production. The signal region with at least one photon and one lepton is comparable to the CMS search~\cite{CMS-SUS-2014-13} and targets wino-like neutralino mixings, where the photon originates from the $\chione\to\gamma\Grav$ NLSP and the lepton is created through a $Z$ or $W$ boson from the \chione NLSP or the \chaone co-NLSP decay, respectively, as shown in Fig.~\ref{fig:feyn_cha1chi2_GammaWGG}. Together with the $\gamma+\MET$ signature~\cite{CMS-SUS-2014-16} the $l+\gamma$ completes the searches for electroweak production of wino-like \chione without photons in the final state with two or three leptons~\cite{ATLAS-SUS-2013-11,CMS-SUS-2013-06} as discussed in Sec.~\ref{sec:weak}.

The photon and lepton event signature~\cite{ATLAS-SUS-2014-01,CMS-SUS-2014-13,CMS-SUS-2014-09} with large transverse momentum is sufficient to efficiently trigger the signal events and to suppress many SM backgrounds. Therefore, no additional requirements with respect to hadronic energy such as jets are necessary, making the analyses particularly sensitive to electroweak direct production of gauginos compared to the previously discussed analyses targeting GMSB, while maintaining sensitivity to strongly produced SUSY involving wino-like NLSPs. The remaining SM background is dominated by electroweak production, e.g. $Z\gamma^*\to ee$ or $W$+jets, where an electron or jet is misreconstructed as photon. 

The photon+\MET analysis~\cite{CMS-SUS-2014-16} targets electroweak production with a bino- or wino-like \chione NLSP with low neutralino and chargino masses and mass differences. The selection with one $\gamma$ and large \MET covers the phase space, where the gravitinos carry away a significant fraction of the energy released by the two gaugino decays. A second photon or a lepton may escape the geometrical or kinematic acceptance.  To select signal events with lowest possible transverse momentum thresholds a special data-set corresponding to $7.4$\,fb$^{-1}$ with relaxed trigger criteria is used, that was recorded during the second half of the 2012 data-taking period but only reconstructed during the Long Shutdown~1 of the LHC as part of the so-called ``parked-data'' program~\cite{CMS-DP-2012-022}. The composition of the \SM background and its estimation is comparable to the photon+lepton analyses.

\begin{figure*}[tb]
\begin{center}
\includegraphics[width=0.7\textwidth]{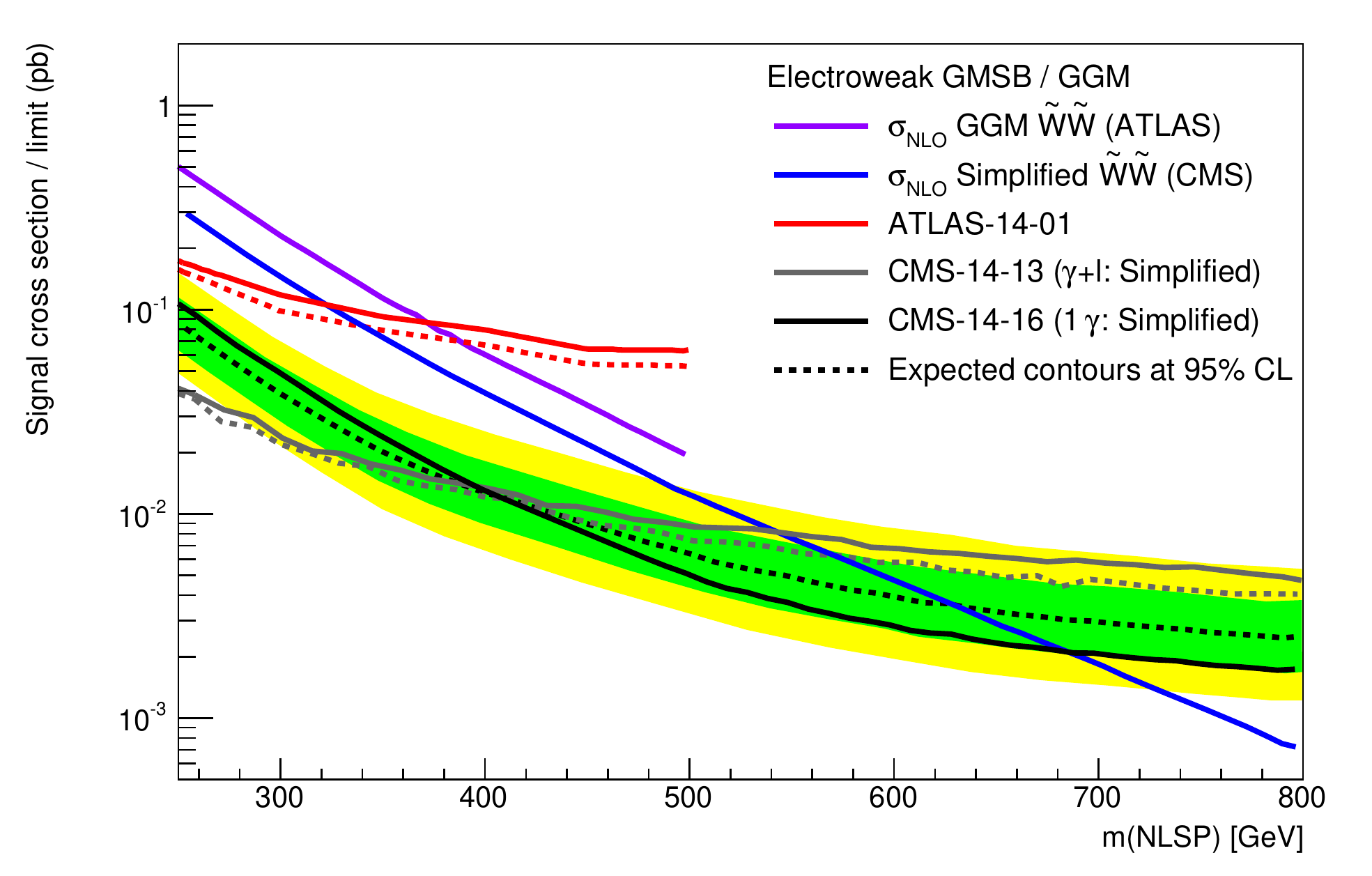}
  \caption{Cross sections and cross section limits at $95\%$~CL for electroweak production of signal events from gauge mediated supersymmetry breaking models for different neutralino mixing assumptions. The legend labels refer to Tab.~\ref{tab:overview2}.}
 \label{fig:GGM_EWK1D}	
\end{center}
\end{figure*}

The results of the searches for electroweak production in GMSB are shown in Fig.~\ref{fig:GGM_EWK1D}. Two different scenarios are studied. The \ATLAS photon+lepton search~\cite{ATLAS-SUS-2014-01} uses a GGM scenarios with three wino-like co-NLSP, including the neutral $\tilde{W}^0$ and the $\tilde{W}^\pm$ wino states. The $\tilde{W}^0$ decays into photon and \Grav for small wino masses below the $Z$-boson mass. The branching fraction approaches $\sin^2\theta_W$ for heavy wino masses above the $Z$-mass. The \CMS lepton+photon analysis~\cite{CMS-SUS-2014-13} and the single-photon analysis~\cite{CMS-SUS-2014-16} assume a simplified wino co-NLSP scenario, where mass-degenerate \chione and \chaone are produced and directly decay into gravitinos and one photon, or $\tilde{G}$ and one $W$-boson, respectively, with $100\%$ branching fraction. Here,
NLSP masses up to $680$\,GeV can be excluded at $95\%$ CL. For the scenario shown in Fig.~\ref{fig:feyn_GMSB}(a,b), where the \chaone and \chitwo masses are only slightly heavier than the \chione NLSP leading to longer decay chains with three-body decays $\tilde{\chi}^{\pm,0}\to ff'\chione$ into the NLSP, neutralino NLSP masses below $570$\,GeV can be excluded.

\section{Resonances and kinematic edges}
\label{sec:edge}
Searches for resonances or kinematic edges in the dilepton invariant mass spectrum~\cite{CMS-SUS-2014-14,ATLAS-SUS-2014-10,ATLAS-SUS-2013-06} are examples for highly specialized supersymmetry analyses. The sensitivity to the targeted signal scenarios is unmatched by the inclusive searches that aim at a broad range of models and signal parameter space. The implicit model dependence limits the searches to signal scenarios, where the resonance or kinematic edge is produced with sufficient probability. 

\subsection{Same-flavor opposite-charge dileptons}\label{sec:edge}

\begin{figure*}[tb]
\hspace*{\fill}%
\subfigure[$\tilde{g}\tilde{g}\to Z$+jets+\MET]{
\includegraphics[scale=1.0]{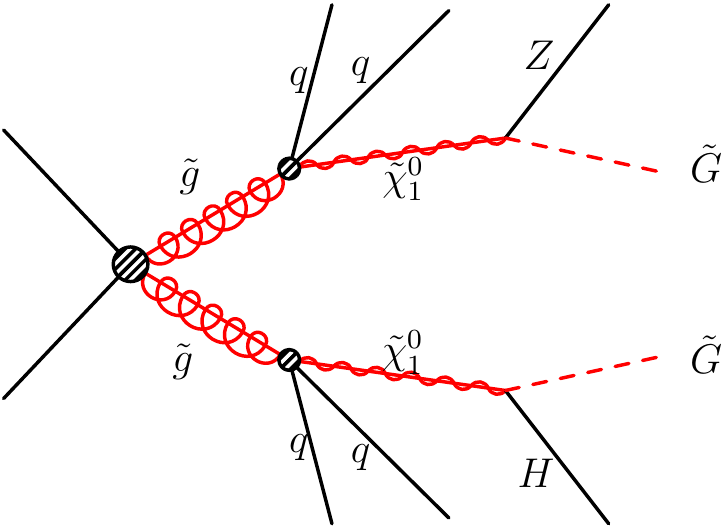}
\label{fig:feyn_glgl_GMSB}
}\hfill
\subfigure[$\tilde{q}\tilde{q}\to l^+l^-$+jets+\MET]{
\includegraphics[scale=1.0]{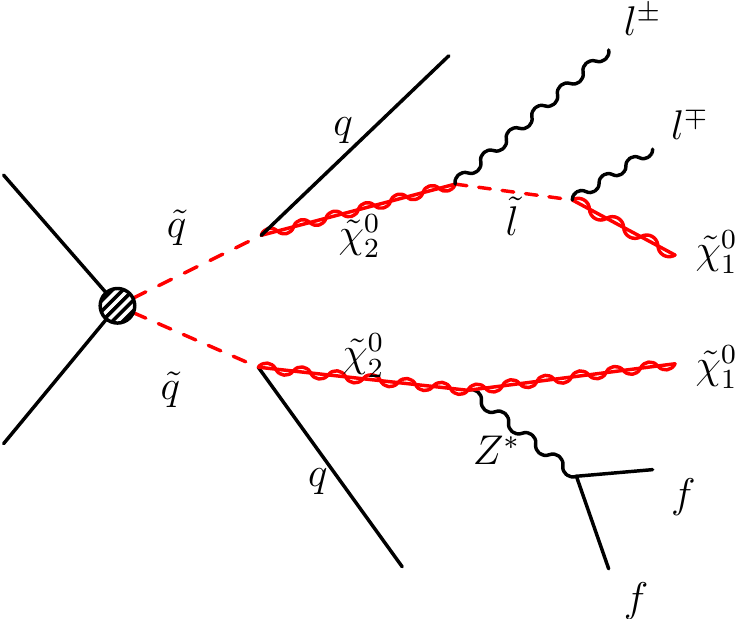}
\label{fig:feyn_sqsq_edge}
}
\hspace*{\fill}%
  \caption{Effective Feynman diagrams for gluino mediated production of dileptons, jets, and \MET through on-shell $Z^0$ decays or Higgs $H$ (a) or through three-body decays of gauginos through sneutrinos or sleptons, or off-shell $Z$ bosons (b).}
 \label{fig:feyn_edge}	
\end{figure*}

A dilepton resonance occurs naturally in models of supersymmetry, when for example an on-shell \SM boson is produced in the cascade decays, e.g. $Z\to\mu\mu$ or $H\to\tau\tau$, as shown in Fig.~\ref{fig:feyn_glgl_GMSB}. Another possibility are $R$-parity violating $LL\bar{E}$ couplings $\lambda_{ijk}$, through which a sneutrino $\tilde{\nu}$ is able to decay into a lepton-pair of opposite-charge and possibly of same-flavor $\tilde{\nu}\to l^+l^-$. More common are kinematic edges in the invariant dilepton mass spectrum, originating from three-body decays of gauginos via sleptons $\tilde{l}$ or sneutrinos $\tilde{\nu}$ e.g. $\chitwo\to\tilde{l}^+l^-\to\chione l^+l^-$, or off-shell $Z$-bosons $\chitwo\to Z^*\chione$, as shown in Fig.~\ref{fig:feyn_sqsq_edge}. The invariant mass of the dilepton system shows a characteristic triangular form. The position of the upper ``edge'' is determined by the maximal energy the dilepton system can obtain, approximately given by the mass difference of the decaying heavy gaugino and the invisible stable gaugino at the end of the decay chain, i.e. $\Delta m(\chitwo,\chione)$ for the given example. The curvature of the lower edge of the triangle is influenced by the mass of the intermediate slepton or sneutrino. 

The analysis~\cite{CMS-SUS-2014-14} received some attention, because an excess of the order of $2.5\,\sigma$ was reported for an edge position at about $80$\,GeV in the $e^+e^-$ and $\mu^+\mu^-$ final states. A feature of the analysis is the relative insensitivity to the details of the production process of the signal and the very reliable and precise estimation of the \SM backgrounds. Jets and a moderate amount of \MET is required in addition to two opposite-sign same-flavor leptons ($e$, or $\mu$). The signal region at low-$m_{ll}$ is dominated by flavor-symmetric backgrounds such as $t\bar{t}$ production, i.e. this type of background produces as many $e^+e^-+\mu^+\mu^-$ events as it has $\mu^+e^-+\mu^-e^+$ events. Therefore, a data control-region with two opposite-charged $e^\pm\mu^\mp$ leptons can be used to model the flavor-symmetric background. The statistical precision is comparable to the irreducible statistical uncertainty of the data in the signal region. Systematic uncertainties arising from potential differences in the reconstruction efficiencies of the electron and muons cancel largely. The remaining correction factor \Rsfof to the opposite-flavor $e\mu$ data control sample can be factorized as:
\begin{equation}
\Rsfof = \frac{1}{2}\left(\sqrt{\frac{N_{\mu\mu}}{N_{ee}}}+\sqrt{\frac{N_{ee}}{N_{\mu\mu}}}\right)\cdot\sqrt{\frac{\epsilon_{ee}\epsilon_{\mu\mu}}{\epsilon_{e\mu}}}
\label{eq:edge}
\end{equation}
where $N_{ee}$ and $N_{\mu\mu}$ are the numbers of selected events in a $e^+e^-$ and $\mu^+\mu^-$ sideband region, and $\epsilon_{ee}$, $\epsilon_{\mu\mu}$ and $\epsilon_{e\mu}$ are the measured trigger efficiencies. \Rsfof is determined by two independent data-driven methods and was found to be consistent with unity with an uncertainty of $4\%$ in the central detector region. This good handle on the dominant background is the reason for the good sensitivity to the studied edge signals. The subdominant resonant background from $Z\to ll$ is also extracted from the data using two methods relying on \MET templates and a variable using the $Z$-momentum balance against jets.

\begin{figure}[tb]
\begin{center}
\includegraphics[width=0.46\textwidth]{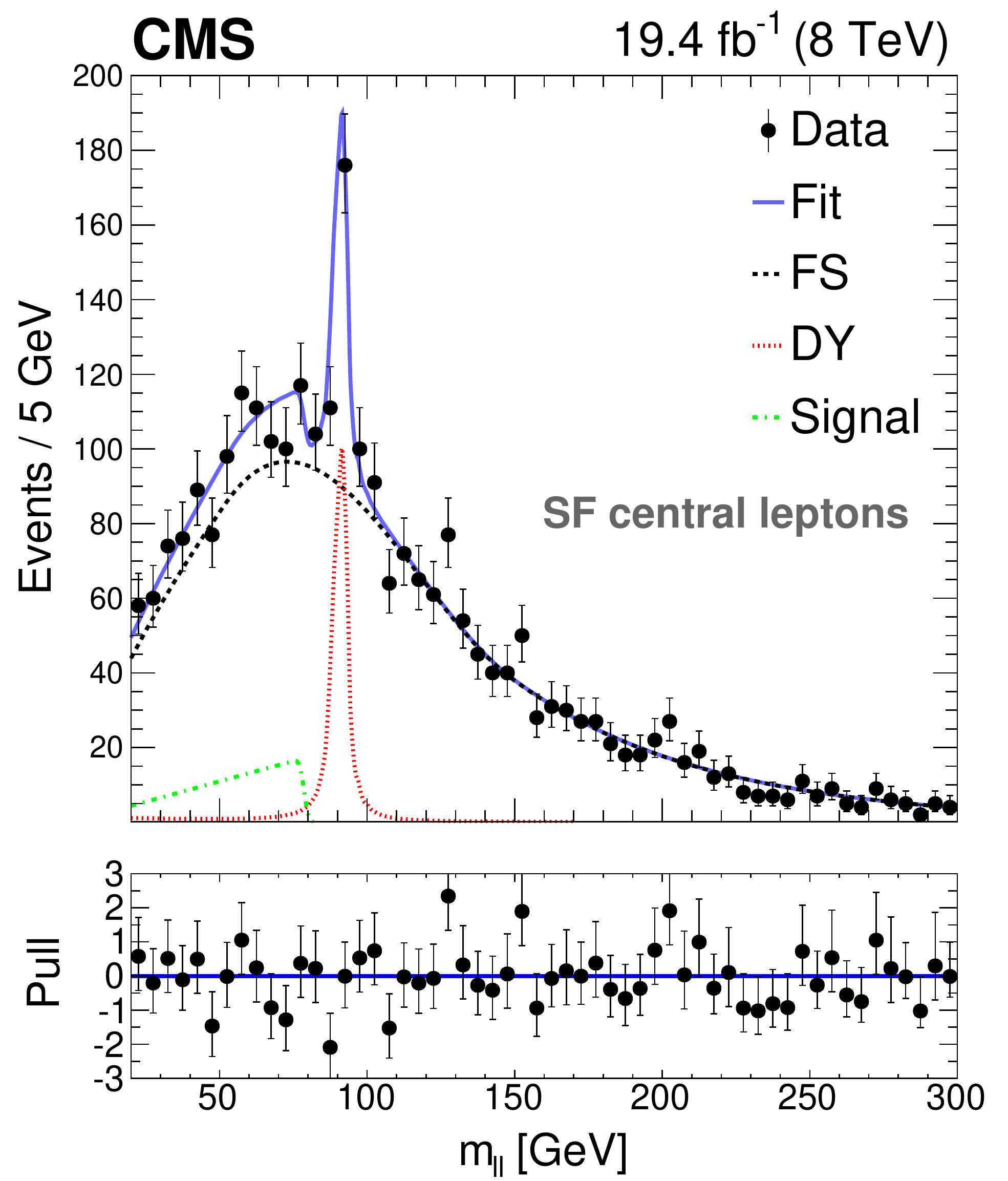}\label{fig:EdgeFit_oR}
  \caption{Fit to the observed data in the central opposite-sign same-flavor signal region of the dilepton edge analysis~\cite{CMS-SUS-2014-14}. The observed local significance of the unbinned fit corresponds to $2.4~\sigma$. A counting experiment in the same invariant dilepton mass region below the $Z$-peak corresponds to a local significance of $2.6~\sigma$.
 \label{fig:theedge}}	
\end{center}
\end{figure}

The analysis~\cite{CMS-SUS-2014-14} proceeds to interpret the observed events with counting-experiments in three regions of $m_{ll}$ and two rapidity regions. For the low-$m_{ll}$ region defined as $20<m_{ll}<70$\,GeV and where both leptons were reconstructed in the central detector $860$~events were observed, while $730\pm40$~events were expected from the combined SM background estimations, corresponding to a local significance of $2.6\sigma$. The five other counting experiments were found to be consistent with the null hypothesis. A second statistical interpretation used an unbinned fit of background  and signal templates simultaneously to all signal and background control regions. Also \Rsfof is allowed to be varied by the fit. In this case $\Rsfof=1.03\pm0.03$ and a signal yield of $126\pm41$~events, corresponding to an edge position at $78.7\pm1.4$\,GeV and a local significance of $2.4$\,$\sigma$ is found, as shown in Fig.~\ref{fig:EdgeFit_oR}, in agreement with the counting experiment result.

While the observed feature in the CMS data~\cite{CMS-SUS-2014-14} is as interesting as it is consistent with the expectation from potential signals~\cite{CMS-SUS-2014-14,Allanach:2014gsa,Huang:2014oza}, the most likely explanation is however a statistical fluctuation of the \SM backgrounds. This is supported by the corresponding \ATLAS analysis~\cite{ATLAS-SUS-2014-10}, which found good agreement with the expectations in a comparable low-$m_{ll}$ selection beneath the $Z$-resonance, and by the preliminary results of the CMS analysis on $2.2$\,fb$^{-1}$ of data collected at center-of-mass energies of $13$\,GeV~\cite{CMS-PAS-SUS-15-011}.

The \ATLAS analysis~\cite{ATLAS-SUS-2014-10} searches for signals from kinematic edges and in a dedicated search region for dileptons ($e$ or $\mu$) from $Z$ decays, aiming for example at models of gauge mediated supersymmetry breaking, where higgsino-like neutralinos can decay like $\chione\to Z\Grav$. The SM background is dominated by flavor symmetric $t\bar{t}$ processes, which are estimated with high precision using opposite-flavor data selections, as discussed previously. The sub-dominant background from SM $Z\to l^+l^-$ is here however of much larger importance, as it mimics the features of a signal resonance. Therefore, also this background is evaluated using the data and validated in other data control regions. In addition, Monte Carlo simulation is used to validate the data-driven prediction. A similar technique as discussed for the QCD-multijet background estimation in all-hadronic searches in Sec.~\ref{sec:inclBKGest} is used, exploiting that also in this case the \MET in the $Z\to l^+l^-$ background is not due to genuine not-interacting particles, but created through a number of effects summarized as jet resolution. Therefore, a data control selection of $Z\to l^+l^-$ events is chosen, which satisfies all signal selection criteria except for \MET, that is required to be small. The jets in these events are then smeared, according to the jet transverse momentum resolution, as a function of the azimuthal angle $\phi$ and the \pt of the jet. The jet resolution function is determined using Monte Carlo simulation and tuned to the data based on a dijet \pt balancing analysis. The \MET is recalculated using the new smeared jets. The data-driven estimation of the resonant $Z\to l^+l^-$ background confirms that this background is negligible in the dilepton on-$Z$ signal region with $\MET>225$\,GeV and a scalar sum of jet and lepton \pt of at least $600$\,GeV, as expected from MC simulation. The remaining small contribution from rare SM backgrounds like diboson-boson or top processes are taken from Monte Carlo simulation and are carefully validated in data-control regions.

The analysis reports an observation of $29$~events in the on-$Z$ signal region, while $10.6\pm3.2$~events were expected from the combined SM background estimation methods, corresponding to an excess with a local significance of $3.0\sigma$. The excess is slightly more pronounced in the di-electron channels compared to di-muons. The on-$Z$ result of the previously discussed CMS analysis is consistent with the \SM expectation, but the selection is much looser defined with respect to \MET and jet+lepton \pt and therefore not directly comparable. The \ATLAS analysis has been repeated on data collected at $13$\,TeV corresponding to $3.2$fb$^{-1}$. The preliminary result~\cite{ATLAS-CONF-2015-082} reports $21$~observed events, while $10.3\pm2.3$~events were expected from the SM background, corresponding to an excess with a local significance of $2.2\sigma$. The corresponding CMS analysis~\cite{CMS-PAS-SUS-15-011} carried out on $2.2$\,fb$^{-1}$ of $13$\,TeV data disfavors a signal and sets an upper limit on $9$~signal events in a comparable signal region.

To summarize, searches for supersymmetry with dilepton resonances or kinematic edges test the data for striking features and reach high sensitivities, as the \SM background can be estimated with high precision. Some interesting fluctuations have been observed in individual search regions at \ATLAS and CMS in $8$\,TeV and $13$\,TeV data, which are --unfortunately-- best explained by statistical fluctuations.

\subsection{Summary of all \CMS SUSY counting experiment results}

\begin{figure}[tb]
\begin{center}
\includegraphics[width=0.5\textwidth]{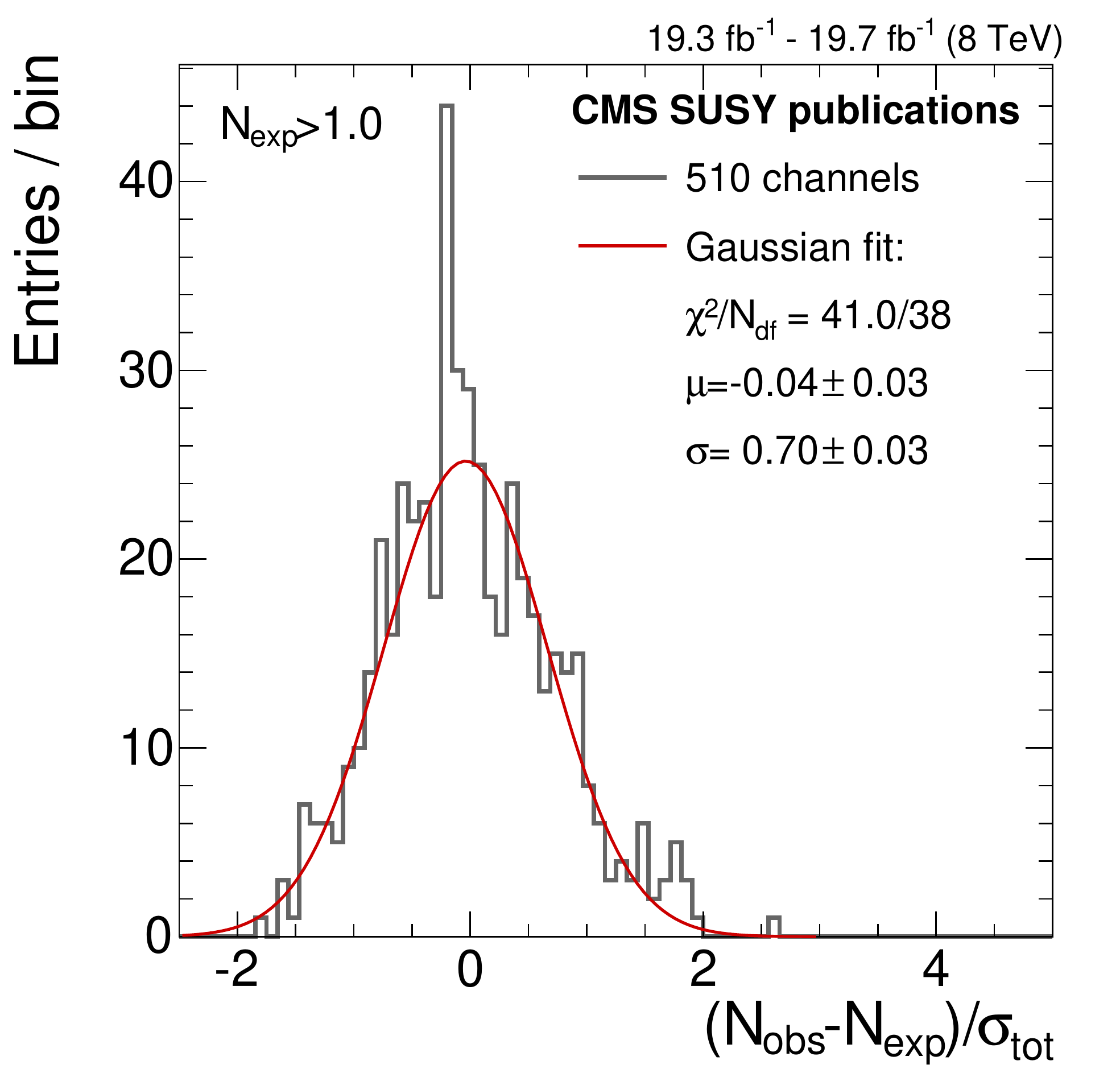}
  \caption{Histogram of the observed local significances in all counting-experiment channels of all published $\sqrt{s}=8$\,TeV CMS SUSY analyses. The largest entry in the histogram at a significance of $+2.6~\sigma$ corresponds to the counting experiment in the invariant dilepton mass region below the $Z$-peak discussed above in Sec.~\ref{sec:edge} and figure~\ref{fig:theedge}. \label{fig:susyResultMy}}
\end{center}
\end{figure}

Figure~\ref{fig:susyResultMy} shows a simplified summary of the observed significances of all counting-experiment channels of all published CMS supersymmetry results at $\sqrt{s}=8$\,TeV until 2015. The significance of a given channel with $N_{\mbox{\tiny obs}}$~observed and $N_{\mbox{\tiny exp}}$~expected events is defined here as $(N_{\mbox{\tiny obs}}-N_{\mbox{\tiny exp}})/\sigma_{\mbox{\tiny tot}}$, where $\sigma_{\mbox{\tiny tot}}$ is a measure of the total uncertainty of the channel, quadratically combining the uncertainties of the background expectation and the data, ignoring all possible correlations. Only channels with $N_{\mbox{\tiny exp}}>1.0$ enter the plot, otherwise the large number of channels, where tiny fractions of background events are expected and no data events are observed, create a sharp peak at very small negative significances in the plot. The obtained distribution has a Gaussian shape with a mean consistent with zero and a width slightly smaller than unity, which could be caused by the neglected correlations. The largest absolute value of the significance in the plot is $+2.6\sigma$ from the low-$m_{ll}$ counting experiment of the edge-analysis discussed above. The plot summarizes nicely the consistency of the observed data in the supersymmetry search channels with the expectations from the \SM background within the expected systematic and statistical fluctuations.

\section{Conclusion and Outlook}
\label{sec:conclusion}

The \ATLAS and \CMS experiments have searched in the Run~I dataset corresponding to a luminosity of about $20$\,fb$^{-1}$ collected at $8$\,TeV center-of-mass energy for supersymmetry, superseding the previous LHC results at $\sqrt{s}=7$\,TeV, and at different colliders such as the Tevatron or LEP. No signs of supersymmetry have been observed in the data. Cross section limits in all relevant final states have been derived and interpreted in different models of supersymmetry. In this article a selected set of the most relevant exclusion contours have been summarized and compared with the help of simplified model scenarios parametrized by the supersymmetric particle masses.

The \ATLAS and \CMS analyses have carried out a large number of different search strategies and in particular used complementary techniques to estimate the background from \SM processes. The discussed analyses cross-checked and validated the background estimation usually with the help of the data, as the precise description of the \SM expectation is of crucial importance. 

The most stringent mass limit is set on the gluino $\tilde{g}$ of up to $1.4$\,TeV in simplified scenarios, where the gluino decays promptly through off-shell squarks of any flavor and where the lightest supersymmetric particle is either a light neutralino \chione or the gravitino \Grav. In the cMSSM, phase space with gluino masses up to $1.8$\,TeV can be excluded.

The squark mass limit in simplified scenarios depends on the number of accessible squarks $\tilde{q}_L$ and $\tilde{q}_R$ with different flavor. In the case of eight mass-degenerate light-flavor squarks, $\tilde{q}$ masses up to approximately $900$\,GeV can be excluded for decoupled gluinos in the limit of massless neutralinos.

Direct stop quark production with both stops directly decaying into neutralinos has been discussed in detail. In this scenario, stop masses up to $800$\,GeV can be excluded for light neutralinos. Depending on the mass difference between the stop and the lightest neutralino, two regions are identified where the experimental sensitivity is small, because the $\tilde{t}\tilde{t}$ pair-production is difficult to distinguish from the \SM $WW$ and $t\bar{t}$ production. In these regions of phase space, natural supersymmetry with very light stop quarks could still be hidden.

In the case of the simplified direct electroweak production of gauginos multiple decay scenarios have been studied with various final states with leptons. In the case of pair- or associate production of mass-degenerate $\chaone/\chitwo$ gauginos, $\chaone$ masses up to $700$\,GeV for $\chione$ masses up to $400$\,GeV can be excluded. 

Analyses targeted at models of gauge mediated supersymmetry models yield comparable mass limits. In these models the almost massless gravitino \Grav is the lightest supersymmetric particle and the lightest neutralino can decay into the \Grav and a photon, a $Z^0$, or a Higgs boson depending on the mixing.

In the constrained MSSM parameter space with $\tan\beta = 30$, $\mu>0$, and $A_0=-2\cdot m_0$, such that the Higgs mass is consistent with the measured value of approximately $125$\,GeV, the universal gaugino mass parameter $m_{1/2}$ is excluded up to $500-800$\,GeV, depending on the value of the universal scalar mass $m_0$. The remaining allowed cMSSM parameter space beyond the direct LHC limits is under pressure with respect to the consistency with \SM precision measurements~\cite{Fittino-2015,Buchmueller:2013rsa}.

The searches for supersymmetry have observed interesting fluctuations in the dilepton spectrum at \ATLAS and \CMS, but the results are still consistent with statistical fluctuations and are not confirmed by the other experiment, respectively.  Considering the large number of analyzed channels the observed data collected at $8$\,TeV are in good agreement with the expectations from the \SM within all uncertainties.

If strongly interacting supersymmetric particles exist at the TeV scale, they should be accessible at
the Large Hadron Collider. The incoming data collected at $\sqrt{s}=13$\,TeV will allow to probe supersymmetry at significantly higher energy scales as before. A discovery of new physics might be imminent, as more data recorded at $13$~TeV becomes available for analyzing.

The \ATLAS and \CMS analyses are developed and ready to discover heavy sparticles beyond the current limits, or to push the exclusion contours to higher mass scales. Specific supersymmetric particles like the stop or the higgsino could hide at lower energy scales. Carefully designed analyses and precision measurements will be necessary, to further constrain the parameter space of compressed scenarios.

\section{Acknowledgments}

The author thanks Lutz Feld, Arnd Meyer, Jory Sonneveld, and Hartmut Stadie for fruitful discussions and their help reviewing this article, and Johannes Lange for his help preparing Fig.~\ref{fig:susyResultMy}.

\begin{table*}[htb]
\centering
\begin{tabularx}{\textwidth}{rr>{\raggedright\arraybackslash}X>{\raggedright\arraybackslash}X}
\hline
Label & Ref. & Analysis \& final state & Model of result interpretation\\ \hline\hline
\multicolumn{4}{c}{Electroweak production of supersymmetry events} \\
ATLAS-13-23 & \cite{ATLAS-SUS-2013-23} & chargino \& neutralino decaying via Higgs & \chicha, \Vdecay  \\
ATLAS-13-14 & \cite{ATLAS-SUS-2013-14} & $2\tau+\MET$  & \chicha, \sldecay \\
ATLAS-13-13 & \cite{ATLAS-SUS-2013-13} & $4$ leptons + \MET & \chicha, \sldecay, \Vdecay \\ 
ATLAS-14-05 & \cite{ATLAS-SUS-2014-05} & $2,3,4$ leptons, $2\tau$ region & \chicha, \sldecay \\ 
CMS-14-02 & \cite{CMS-SUS-2014-02} & \chaone, \chitwo to $HH, HZ, HW$ &  \chicha, \Vdecay \\
CMS-13-06 & \cite{CMS-SUS-2013-06} & $ZW, ZZ, HW$ \& through $\tilde{l}$, $\tilde{\nu}$ &  \chicha, \sldecay, \Vdecay \\
\hline
\multicolumn{4}{c}{Gauge mediated supersymmetry breaking} \\
ATLAS-14-01 & \cite{ATLAS-SUS-2014-01} &  $1,2\gamma$ + jets, $b$-tags, $0,1l$ + \MET & GGM, \TfiveWG, \TChiWg \\
CMS-14-04   & \cite{CMS-SUS-2014-04}  & $\gamma\gamma$ and $1\gamma+\Ht+\MET$, bino- \& wino-like & GGM, \TfiveGG, \TfiveWG\\
ATLAS-13-08 & \cite{ATLAS-SUS-2013-08} &  $\tilde{t}\tilde{t}$, with $Z^0$+$b$-jets & \tonetone, \ttwottwo \\
CMS-13-14   & \cite{CMS-SUS-2013-14} & $\tilde{t}$ and higgsino prod. with $H\to\gamma\gamma$ & GMSB $\footnotesize \tilde{t}\to b\chaone$, $\chaone\to bff'H\Grav$ \\
CMS-14-16   & \cite{CMS-SUS-2014-16} & $1\gamma+\MET$, parked data 7.4~fb$^{-1}$ & GGM, \TChiNg, \TChiWg \\
CMS-14-13    & \cite{CMS-SUS-2014-13} & $1\gamma+1l$ & GGM Wino, \TfiveWG, \TChiWg \\ \hline
\multicolumn{4}{c}{Resonances and kinematic edges} \\
CMS-14-14   & \cite{CMS-SUS-2014-14}  & \multicolumn{2}{l}{$2l$ opposite sign, same flavor: on/off-Z, kinematic edge} \\
ATLAS-14-10 & \cite{ATLAS-SUS-2014-10} & \multicolumn{2}{l}{$2l$ opposite sign, same flavor: on/off-Z, kinematic edge} \\
\hline
\end{tabularx}
\caption{Overview table of the discussed analyses for GMSB supersymmetry and models with lepton resonances or kinematic edges. Generally, the various analyses offer more interpretations in different scenarios, see quoted reference.}
\label{tab:overview2}
\label{tab:overview:ewk}
\end{table*}

%%\clearpage
\nocite{*}
\bibliographystyle{elsarticle-num}
\bibliography{autermannPPNP}

\begin{thebibliography}{100}
\expandafter\ifx\csname url\endcsname\relax
  \def\url#1{\texttt{#1}}\fi
\expandafter\ifx\csname urlprefix\endcsname\relax\def\urlprefix{URL }\fi
\expandafter\ifx\csname href\endcsname\relax
  \def\href#1#2{#2} \def\path#1{#1}\fi

\bibitem{Aad:2012tfa}
{ATLAS Collaboration}, {Observation of a new particle in the search for the
  Standard Model Higgs boson with the ATLAS detector at the LHC}, Phys. Lett. B
  716 (2013) 1.
\newblock \href {http://arxiv.org/abs/1207.7214} {\path{arXiv:1207.7214}},
  \href {http://dx.doi.org/10.1016/j.physletb.2012.08.020}
  {\path{doi:10.1016/j.physletb.2012.08.020}}.

\bibitem{Chatrchyan:2012xdj}
{CMS Collaboration}, {Observation of a new boson at a mass of 125 GeV with the
  CMS experiment at the LHC}, Phys. Lett. B 716 (2012) 30.
\newblock \href {http://arxiv.org/abs/1207.7235} {\path{arXiv:1207.7235}},
  \href {http://dx.doi.org/10.1016/j.physletb.2012.08.021}
  {\path{doi:10.1016/j.physletb.2012.08.021}}.

\bibitem{Ramond}
P.~Ramond, {Dual theory for free fermions}, Phys. Rev. D 3 (1971) 2415.
\newblock \href {http://dx.doi.org/10.1103/PhysRevD.3.2415}
  {\path{doi:10.1103/PhysRevD.3.2415}}.

\bibitem{Ramond:1971kx}
P.~Ramond, {An interpretation of dual theories}, Nuovo Cim. A 4 (1971) 544.
\newblock \href {http://dx.doi.org/10.1007/BF02731370}
  {\path{doi:10.1007/BF02731370}}.

\bibitem{Golfand}
Y.~A. Golfand, E.~P. Likhtman,
  \href{http://www.jetpletters.ac.ru/ps/1584/article_24309.pdf}{{Extension of
  the algebra of Poincar{\'e} group generators and violation of P invariance}},
  JETP Lett. 13 (1971) 323.
\newline\urlprefix\url{http://www.jetpletters.ac.ru/ps/1584/article_24309.pdf}

\bibitem{Volkov}
D.~V. Volkov, V.~P. Akulov, {Possible universal neutrino interaction}, JETP
  Lett. 16 (1972) 438.
\newblock \href {http://dx.doi.org/10.1007/BFb0105270}
  {\path{doi:10.1007/BFb0105270}}.

\bibitem{Wess:1974tw}
J.~Wess, B.~Zumino, {Supergauge transformations in four-dimensions}, Nucl.
  Phys. B 70 (1974) 39.
\newblock \href {http://dx.doi.org/10.1016/0550-3213(74)90355-1}
  {\path{doi:10.1016/0550-3213(74)90355-1}}.

\bibitem{Freedman:1976xh}
D.~Z. Freedman, P.~van Nieuwenhuizen, S.~Ferrara, {Progress toward a theory of
  supergravity}, Phys. Rev. D 13 (1976) 3214.
\newblock \href {http://dx.doi.org/10.1103/PhysRevD.13.3214}
  {\path{doi:10.1103/PhysRevD.13.3214}}.

\bibitem{Deser:1976eh}
S.~Deser, B.~Zumino, {Consistent supergravity}, Phys. Lett. B 62 (1976) 335.
\newblock \href {http://dx.doi.org/10.1016/0370-2693(76)90089-7}
  {\path{doi:10.1016/0370-2693(76)90089-7}}.

\bibitem{Ferrara:1976fu}
S.~Ferrara, P.~van Nieuwenhuizen, {Consistent supergravity with complex spin
  3/2 gauge fields}, Phys. Rev. Lett. 37 (1976) 1669.
\newblock \href {http://dx.doi.org/10.1103/PhysRevLett.37.1669}
  {\path{doi:10.1103/PhysRevLett.37.1669}}.

\bibitem{Fayet}
P.~Fayet, {Supergauge invariant extension of the Higgs mechanism and a model
  for the electron and its neutrino}, Nucl. Phys. B 90 (1975) 104.
\newblock \href {http://dx.doi.org/10.1016/0550-3213(75)90636-7}
  {\path{doi:10.1016/0550-3213(75)90636-7}}.

\bibitem{Kane}
G.~L. Kane, C.~F. Kolda, L.~Roszkowski, J.~D. Wells, {Study of constrained
  minimal supersymmetry}, Phys. Rev. D 49 (1994) 6173.
\newblock \href {http://arxiv.org/abs/hep-ph/9312272}
  {\path{arXiv:hep-ph/9312272}}, \href
  {http://dx.doi.org/10.1103/PhysRevD.49.6173}
  {\path{doi:10.1103/PhysRevD.49.6173}}.

\bibitem{PrimerMartin}
S.~P. Martin, {A Supersymmetry primer}[Adv. Ser. Direct. High Energy
  Phys.18,1(1998)].
\newblock \href {http://arxiv.org/abs/hep-ph/9709356}
  {\path{arXiv:hep-ph/9709356}}, \href
  {http://dx.doi.org/10.1142/9789812839657\_0001, 10.1142/9789814307505\_0001}
  {\path{doi:10.1142/9789812839657\_0001, 10.1142/9789814307505\_0001}}.

\bibitem{Feng:2009te}
J.~L. Feng, J.-F. Grivaz, J.~Nachtman, {Searches for Supersymmetry at
  High-Energy Colliders}, Rev. Mod. Phys. 82 (2010) 699--727, [Reprint: Adv.
  Ser. Direct. High Energy Phys.21,351(2010)].
\newblock \href {http://arxiv.org/abs/0903.0046} {\path{arXiv:0903.0046}},
  \href {http://dx.doi.org/10.1142/9789814307505_0009,
  10.1103/RevModPhys.82.699} {\path{doi:10.1142/9789814307505_0009,
  10.1103/RevModPhys.82.699}}.

\bibitem{Barbieri198863}
R.~Barbieri, G.~F. Giudice, Upper bounds on supersymmetric particle masses,
  Nucl. Phys. B 306 (1988) 63.
\newblock \href {http://dx.doi.org/10.1016/0550-3213(88)90171-X}
  {\path{doi:10.1016/0550-3213(88)90171-X}}.

\bibitem{Witten1981513}
E.~Witten,
  \href{http://www.sciencedirect.com/science/article/pii/0550321381900067}{Dynamical
  breaking of supersymmetry}, Nuclear Physics B 188~(3) (1981) 513 -- 554.
\newblock \href
  {http://dx.doi.org/http://dx.doi.org/10.1016/0550-3213(81)90006-7}
  {\path{doi:http://dx.doi.org/10.1016/0550-3213(81)90006-7}}.
\newline\urlprefix\url{http://www.sciencedirect.com/science/article/pii/0550321381900067}

\bibitem{Dimopoulos1981150}
S.~Dimopoulos, H.~Georgi,
  \href{http://www.sciencedirect.com/science/article/pii/0550321381905228}{{Softly
  broken supersymmetry and SU(5)}}, Nuclear Physics B 193~(1) (1981) 150 --
  162.
\newblock \href
  {http://dx.doi.org/http://dx.doi.org/10.1016/0550-3213(81)90522-8}
  {\path{doi:http://dx.doi.org/10.1016/0550-3213(81)90522-8}}.
\newline\urlprefix\url{http://www.sciencedirect.com/science/article/pii/0550321381905228}

\bibitem{PhysRevLett.49.970}
A.~H. Chamseddine, R.~Arnowitt, P.~Nath,
  \href{http://link.aps.org/doi/10.1103/PhysRevLett.49.970}{Locally
  supersymmetric grand unification}, Phys. Rev. Lett. 49 (1982) 970--974.
\newblock \href {http://dx.doi.org/10.1103/PhysRevLett.49.970}
  {\path{doi:10.1103/PhysRevLett.49.970}}.
\newline\urlprefix\url{http://link.aps.org/doi/10.1103/PhysRevLett.49.970}

\bibitem{Barbieri}
R.~Barbieri, S.~Ferrara, C.~A. Savoy, {Gauge models with spontaneously broken
  local supersymmetry}, Phys. Lett. B 119 (1982) 343.
\newblock \href {http://dx.doi.org/10.1016/0370-2693(82)90685-2}
  {\path{doi:10.1016/0370-2693(82)90685-2}}.

\bibitem{Hall}
L.~J. Hall, J.~D. Lykken, S.~Weinberg, {Supergravity as the messenger of
  supersymmetry breaking}, Phys. Rev. D 27 (1983) 2359.
\newblock \href {http://dx.doi.org/10.1103/PhysRevD.27.2359}
  {\path{doi:10.1103/PhysRevD.27.2359}}.

\bibitem{NILLES19841}
H.~Nilles,
  \href{http://www.sciencedirect.com/science/article/pii/0370157384900085}{Supersymmetry,
  supergravity and particle physics}, Physics Reports 110~(1) (1984) 1 -- 162.
\newblock \href
  {http://dx.doi.org/http://dx.doi.org/10.1016/0370-1573(84)90008-5}
  {\path{doi:http://dx.doi.org/10.1016/0370-1573(84)90008-5}}.
\newline\urlprefix\url{http://www.sciencedirect.com/science/article/pii/0370157384900085}

\bibitem{GGMa}
P.~Fayet, Mixing between gravitational and weak interactions through the
  massive gravitino, Phys. Lett. B 70 (1977) 461.
\newblock \href {http://dx.doi.org/10.1016/0370-2693(77)90414-2}
  {\path{doi:10.1016/0370-2693(77)90414-2}}.

\bibitem{GGMd2}
H.~Baer, M.~Brhlik, C.~H. Chen, X.~Tata, {Signals for the minimal
  gauge-mediated supersymmetry breaking model at the Fermilab Tevatron
  collider}, Phys. Rev. D 55 (1997) 4463.
\newblock \href {http://arxiv.org/abs/hep-ph/9610358}
  {\path{arXiv:hep-ph/9610358}}, \href
  {http://dx.doi.org/10.1103/PhysRevD.55.4463}
  {\path{doi:10.1103/PhysRevD.55.4463}}.

\bibitem{GGMd3}
H.~Baer, P.~G. Mercadante, X.~Tata, Y.~L. Wang, {Reach of Tevatron upgrades in
  gauge-mediated supersymmetry breaking models}, Phys. Rev. D 60 (1999) 055001.
\newblock \href {http://arxiv.org/abs/hep-ph/9903333}
  {\path{arXiv:hep-ph/9903333}}, \href
  {http://dx.doi.org/10.1103/PhysRevD.60.055001}
  {\path{doi:10.1103/PhysRevD.60.055001}}.

\bibitem{GGMd4}
S.~Dimopoulos, S.~Thomas, J.~D. Wells, {Sparticle spectroscopy and electroweak
  symmetry breaking with gauge-mediated supersymmetry breaking}, Nucl. Phys. B
  488 (1997) 39.
\newblock \href {http://arxiv.org/abs/hep-ph/9609434}
  {\path{arXiv:hep-ph/9609434}}, \href
  {http://dx.doi.org/10.1016/S0550-3213(97)00030-8}
  {\path{doi:10.1016/S0550-3213(97)00030-8}}.

\bibitem{GGMd5}
J.~R. Ellis, J.~L. Lopez, D.~V. Nanopoulos, {Analysis of LEP constraints on
  supersymmetric models with a light gravitino}, Phys. Lett. B 394 (1997) 354.
\newblock \href {http://arxiv.org/abs/hep-ph/9610470}
  {\path{arXiv:hep-ph/9610470}}, \href
  {http://dx.doi.org/10.1016/S0370-2693(97)00019-1}
  {\path{doi:10.1016/S0370-2693(97)00019-1}}.

\bibitem{GGMd1}
M.~Dine, A.~E. Nelson, Y.~Nir, Y.~Shirman, {New tools for low energy dynamical
  supersymmetry breaking}, Phys. Rev. D 53 (1996) 2658.
\newblock \href {http://arxiv.org/abs/hep-ph/9507378}
  {\path{arXiv:hep-ph/9507378}}, \href
  {http://dx.doi.org/10.1103/PhysRevD.53.2658}
  {\path{doi:10.1103/PhysRevD.53.2658}}.

\bibitem{GGMd}
G.~F. Giudice, R.~Rattazzi, Gauge-mediated supersymmetry breaking, in:
  Perspectives on supersymmetry, World Scientific, Singapore, 1998, p. 355.

\bibitem{Randall:1998uk}
L.~Randall, R.~Sundrum, {Out of this world supersymmetry breaking}, Nucl. Phys.
  B557 (1999) 79--118.
\newblock \href {http://arxiv.org/abs/hep-th/9810155}
  {\path{arXiv:hep-th/9810155}}, \href
  {http://dx.doi.org/10.1016/S0550-3213(99)00359-4}
  {\path{doi:10.1016/S0550-3213(99)00359-4}}.

\bibitem{Giudice:1998xp}
G.~F. Giudice, M.~A. Luty, H.~Murayama, R.~Rattazzi, {Gaugino mass without
  singlets}, JHEP 12 (1998) 027.
\newblock \href {http://arxiv.org/abs/hep-ph/9810442}
  {\path{arXiv:hep-ph/9810442}}, \href
  {http://dx.doi.org/10.1088/1126-6708/1998/12/027}
  {\path{doi:10.1088/1126-6708/1998/12/027}}.

\bibitem{Barbieri:2000gf}
R.~Barbieri, A.~Strumia,
  \href{http://alice.cern.ch/format/showfull?sysnb=2199577}{{The 'LEP
  paradox'}}, in: {4th Rencontres du Vietnam: Physics at Extreme Energies
  (Particle Physics and Astrophysics) Hanoi, Vietnam, July 19-25, 2000}, 2000.
\newblock \href {http://arxiv.org/abs/hep-ph/0007265}
  {\path{arXiv:hep-ph/0007265}}.
\newline\urlprefix\url{http://alice.cern.ch/format/showfull?sysnb=2199577}

\bibitem{Papucci:2011wy}
M.~Papucci, J.~T. Ruderman, A.~Weiler, {Natural SUSY Endures}, JHEP 09 (2012)
  035.
\newblock \href {http://arxiv.org/abs/1110.6926} {\path{arXiv:1110.6926}},
  \href {http://dx.doi.org/10.1007/JHEP09(2012)035}
  {\path{doi:10.1007/JHEP09(2012)035}}.

\bibitem{FARRAR1978575}
G.~R. Farrar, P.~Fayet,
  \href{http://www.sciencedirect.com/science/article/pii/0370269378908584}{Phenomenology
  of the production, decay, and detection of new hadronic states associated
  with supersymmetry}, Physics Letters B 76~(5) (1978) 575 -- 579.
\newblock \href
  {http://dx.doi.org/http://dx.doi.org/10.1016/0370-2693(78)90858-4}
  {\path{doi:http://dx.doi.org/10.1016/0370-2693(78)90858-4}}.
\newline\urlprefix\url{http://www.sciencedirect.com/science/article/pii/0370269378908584}

\bibitem{Rparity}
R.~Barbier, et~al., {R-parity violating supersymmetry}, Phys. Rept. 420 (2005)
  1--202.
\newblock \href {http://arxiv.org/abs/hep-ph/0406039}
  {\path{arXiv:hep-ph/0406039}}, \href
  {http://dx.doi.org/10.1016/j.physrep.2005.08.006}
  {\path{doi:10.1016/j.physrep.2005.08.006}}.

\bibitem{Djouadi:1998di}
A.~Djouadi, et~al., {The minimal supersymmetric standard model: Group summary
  report}, in: {GDR (Groupement De Recherche) - Supersymetrie Montpellier,
  France, April 15-17, 1998}, 1998.
\newblock \href {http://arxiv.org/abs/hep-ph/9901246}
  {\path{arXiv:hep-ph/9901246}}.

\bibitem{CMS-PAS-SUS-2015-10}
{CMS Collaboration}, {Phenomenological MSSM interpretation of CMS results at
  $\sqrt{s}=$ 7 and 8 TeV}, CMS-PAS-SUS-15-010, CERN (2015).

\bibitem{ATLAS-SUS-2014-08}
{ATLAS Collaboration}, {Summary of the ATLAS experiment’s sensitivity to
  supersymmetry after LHC Run 1 — interpreted in the phenomenological MSSM},
  JHEP 10 (2015) 134.
\newblock \href {http://arxiv.org/abs/1508.06608} {\path{arXiv:1508.06608}},
  \href {http://dx.doi.org/10.1007/JHEP10(2015)134}
  {\path{doi:10.1007/JHEP10(2015)134}}.

\bibitem{Fittino-2015}
P.~Bechtle, et~al., {Killing the cMSSM softly, }\href
  {http://arxiv.org/abs/1508.05951} {\path{arXiv:1508.05951}}.

\bibitem{Buchmueller:2013rsa}
O.~Buchmueller, et~al., {The CMSSM and NUHM1 after LHC Run 1}, Eur. Phys. J.
  C74~(6) (2014) 2922.
\newblock \href {http://arxiv.org/abs/1312.5250} {\path{arXiv:1312.5250}},
  \href {http://dx.doi.org/10.1140/epjc/s10052-014-2922-3}
  {\path{doi:10.1140/epjc/s10052-014-2922-3}}.

\bibitem{Dimopoulos:1995ju}
S.~Dimopoulos, D.~W. Sutter, {The Supersymmetric flavor problem}, Nucl. Phys.
  B452 (1995) 496--512.
\newblock \href {http://arxiv.org/abs/hep-ph/9504415}
  {\path{arXiv:hep-ph/9504415}}, \href
  {http://dx.doi.org/10.1016/0550-3213(95)00421-N}
  {\path{doi:10.1016/0550-3213(95)00421-N}}.

\bibitem{Steffen:2006hw}
F.~D. Steffen, {Gravitino dark matter and cosmological constraints}, JCAP 0609
  (2006) 001.
\newblock \href {http://arxiv.org/abs/hep-ph/0605306}
  {\path{arXiv:hep-ph/0605306}}, \href
  {http://dx.doi.org/10.1088/1475-7516/2006/09/001}
  {\path{doi:10.1088/1475-7516/2006/09/001}}.

\bibitem{MOROI1993289}
T.~Moroi, H.~Murayama, M.~Yamaguchi, Cosmological constraints on the light
  stable gravitino, Physics Letters B 303~(3) (1993) 289 -- 294.
\newblock \href
  {http://dx.doi.org/http://dx.doi.org/10.1016/0370-2693(93)91434-O}
  {\path{doi:http://dx.doi.org/10.1016/0370-2693(93)91434-O}}.

\bibitem{GGMe}
P.~Meade, N.~Seiberg, D.~Shih, General gauge mediation, Prog. Theor. Phys.
  Suppl. 177 (2009) 143.
\newblock \href {http://arxiv.org/abs/0801.3278} {\path{arXiv:0801.3278}},
  \href {http://dx.doi.org/10.1143/PTPS.177.143}
  {\path{doi:10.1143/PTPS.177.143}}.

\bibitem{GGMf}
M.~Buican, P.~Meade, N.~Seiberg, D.~Shih, Exploring general gauge mediation,
  JHEP 03 (2009) 016.
\newblock \href {http://arxiv.org/abs/0812.3668} {\path{arXiv:0812.3668}},
  \href {http://dx.doi.org/10.1088/1126-6708/2009/03/016}
  {\path{doi:10.1088/1126-6708/2009/03/016}}.

\bibitem{Ruderman:2011vv}
J.~T. Ruderman, D.~Shih, {General Neutralino NLSPs at the Early LHC}, JHEP 08
  (2012) 159.
\newblock \href {http://arxiv.org/abs/1103.6083} {\path{arXiv:1103.6083}},
  \href {http://dx.doi.org/10.1007/JHEP08(2012)159}
  {\path{doi:10.1007/JHEP08(2012)159}}.

\bibitem{Kats:2011qh}
Y.~Kats, P.~Meade, M.~Reece, D.~Shih, {The status of GMSB after 1/fb at the
  LHC}, JHEP 02 (2012) 115.
\newblock \href {http://arxiv.org/abs/1110.6444} {\path{arXiv:1110.6444}},
  \href {http://dx.doi.org/10.1007/JHEP02(2012)115}
  {\path{doi:10.1007/JHEP02(2012)115}}.

\bibitem{Kats:2012ym}
Y.~Kats, M.~J. Strassler, {Probing colored particles with photons, leptons, and
  jets}, JHEP 11 (2012) 097.
\newblock \href {http://arxiv.org/abs/1204.1119} {\path{arXiv:1204.1119}},
  \href {http://dx.doi.org/10.1007/JHEP11(2012)097}
  {\path{doi:10.1007/JHEP11(2012)097}}.

\bibitem{ArkaniHamed:2007fw}
N.~Arkani-Hamed, P.~Schuster, N.~Toro, J.~Thaler, L.-T. Wang, et~al.,
  {MARMOSET: The path from LHC data to the new standard model via on-shell
  effective theories} (2007).
\newblock \href {http://arxiv.org/abs/hep-ph/0703088}
  {\path{arXiv:hep-ph/0703088}}.

\bibitem{Alwall:2008ag}
J.~Alwall, P.~Schuster, N.~Toro, {Simplified models for a first
  characterization of new physics at the LHC}, Phys.~Rev. D 79 (2009) 075020.
\newblock \href {http://arxiv.org/abs/0810.3921} {\path{arXiv:0810.3921}},
  \href {http://dx.doi.org/10.1103/PhysRevD.79.075020}
  {\path{doi:10.1103/PhysRevD.79.075020}}.

\bibitem{Alves:2011wf}
D.~Alves, et~al., {Simplified models for LHC new physics searches}, J.~Phys. G
  39 (2012) 105005.
\newblock \href {http://arxiv.org/abs/1105.2838} {\path{arXiv:1105.2838}},
  \href {http://dx.doi.org/10.1088/0954-3899/39/10/105005}
  {\path{doi:10.1088/0954-3899/39/10/105005}}.

\bibitem{SMS_ATLAS}
H.~Okawa,
  \href{http://inspirehep.net/record/930327/files/arXiv:1110.0282.pdf}{{Interpretations
  of SUSY Searches in ATLAS with Simplified Models}}, in: {Particles and
  fields. Proceedings, Meeting of the Division of the American Physical
  Society, DPF 2011, Providence, USA, August 9-13, 2011}, 2011.
\newblock \href {http://arxiv.org/abs/1110.0282} {\path{arXiv:1110.0282}}.
\newline\urlprefix\url{http://inspirehep.net/record/930327/files/arXiv:1110.0282.pdf}

\bibitem{SMS_CMS}
{CMS Collaboration}, {Interpretation of Searches for Supersymmetry with
  simplified Models}, Phys. Rev. D88~(5) (2013) 052017.
\newblock \href {http://arxiv.org/abs/1301.2175} {\path{arXiv:1301.2175}},
  \href {http://dx.doi.org/10.1103/PhysRevD.88.052017}
  {\path{doi:10.1103/PhysRevD.88.052017}}.

\bibitem{SModelS}
S.~Kraml, S.~Kulkarni, U.~Laa, A.~Lessa, W.~Magerl, D.~Proschofsky-Spindler,
  W.~Waltenberger, {SModelS: a tool for interpreting simplified-model results
  from the LHC and its application to supersymmetry}, Eur. Phys. J. C74 (2014)
  2868.
\newblock \href {http://arxiv.org/abs/1312.4175} {\path{arXiv:1312.4175}},
  \href {http://dx.doi.org/10.1140/epjc/s10052-014-2868-5}
  {\path{doi:10.1140/epjc/s10052-014-2868-5}}.

\bibitem{Golling:2016thc}
T.~Golling, {LHC searches for exotic new particles}, Prog. Part. Nucl. Phys. 90
  (2016) 156--200.
\newblock \href {http://dx.doi.org/10.1016/j.ppnp.2016.05.001}
  {\path{doi:10.1016/j.ppnp.2016.05.001}}.

\bibitem{ATLAS}
{ATLAS Collaboration}, {The ATLAS experiment at the CERN Large Hadron
  Collider}, JINST 3 (2008) S08003.
\newblock \href {http://dx.doi.org/10.1088/1748-0221/3/08/S08003}
  {\path{doi:10.1088/1748-0221/3/08/S08003}}.

\bibitem{CMS}
{CMS Collaboration},
  \href{http://stacks.iop.org/1748-0221/3/i=08/a=S08004}{{The CMS experiment at
  the CERN LHC}}, JINST 3~(08) (2008) S08004.
\newblock \href {http://dx.doi.org/10.1088/1748-0221/3/08/S08004}
  {\path{doi:10.1088/1748-0221/3/08/S08004}}.
\newline\urlprefix\url{http://stacks.iop.org/1748-0221/3/i=08/a=S08004}

\bibitem{CMS-PAS-PFT-09-001}
{CMS Collaboration}, \href{https://cds.cern.ch/record/1194487}{{Particle-Flow
  Event Reconstruction in CMS and Performance for Jets, Taus, and MET}},
  CMS-PAS-PFT-09-001, CERN (Apr 2009).
\newline\urlprefix\url{https://cds.cern.ch/record/1194487}

\bibitem{mt2_lester}
C.~G. Lester, D.~J. Summers, {Measuring masses of semiinvisibly decaying
  particles pair produced at hadron colliders}, Phys. Lett. B463 (1999)
  99--103.
\newblock \href {http://arxiv.org/abs/hep-ph/9906349}
  {\path{arXiv:hep-ph/9906349}}, \href
  {http://dx.doi.org/10.1016/S0370-2693(99)00945-4}
  {\path{doi:10.1016/S0370-2693(99)00945-4}}.

\bibitem{mt2_barr}
A.~Barr, C.~Lester, P.~Stephens, {m(T2): The Truth behind the glamour}, J.
  Phys. G29 (2003) 2343--2363.
\newblock \href {http://arxiv.org/abs/hep-ph/0304226}
  {\path{arXiv:hep-ph/0304226}}, \href
  {http://dx.doi.org/10.1088/0954-3899/29/10/304}
  {\path{doi:10.1088/0954-3899/29/10/304}}.

\bibitem{hemisphere}
{CMS Collaboration}, \href{http://stacks.iop.org/0954-3899/34/i=6/a=S01}{{CMS
  Physics Technical Design Report, Volume II: Physics Performance}}, {Journal
  of Physics G: Nuclear and Particle Physics} {34}~({6}) ({2007}) {995}.
\newline\urlprefix\url{http://stacks.iop.org/0954-3899/34/i=6/a=S01}

\bibitem{alphat_randall}
L.~Randall, D.~Tucker-Smith,
  \href{http://link.aps.org/doi/10.1103/PhysRevLett.101.221803}{Dijet searches
  for supersymmetry at the large hadron collider}, Phys. Rev. Lett. 101 (2008)
  221803.
\newblock \href {http://dx.doi.org/10.1103/PhysRevLett.101.221803}
  {\path{doi:10.1103/PhysRevLett.101.221803}}.
\newline\urlprefix\url{http://link.aps.org/doi/10.1103/PhysRevLett.101.221803}

\bibitem{Razor}
C.~Rogan, {Kinematical variables towards new dynamics at the LHC, }\href
  {http://arxiv.org/abs/1006.2727} {\path{arXiv:1006.2727}}.

\bibitem{Junk}
T.~Junk, {Confidence level computation for combining searches with small
  statistics}, Nucl. Instrum. Meth. A 434 (1999) 435.
\newblock \href {http://arxiv.org/abs/hep-ex/9902006}
  {\path{arXiv:hep-ex/9902006}}, \href
  {http://dx.doi.org/10.1016/S0168-9002(99)00498-2}
  {\path{doi:10.1016/S0168-9002(99)00498-2}}.

\bibitem{Read}
A.~L. Read, Presentation of search results: the cls technique, J. Phys. G 28
  (2002) 2693.
\newblock \href {http://dx.doi.org/10.1088/0954-3899/28/10/313}
  {\path{doi:10.1088/0954-3899/28/10/313}}.

\bibitem{NeymanPearson}
{J. Neyman, E. S. Pearson}, \href{http://www.jstor.org/stable/91247}{{On the
  problem of the most efficient tests of statistical hypotheses}},
  {Philosophical Transactions of the Royal Society of London. Series A,
  Containing Papers of a Mathematical or Physical Character} 231 (1933)
  289--337.
\newline\urlprefix\url{http://www.jstor.org/stable/91247}

\bibitem{CMS-NOTE-2011-005}
\href{http://cds.cern.ch/record/1379837}{{Procedure for the LHC Higgs boson
  search combination in Summer 2011}}, CMS-NOTE-2011-005, ATL-PHYS-PUB-2011-11,
  CERN (Aug 2011).
\newline\urlprefix\url{http://cds.cern.ch/record/1379837}

\bibitem{Beenakker:1996ch}
W.~Beenakker, R.~Hopker, M.~Spira, P.~M. Zerwas, {Squark and gluino production
  at hadron colliders}, Nucl. Phys. B492 (1997) 51--103.
\newblock \href {http://arxiv.org/abs/hep-ph/9610490}
  {\path{arXiv:hep-ph/9610490}}, \href
  {http://dx.doi.org/10.1016/S0550-3213(97)80027-2}
  {\path{doi:10.1016/S0550-3213(97)80027-2}}.

\bibitem{Kramer:2012bx}
M.~Kramer, A.~Kulesza, R.~van~der Leeuw, M.~Mangano, S.~Padhi, T.~Plehn,
  X.~Portell, {Supersymmetry production cross sections in $pp$ collisions at
  $\sqrt{s}=7$ TeV}\href {http://arxiv.org/abs/1206.2892}
  {\path{arXiv:1206.2892}}.

\bibitem{ATLAS-SUS-2013-02}
{ATLAS Collaboration}, {Search for squarks and gluinos with the ATLAS detector
  in final states with jets and missing transverse momentum using $\sqrt{s}=8$
  TeV proton--proton collision data}, JHEP 09 (2014) 176.
\newblock \href {http://arxiv.org/abs/1405.7875} {\path{arXiv:1405.7875}},
  \href {http://dx.doi.org/10.1007/JHEP09(2014)176}
  {\path{doi:10.1007/JHEP09(2014)176}}.

\bibitem{ATLAS-SUS-2013-04}
{ATLAS Collaboration}, {Search for new phenomena in final states with large jet
  multiplicities and missing transverse momentum at $\sqrt{s}$=8 TeV
  proton-proton collisions using the ATLAS experiment}, JHEP 10 (2013) 130,
  [Erratum: JHEP01,109(2014)].
\newblock \href {http://arxiv.org/abs/1308.1841} {\path{arXiv:1308.1841}},
  \href {http://dx.doi.org/10.1007/JHEP10(2013)130, 10.1007/JHEP01(2014)109}
  {\path{doi:10.1007/JHEP10(2013)130, 10.1007/JHEP01(2014)109}}.

\bibitem{CMS-SUS-2013-12}
{CMS Collaboration}, {Search for new physics in the multijet and missing
  transverse momentum final state in proton-proton collisions at $\sqrt{s}$= 8
  TeV}, JHEP 06 (2014) 055.
\newblock \href {http://arxiv.org/abs/1402.4770} {\path{arXiv:1402.4770}},
  \href {http://dx.doi.org/10.1007/JHEP06(2014)055}
  {\path{doi:10.1007/JHEP06(2014)055}}.

\bibitem{CMS-SUS-2013-19}
{CMS Collaboration}, {Searches for Supersymmetry using the $M_{T2}$ Variable in
  Hadronic Events Produced in pp Collisions at 8 TeV}, JHEP 05 (2015) 078.
\newblock \href {http://arxiv.org/abs/1502.04358} {\path{arXiv:1502.04358}},
  \href {http://dx.doi.org/10.1007/JHEP05(2015)078}
  {\path{doi:10.1007/JHEP05(2015)078}}.

\bibitem{ATLAS-SUS-2014-07}
{ATLAS Collaboration}, {ATLAS Run 1 searches for direct pair production of
  third-generation squarks at the Large Hadron Collider}, Eur. Phys. J.
  C75~(10) (2015) 510.
\newblock \href {http://arxiv.org/abs/1506.08616} {\path{arXiv:1506.08616}},
  \href {http://dx.doi.org/10.1140/epjc/s10052-015-3726-9}
  {\path{doi:10.1140/epjc/s10052-015-3726-9}}.

\bibitem{CMS-SUS-2012-28}
{CMS Collaboration}, {Search for supersymmetry in hadronic final states with
  missing transverse energy using the variables $\alpha_T$ and b-quark
  multiplicity in pp collisions at $\sqrt {s}=8$ TeV}, Eur. Phys. J. C73~(9)
  (2013) 2568.
\newblock \href {http://arxiv.org/abs/1303.2985} {\path{arXiv:1303.2985}},
  \href {http://dx.doi.org/10.1140/epjc/s10052-013-2568-6}
  {\path{doi:10.1140/epjc/s10052-013-2568-6}}.

\bibitem{ATLAS_7TEV_Razor}
{ATLAS Collaboration}, {Multi-channel search for squarks and gluinos in
  $\sqrt{s}=7$ TeV $pp$ collisions with the ATLAS detector}, Eur. Phys. J.
  C73~(3) (2013) 2362.
\newblock \href {http://arxiv.org/abs/1212.6149} {\path{arXiv:1212.6149}},
  \href {http://dx.doi.org/10.1140/epjc/s10052-013-2362-5}
  {\path{doi:10.1140/epjc/s10052-013-2362-5}}.

\bibitem{ATLAS-SUS-2013-20}
{ATLAS Collaboration}, {Search for squarks and gluinos in events with isolated
  leptons, jets and missing transverse momentum at $\sqrt{s}=8$ TeV with the
  ATLAS detector}, JHEP 04 (2015) 116.
\newblock \href {http://arxiv.org/abs/1501.03555} {\path{arXiv:1501.03555}},
  \href {http://dx.doi.org/10.1007/JHEP04(2015)116}
  {\path{doi:10.1007/JHEP04(2015)116}}.

\bibitem{CMS-SUS-2013-04}
{CMS Collaboration}, {Search for supersymmetry using razor variables in events
  with $b$-tagged jets at $\sqrt{s} =$ 8 TeV}, Phys. Rev. D91 (2015) 052018.
\newblock \href {http://arxiv.org/abs/1502.00300} {\path{arXiv:1502.00300}},
  \href {http://dx.doi.org/10.1103/PhysRevD.91.052018}
  {\path{doi:10.1103/PhysRevD.91.052018}}.

\bibitem{CMS-PAS-SUS-2014-11}
{CMS Collaboration}, \href{https://cds.cern.ch/record/1745586}{{Exclusion
  limits on gluino and top-squark pair production in natural SUSY scenarios
  with inclusive razor and exclusive single-lepton searches at 8 TeV.}},
  CMS-PAS-SUS-14-011, CERN (2014).
\newline\urlprefix\url{https://cds.cern.ch/record/1745586}

\bibitem{ATLAS-SUS-2014-06}
{ATLAS Collaboration}, {Summary of the searches for squarks and gluinos using $
  \sqrt{s}=8 $ TeV pp collisions with the ATLAS experiment at the LHC}, JHEP 10
  (2015) 054.
\newblock \href {http://arxiv.org/abs/1507.05525} {\path{arXiv:1507.05525}},
  \href {http://dx.doi.org/10.1007/JHEP10(2015)054}
  {\path{doi:10.1007/JHEP10(2015)054}}.

\bibitem{CMS-SUS-2013-13}
{CMS Collaboration}, {Search for new physics in events with same-sign dileptons
  and jets in pp collisions at $\sqrt{s}$ = 8 TeV}, JHEP 01 (2014) 163,
  [Erratum: JHEP01,014(2015)].
\newblock \href {http://arxiv.org/abs/1311.6736} {\path{arXiv:1311.6736}},
  \href {http://dx.doi.org/10.1007/JHEP01(2015)014, 10.1007/JHEP01(2014)163}
  {\path{doi:10.1007/JHEP01(2015)014, 10.1007/JHEP01(2014)163}}.

\bibitem{MassDiff}
R.~Barbieri, D.~Pappadopulo,
  \href{http://stacks.iop.org/1126-6708/2009/i=10/a=061}{S-particles at their
  naturalness limits}, Journal of High Energy Physics 2009~(10) (2009) 061.
\newline\urlprefix\url{http://stacks.iop.org/1126-6708/2009/i=10/a=061}

\bibitem{CMS-SUS-2012-24}
{CMS Collaboration}, {Search for gluino mediated bottom- and top-squark
  production in multijet final states in pp collisions at 8 TeV}, Phys. Lett.
  B725 (2013) 243--270.
\newblock \href {http://arxiv.org/abs/1305.2390} {\path{arXiv:1305.2390}},
  \href {http://dx.doi.org/10.1016/j.physletb.2013.06.058}
  {\path{doi:10.1016/j.physletb.2013.06.058}}.

\bibitem{CMS-SUS-2013-07}
{CMS Collaboration}, {Search for supersymmetry in pp collisions at $\sqrt{s}$=8
  TeV in events with a single lepton, large jet multiplicity, and multiple b
  jets}, Phys. Lett. B733 (2014) 328--353.
\newblock \href {http://arxiv.org/abs/1311.4937} {\path{arXiv:1311.4937}},
  \href {http://dx.doi.org/10.1016/j.physletb.2014.04.023}
  {\path{doi:10.1016/j.physletb.2014.04.023}}.

\bibitem{ATLAS-SUS-2013-09}
{ATLAS Collaboration}, {Search for supersymmetry at $\sqrt{s}$=8 TeV in final
  states with jets and two same-sign leptons or three leptons with the ATLAS
  detector}, JHEP 06 (2014) 035.
\newblock \href {http://arxiv.org/abs/1404.2500} {\path{arXiv:1404.2500}},
  \href {http://dx.doi.org/10.1007/JHEP06(2014)035}
  {\path{doi:10.1007/JHEP06(2014)035}}.

\bibitem{ATLAS-SUS-2013-18}
{ATLAS Collaboration}, {Search for strong production of supersymmetric
  particles in final states with missing transverse momentum and at least three
  $b$-jets at $\sqrt{s}$= 8 TeV proton-proton collisions with the ATLAS
  detector}, JHEP 10 (2014) 24.
\newblock \href {http://arxiv.org/abs/1407.0600} {\path{arXiv:1407.0600}},
  \href {http://dx.doi.org/10.1007/JHEP10(2014)024}
  {\path{doi:10.1007/JHEP10(2014)024}}.

\bibitem{StopXsec}
W.~Beenakker, S.~Brensing, M.~{Kr\"amer}, A.~Kulesza, E.~Laenen, L.~Motyka,
  I.~Niessen, Squark and gluino hadroproduction, International Journal of
  Modern Physics A 26~(16) (2011) 2637--2664.
\newblock \href {http://dx.doi.org/10.1142/S0217751X11053560}
  {\path{doi:10.1142/S0217751X11053560}}.

\bibitem{PhysRevLett.42.1117}
C.~B. Dover, T.~K. Gaisser, G.~Steigman,
  \href{http://link.aps.org/doi/10.1103/PhysRevLett.42.1117}{Cosmological
  constraints on new stable hadrons}, Phys. Rev. Lett. 42 (1979) 1117--1120.
\newblock \href {http://dx.doi.org/10.1103/PhysRevLett.42.1117}
  {\path{doi:10.1103/PhysRevLett.42.1117}}.
\newline\urlprefix\url{http://link.aps.org/doi/10.1103/PhysRevLett.42.1117}

\bibitem{ATLAS-SUS-2013-21}
{ATLAS Collaboration}, {Search for pair-produced third-generation squarks
  decaying via charm quarks or in compressed supersymmetric scenarios in $pp$
  collisions at $\sqrt{s}=8~$TeV with the ATLAS detector}, Phys. Rev. D90~(5)
  (2014) 052008.
\newblock \href {http://arxiv.org/abs/1407.0608} {\path{arXiv:1407.0608}},
  \href {http://dx.doi.org/10.1103/PhysRevD.90.052008}
  {\path{doi:10.1103/PhysRevD.90.052008}}.

\bibitem{CMS-SUS-2014-01}
{CMS Collaboration}, {Searches for third-generation squark production in fully
  hadronic final states in proton-proton collisions at $ \sqrt{s} = 8$ TeV},
  JHEP 06 (2015) 116.
\newblock \href {http://arxiv.org/abs/1503.08037} {\path{arXiv:1503.08037}},
  \href {http://dx.doi.org/10.1007/JHEP06(2015)116}
  {\path{doi:10.1007/JHEP06(2015)116}}.

\bibitem{ATLAS-EXO-2013-13}
{ATLAS Collaboration}, {Search for new phenomena in final states with an
  energetic jet and large missing transverse momentum in pp collisions at
  $\sqrt{s}=8$ TeV with the ATLAS detector}, Eur. Phys. J. C75~(7) (2015) 299,
  [Erratum: Eur. Phys. J.C75,no.9,408(2015)].
\newblock \href {http://arxiv.org/abs/1502.01518} {\path{arXiv:1502.01518}},
  \href {http://dx.doi.org/10.1140/epjc/s10052-015-3517-3,
  10.1140/epjc/s10052-015-3639-7} {\path{doi:10.1140/epjc/s10052-015-3517-3,
  10.1140/epjc/s10052-015-3639-7}}.

\bibitem{ATLAS-SUS-2014-03}
{ATLAS Collaboration}, {Search for scalar charm quark pair production in $pp$
  collisions at $\sqrt{s}=$ 8  TeV with the ATLAS detector}, Phys. Rev.
  Lett. 114~(16) (2015) 161801.
\newblock \href {http://arxiv.org/abs/1501.01325} {\path{arXiv:1501.01325}},
  \href {http://dx.doi.org/10.1103/PhysRevLett.114.161801}
  {\path{doi:10.1103/PhysRevLett.114.161801}}.

\bibitem{ATLAS-SUS-2013-15}
{ATLAS Collaboration}, {Search for top squark pair production in final states
  with one isolated lepton, jets, and missing transverse momentum in $\sqrt s
  =$8 TeV $pp$ collisions with the ATLAS detector}, JHEP 11 (2014) 118.
\newblock \href {http://arxiv.org/abs/1407.0583} {\path{arXiv:1407.0583}},
  \href {http://dx.doi.org/10.1007/JHEP11(2014)118}
  {\path{doi:10.1007/JHEP11(2014)118}}.

\bibitem{CMS-SUS-2014-21}
{CMS Collaboration}, {Search for supersymmetry in events with soft leptons, low
  jet multiplicity, and missing transverse momentum in proton-proton collisions
  at $\sqrt{s}=8$\,TeV}\href {http://arxiv.org/abs/1512.08002}
  {\path{arXiv:1512.08002}}.

\bibitem{AtlasWW8}
{ATLAS Collaboration}, \href{https://cds.cern.ch/record/2136855}{{Measurement
  of total and differential $W^+W^-$ production cross sections in proton-proton
  collisions at $\sqrt{s}=$ 8 TeV with the ATLAS detector and limits on
  anomalous triple-gauge-boson couplings}}, CERN-PH-EP-2015-323,~\href
  {http://arxiv.org/abs/1603.01702} {\path{arXiv:1603.01702}}.
\newline\urlprefix\url{https://cds.cern.ch/record/2136855}

\bibitem{CMSWW8}
{CMS Collaboration}, {Measurement of W+W- and ZZ production cross sections in
  pp collisions at $\sqrt{s}=8$\,TeV}, Phys. Lett. B721 (2013) 190--211.
\newblock \href {http://arxiv.org/abs/1301.4698} {\path{arXiv:1301.4698}},
  \href {http://dx.doi.org/10.1016/j.physletb.2013.03.027}
  {\path{doi:10.1016/j.physletb.2013.03.027}}.

\bibitem{Kim:2014eva}
J.~S. Kim, K.~Rolbiecki, K.~Sakurai, J.~Tattersall, {'Stop' that ambulance! New
  physics at the LHC?}, JHEP 12 (2014) 010.
\newblock \href {http://arxiv.org/abs/1406.0858} {\path{arXiv:1406.0858}},
  \href {http://dx.doi.org/10.1007/JHEP12(2014)010}
  {\path{doi:10.1007/JHEP12(2014)010}}.

\bibitem{Curtin:2014zua}
D.~Curtin, P.~Meade, P.-J. Tien, {Natural SUSY in plain sight}, Phys. Rev.
  D90~(11) (2014) 115012.
\newblock \href {http://arxiv.org/abs/1406.0848} {\path{arXiv:1406.0848}},
  \href {http://dx.doi.org/10.1103/PhysRevD.90.115012}
  {\path{doi:10.1103/PhysRevD.90.115012}}.

\bibitem{CMS-SUS-2013-11}
{CMS Collaboration}, {Search for top-squark pair production in the
  single-lepton final state in pp collisions at $\sqrt{s}$ = 8 TeV}, Eur. Phys.
  J. C73~(12) (2013) 2677.
\newblock \href {http://arxiv.org/abs/1308.1586} {\path{arXiv:1308.1586}},
  \href {http://dx.doi.org/10.1140/epjc/s10052-013-2677-2}
  {\path{doi:10.1140/epjc/s10052-013-2677-2}}.

\bibitem{CMS-SUS-2014-15}
{CMS Collaboration}, {Search for direct pair production of scalar top quarks in
  the single- and dilepton channels in proton-proton collisions at $\sqrt{s}$ =
  8 TeV}, Submitted to: JHEP\href {http://arxiv.org/abs/1602.03169}
  {\path{arXiv:1602.03169}}.

\bibitem{ATLAS-SUS-2013-19}
{ATLAS Collaboration}, {Search for direct top-squark pair production in final
  states with two leptons in pp collisions at $\sqrt{s} =$ 8TeV with the ATLAS
  detector}, JHEP 06 (2014) 124.
\newblock \href {http://arxiv.org/abs/1403.4853} {\path{arXiv:1403.4853}},
  \href {http://dx.doi.org/10.1007/JHEP06(2014)124}
  {\path{doi:10.1007/JHEP06(2014)124}}.

\bibitem{ATLAS-TOP-2013-04}
{ATLAS Collaboration}, {Measurement of the $t\overline{t}$ production
  cross-section using $e\mu $ events with $b$ -tagged jets in $pp$ collisions
  at $\sqrt{s}=7$ and 8 TeV with the ATLAS detector}, Eur. Phys. J. C74~(10)
  (2014) 3109.
\newblock \href {http://arxiv.org/abs/1406.5375} {\path{arXiv:1406.5375}},
  \href {http://dx.doi.org/10.1140/epjc/s10052-014-3109-7}
  {\path{doi:10.1140/epjc/s10052-014-3109-7}}.

\bibitem{Khachatryan:2016mqs}
{CMS Collaboration}, {Measurement of the t-tbar production cross section in the
  e-mu channel in proton-proton collisions at $\sqrt{s}=7$ and $8$\,TeV}\href
  {http://arxiv.org/abs/1603.02303} {\path{arXiv:1603.02303}}.

\bibitem{Khachatryan:2015fwh}
{CMS Collaboration}, {Measurement of the $\mathrm{t}\overline{{\mathrm{t}}}$
  production cross section in the all-jets final state in pp collisions at
  $\sqrt{s}=8$ $\,\mbox{TeV}$}, Eur. Phys. J. C76~(3) (2016) 128.
\newblock \href {http://arxiv.org/abs/1509.06076} {\path{arXiv:1509.06076}},
  \href {http://dx.doi.org/10.1140/epjc/s10052-016-3956-5}
  {\path{doi:10.1140/epjc/s10052-016-3956-5}}.

\bibitem{ATLAS-TOP-2014-07}
{ATLAS Collaboration}, {Measurement of spin correlation in top-antitop quark
  events and search for top squark pair production in pp collisions at
  $\sqrt{s}=8$ TeV using the ATLAS detector}, Phys. Rev. Lett. 114~(14) (2015)
  142001.
\newblock \href {http://arxiv.org/abs/1412.4742} {\path{arXiv:1412.4742}},
  \href {http://dx.doi.org/10.1103/PhysRevLett.114.142001}
  {\path{doi:10.1103/PhysRevLett.114.142001}}.

\bibitem{CMS-TOP-2014-23}
{CMS Collaboration}, {Measurements of t t-bar spin correlations and top quark
  polarization using dilepton final states in pp collisions at
  $\sqrt{s}=8$\,TeV}, Phys. Rev. D93~(5) (2016) 052007.
\newblock \href {http://arxiv.org/abs/1601.01107} {\path{arXiv:1601.01107}},
  \href {http://dx.doi.org/10.1103/PhysRevD.93.052007}
  {\path{doi:10.1103/PhysRevD.93.052007}}.

\bibitem{Mahlon:2010gw}
G.~Mahlon, S.~J. Parke, {Spin Correlation Effects in Top Quark Pair Production
  at the LHC}, Phys. Rev. D81 (2010) 074024.
\newblock \href {http://arxiv.org/abs/1001.3422} {\path{arXiv:1001.3422}},
  \href {http://dx.doi.org/10.1103/PhysRevD.81.074024}
  {\path{doi:10.1103/PhysRevD.81.074024}}.

\bibitem{Bernreuther:2013aga}
W.~Bernreuther, Z.-G. Si, {Top quark spin correlations and polarization at the
  LHC: standard model predictions and effects of anomalous top chromo moments},
  Phys. Lett. B725 (2013) 115--122, [Erratum: Phys. Lett.B744,413(2015)].
\newblock \href {http://arxiv.org/abs/1305.2066} {\path{arXiv:1305.2066}},
  \href {http://dx.doi.org/10.1016/j.physletb.2013.06.051,
  10.1016/j.physletb.2015.03.035} {\path{doi:10.1016/j.physletb.2013.06.051,
  10.1016/j.physletb.2015.03.035}}.

\bibitem{ATLAS-SUS-2013-08}
{ATLAS Collaboration}, {Search for direct top squark pair production in events
  with a Z boson, b-jets and missing transverse momentum in $\sqrt{s}=8$ TeV pp
  collisions with the ATLAS detector}, Eur. Phys. J. C74~(6) (2014) 2883.
\newblock \href {http://arxiv.org/abs/1403.5222} {\path{arXiv:1403.5222}},
  \href {http://dx.doi.org/10.1140/epjc/s10052-014-2883-6}
  {\path{doi:10.1140/epjc/s10052-014-2883-6}}.

\bibitem{ATLAS-SUS-2013-16}
{ATLAS Collaboration}, {Search for direct pair production of the top squark in
  all-hadronic final states in proton-proton collisions at $\sqrt{s}=8$\,TeV
  with the ATLAS detector}, JHEP 09 (2014) 015.
\newblock \href {http://arxiv.org/abs/1406.1122} {\path{arXiv:1406.1122}},
  \href {http://dx.doi.org/10.1007/JHEP09(2014)015}
  {\path{doi:10.1007/JHEP09(2014)015}}.

\bibitem{ATLAS-SUS-2013-05}
{ATLAS Collaboration}, {Search for direct third-generation squark pair
  production in final states with missing transverse momentum and two $b$-jets
  in $\sqrt{s} =$ 8 TeV $pp$ collisions with the ATLAS detector}, JHEP 10
  (2013) 189.
\newblock \href {http://arxiv.org/abs/1308.2631} {\path{arXiv:1308.2631}},
  \href {http://dx.doi.org/10.1007/JHEP10(2013)189}
  {\path{doi:10.1007/JHEP10(2013)189}}.

\bibitem{CMS-SUS-2013-23}
{CMS Collaboration}, {Search for direct pair production of supersymmetric top
  quarks decaying to all-hadronic final states in pp collisions at
  $\sqrt{s}=8$\,TeV}\href {http://arxiv.org/abs/1603.00765}
  {\path{arXiv:1603.00765}}.

\bibitem{CMS-SUS-2014-10}
{CMS Collaboration}, {Searches for supersymmetry based on events with b jets
  and four W bosons in pp collisions at 8 TeV}, Phys. Lett. B745 (2015) 5--28.
\newblock \href {http://arxiv.org/abs/1412.4109} {\path{arXiv:1412.4109}},
  \href {http://dx.doi.org/10.1016/j.physletb.2015.04.002}
  {\path{doi:10.1016/j.physletb.2015.04.002}}.

\bibitem{CMS-SUS-2013-24}
{CMS Collaboration}, {Search for top-squark pairs decaying into Higgs or Z
  bosons in pp collisions at $\sqrt{s}$=8 TeV}, Phys. Lett. B736 (2014)
  371--397.
\newblock \href {http://arxiv.org/abs/1405.3886} {\path{arXiv:1405.3886}},
  \href {http://dx.doi.org/10.1016/j.physletb.2014.07.053}
  {\path{doi:10.1016/j.physletb.2014.07.053}}.

\bibitem{ATLAS-SUS-2014-05}
{ATLAS Collaboration}, {Search for the electroweak production of supersymmetric
  particles in $\sqrt{s}$=8 TeV $pp$ collisions with the ATLAS detector},
  CERN-PH-EP-2015-218,~\href {http://arxiv.org/abs/1509.07152}
  {\path{arXiv:1509.07152}}.

\bibitem{pdg2014}
K.~A. Olive, et~al., {Review of Particle Physics}, Chin. Phys. C38 (2014)
  090001.
\newblock \href {http://dx.doi.org/10.1088/1674-1137/38/9/090001}
  {\path{doi:10.1088/1674-1137/38/9/090001}}.

\bibitem{relicdensity}
G.~Hinshaw, et~al.,
  \href{http://stacks.iop.org/0067-0049/208/i=2/a=19}{{Nine-year Wilkinson
  Microwave Anisotropy Probe (WMAP) Observations: Cosmological Parameter
  Results}}, {The Astrophysical Journal Supplement Series} 208~(2) (2013) 19.
\newline\urlprefix\url{http://stacks.iop.org/0067-0049/208/i=2/a=19}

\bibitem{coannihilation}
K.~Griest, D.~Seckel,
  \href{http://link.aps.org/doi/10.1103/PhysRevD.43.3191}{Three exceptions in
  the calculation of relic abundances}, Phys. Rev. D 43 (1991) 3191--3203.
\newblock \href {http://dx.doi.org/10.1103/PhysRevD.43.3191}
  {\path{doi:10.1103/PhysRevD.43.3191}}.
\newline\urlprefix\url{http://link.aps.org/doi/10.1103/PhysRevD.43.3191}

\bibitem{ATLAS-SUS-2013-11}
{ATLAS Collaboration}, {Search for direct production of charginos, neutralinos
  and sleptons in final states with two leptons and missing transverse momentum
  in $pp$ collisions at $\sqrt{s} =$ 8 TeV with the ATLAS detector}, JHEP 05
  (2014) 071.
\newblock \href {http://arxiv.org/abs/1403.5294} {\path{arXiv:1403.5294}},
  \href {http://dx.doi.org/10.1007/JHEP05(2014)071}
  {\path{doi:10.1007/JHEP05(2014)071}}.

\bibitem{CMS-SUS-2013-06}
{CMS Collaboration}, {Searches for electroweak production of charginos,
  neutralinos, and sleptons decaying to leptons and W, Z, and Higgs bosons in
  pp collisions at 8 TeV}, Eur. Phys. J. C74~(9) (2014) 3036.
\newblock \href {http://arxiv.org/abs/1405.7570} {\path{arXiv:1405.7570}},
  \href {http://dx.doi.org/10.1140/epjc/s10052-014-3036-7}
  {\path{doi:10.1140/epjc/s10052-014-3036-7}}.

\bibitem{ATLAS-SUS-2013-12}
{ATLAS Collaboration}, {Search for direct production of charginos and
  neutralinos in events with three leptons and missing transverse momentum in
  $\sqrt{s} =8$ TeV $pp$ collisions with the ATLAS detector}, JHEP 04 (2014)
  169.
\newblock \href {http://arxiv.org/abs/1402.7029} {\path{arXiv:1402.7029}},
  \href {http://dx.doi.org/10.1007/JHEP04(2014)169}
  {\path{doi:10.1007/JHEP04(2014)169}}.

\bibitem{ATLAS-SUS-2013-14}
{ATLAS Collaboration}, {Search for the direct production of charginos,
  neutralinos and staus in final states with at least two hadronically decaying
  taus and missing transverse momentum in $pp$ collisions at $\sqrt{s}$ = 8 TeV
  with the ATLAS detector}, JHEP 10 (2014) 96.
\newblock \href {http://arxiv.org/abs/1407.0350} {\path{arXiv:1407.0350}},
  \href {http://dx.doi.org/10.1007/JHEP10(2014)096}
  {\path{doi:10.1007/JHEP10(2014)096}}.

\bibitem{ATLAS-SUS-2013-13}
{ATLAS Collaboration}, {Search for supersymmetry in events with four or more
  leptons in $\sqrt{s}$ = 8 TeV pp collisions with the ATLAS detector}, Phys.
  Rev. D90~(5) (2014) 052001.
\newblock \href {http://arxiv.org/abs/1405.5086} {\path{arXiv:1405.5086}},
  \href {http://dx.doi.org/10.1103/PhysRevD.90.052001}
  {\path{doi:10.1103/PhysRevD.90.052001}}.

\bibitem{ATLAS-SUS-2013-23}
{ATLAS Collaboration}, {Search for direct pair production of a chargino and a
  neutralino decaying to the 125 GeV Higgs boson in $\sqrt{s} = 8$ TeV ${pp}$
  collisions with the ATLAS detector}, Eur. Phys. J. C75~(5) (2015) 208.
\newblock \href {http://arxiv.org/abs/1501.07110} {\path{arXiv:1501.07110}},
  \href {http://dx.doi.org/10.1140/epjc/s10052-015-3408-7}
  {\path{doi:10.1140/epjc/s10052-015-3408-7}}.

\bibitem{CMS-SUS-2014-02}
{CMS Collaboration}, {Searches for electroweak neutralino and chargino
  production in channels with Higgs, Z, and W bosons in pp collisions at 8
  TeV}, Phys. Rev. D90~(9) (2014) 092007.
\newblock \href {http://arxiv.org/abs/1409.3168} {\path{arXiv:1409.3168}},
  \href {http://dx.doi.org/10.1103/PhysRevD.90.092007}
  {\path{doi:10.1103/PhysRevD.90.092007}}.

\bibitem{CMS-SUS-2014-04}
{CMS Collaboration}, {Search for supersymmetry with photons in pp collisions at
  $\sqrt{s}$=8  TeV}, Phys. Rev. D92~(7) (2015) 072006.
\newblock \href {http://arxiv.org/abs/1507.02898} {\path{arXiv:1507.02898}},
  \href {http://dx.doi.org/10.1103/PhysRevD.92.072006}
  {\path{doi:10.1103/PhysRevD.92.072006}}.

\bibitem{ATLAS-SUS-2014-01}
{ATLAS Collaboration}, {Search for photonic signatures of gauge-mediated
  supersymmetry in 8 TeV pp collisions with the ATLAS detector}, Phys. Rev.
  D92~(7) (2015) 072001.
\newblock \href {http://arxiv.org/abs/1507.05493} {\path{arXiv:1507.05493}},
  \href {http://dx.doi.org/10.1103/PhysRevD.92.072001}
  {\path{doi:10.1103/PhysRevD.92.072001}}.

\bibitem{ATLAS-EXO-2014-06}
{ATLAS Collaboration}, {Search for new phenomena in events with a photon and
  missing transverse momentum in $pp$ collisions at $\sqrt{s}=8$ TeV with the
  ATLAS detector}, Phys. Rev. D91~(1) (2015) 012008, [Erratum: Phys.
  Rev.D92,no.5,059903(2015)].
\newblock \href {http://arxiv.org/abs/1411.1559} {\path{arXiv:1411.1559}},
  \href {http://dx.doi.org/10.1103/PhysRevD.92.059903,
  10.1103/PhysRevD.91.012008} {\path{doi:10.1103/PhysRevD.92.059903,
  10.1103/PhysRevD.91.012008}}.

\bibitem{ATLAS-SUS-2013-10}
{ATLAS Collaboration}, {Search for supersymmetry in events with large missing
  transverse momentum, jets, and at least one tau lepton in 20 fb$^{-1}$ of
  $\sqrt{s}=$ 8 TeV proton-proton collision data with the ATLAS detector}, JHEP
  09 (2014) 103.
\newblock \href {http://arxiv.org/abs/1407.0603} {\path{arXiv:1407.0603}},
  \href {http://dx.doi.org/10.1007/JHEP09(2014)103}
  {\path{doi:10.1007/JHEP09(2014)103}}.

\bibitem{CMS-SUS-2013-14}
{CMS Collaboration}, {Search for top squark and higgsino production using
  diphoton Higgs boson decays}, Phys. Rev. Lett. 112 (2014) 161802.
\newblock \href {http://arxiv.org/abs/1312.3310} {\path{arXiv:1312.3310}},
  \href {http://dx.doi.org/10.1103/PhysRevLett.112.161802}
  {\path{doi:10.1103/PhysRevLett.112.161802}}.

\bibitem{CMS-SUS-2014-13}
{CMS Collaboration}, {Search for supersymmetry with a photon, a lepton, and
  missing transverse momentum in pp collisions at $\sqrt{s}$ = 8 TeV}\href
  {http://arxiv.org/abs/1508.01218} {\path{arXiv:1508.01218}}.

\bibitem{CMS-SUS-2014-16}
{CMS Collaboration}, \href{https://cds.cern.ch/record/2134960}{{Search for
  supersymmetry in electroweak production with photons and large missing
  transverse energy in pp collisions at $\sqrt{s} = $ 8 TeV}},
  CERN-EP-2016-012. CMS-SUS-14-016,~Submitted to Phys. Lett. B.
\newblock \href {http://arxiv.org/abs/1602.08772} {\path{arXiv:1602.08772}}.
\newline\urlprefix\url{https://cds.cern.ch/record/2134960}

\bibitem{CMS-SUS-2014-09}
{CMS Collaboration}, {Search for stealth supersymmetry in events with jets,
  either photons or leptons, and low missing transverse momentum in pp
  collisions at 8 TeV}, Phys. Lett. B743 (2015) 503--525.
\newblock \href {http://arxiv.org/abs/1411.7255} {\path{arXiv:1411.7255}},
  \href {http://dx.doi.org/10.1016/j.physletb.2015.03.017}
  {\path{doi:10.1016/j.physletb.2015.03.017}}.

\bibitem{CMS-DP-2012-022}
{CMS Collaboration}, \href{https://cds.cern.ch/record/1480607}{{Data Parking
  and Data Scouting at the CMS Experiment}}, CMS-DP-2012-022.
\newline\urlprefix\url{https://cds.cern.ch/record/1480607}

\bibitem{CMS-SUS-2014-14}
{CMS Collaboration}, {Search for Physics Beyond the Standard Model in Events
  with Two Leptons, Jets, and Missing Transverse Momentum in pp Collisions at
  $\sqrt{s}$ = 8 TeV}, JHEP 04 (2015) 124.
\newblock \href {http://arxiv.org/abs/1502.06031} {\path{arXiv:1502.06031}},
  \href {http://dx.doi.org/10.1007/JHEP04(2015)124}
  {\path{doi:10.1007/JHEP04(2015)124}}.

\bibitem{ATLAS-SUS-2014-10}
{ATLAS Collaboration}, {Search for supersymmetry in events containing a
  same-flavour opposite-sign dilepton pair, jets, and large missing transverse
  momentum in $\sqrt{s}=8$ TeV pp collisions with the ATLAS detector}, Eur.
  Phys. J. C75~(7) (2015) 318, [Erratum: Eur. Phys. J.C75,no.10,463(2015)].
\newblock \href {http://arxiv.org/abs/1503.03290} {\path{arXiv:1503.03290}},
  \href {http://dx.doi.org/10.1140/epjc/s10052-015-3661-9,
  10.1140/epjc/s10052-015-3518-2} {\path{doi:10.1140/epjc/s10052-015-3661-9,
  10.1140/epjc/s10052-015-3518-2}}.

\bibitem{ATLAS-SUS-2013-06}
{ATLAS Collaboration}, {Search for a Heavy Neutral Particle Decaying to $e\mu$,
  $e\tau$, or $\mu\tau$ in $pp$ Collisions at $\sqrt{s}=8$ TeV with the ATLAS
  Detector}, Phys. Rev. Lett. 115~(3) (2015) 031801.
\newblock \href {http://arxiv.org/abs/1503.04430} {\path{arXiv:1503.04430}},
  \href {http://dx.doi.org/10.1103/PhysRevLett.115.031801}
  {\path{doi:10.1103/PhysRevLett.115.031801}}.

\bibitem{Allanach:2014gsa}
B.~Allanach, A.~R. Raklev, A.~Kvellestad, {Interpreting a CMS excess in
  $lljj+missing$-transverse-momentum with the golden cascade of the minimal
  supersymmetric standard model}, Phys. Rev. D91~(11) (2015) 115022.
\newblock \href {http://arxiv.org/abs/1409.3532} {\path{arXiv:1409.3532}},
  \href {http://dx.doi.org/10.1103/PhysRevD.91.115022}
  {\path{doi:10.1103/PhysRevD.91.115022}}.

\bibitem{Huang:2014oza}
P.~Huang, C.~E.~M. Wagner, {CMS kinematic edge from sbottoms}, Phys. Rev.
  D91~(1) (2015) 015014.
\newblock \href {http://arxiv.org/abs/1410.4998} {\path{arXiv:1410.4998}},
  \href {http://dx.doi.org/10.1103/PhysRevD.91.015014}
  {\path{doi:10.1103/PhysRevD.91.015014}}.

\bibitem{CMS-PAS-SUS-15-011}
{CMS Collaboration}, \href{https://cds.cern.ch/record/2114811}{{Search for new
  physics in final states with two opposite-sign same-flavor leptons, jets and
  missing transverse momentum in pp collisions at $\sqrt{s}=13$\,TeV}},
  CMS-PAS-SUS-15-011, CERN (2015).
\newline\urlprefix\url{https://cds.cern.ch/record/2114811}

\bibitem{ATLAS-CONF-2015-082}
{ATLAS Collaboration}, \href{https://cds.cern.ch/record/2114854}{{A search for
  Supersymmetry in events containing a leptonically decaying $Z$ boson, jets
  and missing transverse momentum in $\sqrt{s}=13~$TeV $pp$ collisions with the
  ATLAS detector}}, ATLAS-CONF-2015-082, CERN (Dec 2015).
\newline\urlprefix\url{https://cds.cern.ch/record/2114854}

\bibitem{ATLAS-CONF-2015-81}
{ATLAS Collaboration}, \href{http://cds.cern.ch/record/2114853}{{Search for
  resonances decaying to photon pairs in 3.2 fb$^{-1}$ of $pp$ collisions at
  $\sqrt{s}$ = 13 TeV with the ATLAS detector}}, ATLAS-CONF-2015-081, CERN
  (2015).
\newline\urlprefix\url{http://cds.cern.ch/record/2114853}

\bibitem{CMS-PAS-EXO-15-004}
{CMS Collaboration}, \href{https://cds.cern.ch/record/2114808}{{Search for new
  physics in high mass diphoton events in proton-proton collisions at $\sqrt{s}
  = 13$ TeV}}, CMS-PAS-EXO-15-004, CERN (2015).
\newline\urlprefix\url{https://cds.cern.ch/record/2114808}

\end{thebibliography}

\end{document}